\documentclass[superscriptaddress,amsmath,amssymb,aps,prx,twocolumn,longbibliography]{revtex4-2}
\usepackage{braket}
\usepackage{graphicx}
\usepackage{amsmath}
\usepackage{amsfonts}
\usepackage{amssymb}
\usepackage{xcolor}
\usepackage{hyperref}
\usepackage{dsfont}
\usepackage{amsthm}
\usepackage{enumitem}
\usepackage{dsfont}
\usepackage{tabularx}

\hypersetup{colorlinks = true, urlcolor = blue, linkcolor=blue, citecolor = blue}
\newtheorem{definition}{Definition}
\newtheorem{theorem}{Theorem}
\newtheorem{lemma}{Lemma}
\newcommand{\pth}{p_{\mathrm{th}}}

\newcommand{\cmmnt}[1]{}
\newcommand{\deff}{d_{\mathrm{eff}}}
\newcommand{\ZDGbulk}{\overline{\mathrm{ZDG}}}
\newcommand{\ZDGbulkC}{\overline{\mathrm{ZDG}}_C}
\newcommand{\calD}{\mathcal{D}}

\begin{document}
\title{Tailoring Dynamical Codes for Biased Noise: The X\texorpdfstring{$^3$}{}Z\texorpdfstring{$^3$}{} Floquet Code}
\author{F. Setiawan}
\email{setiawan.wenming@gmail.com}
\affiliation{Riverlane Research Inc., Cambridge, Massachusetts 02142, USA}
\author{Campbell McLauchlan}\email{campbell.mclauchlan@gmail.com}
\affiliation{Centre for Engineered Quantum Systems, School of Physics, The University of Sydney, Sydney, NSW 2006, Australia}
\affiliation{Riverlane, Cambridge, CB2 3BZ, UK}

\begin{abstract}
We propose the X$^3$Z$^3$ Floquet code, a dynamical code with improved performance under biased noise compared to other Floquet codes. The enhanced performance is attributed to a simplified decoding problem resulting from a persistent stabiliser-product symmetry, which surprisingly exists in a code without constant stabilisers. Even if such a symmetry is allowed, we prove that general dynamical codes with two-qubit parity measurements 
cannot admit one-dimensional decoding graphs, a key feature responsible for the high performance of bias-tailored stabiliser codes. 
Despite this, our comprehensive  simulations show that the symmetry of the X$^3$Z$^3$ Floquet code renders its performance under biased noise far better than several leading Floquet codes. To maintain high-performance implementation in hardware without native two-qubit parity measurements, we introduce ancilla-assisted bias-preserving parity measurement circuits. Our work establishes the X$^3$Z$^3$ code as a prime quantum error-correcting code, particularly for devices with reduced connectivity, such as the honeycomb and heavy-hexagonal architectures.
\end{abstract}

\maketitle

\section{Introduction}

Quantum error correction (QEC)~\cite{Shor1995Scheme,shor1996fault,Calderbank1996Good,devitt2013quantum} should be understood as occurring both in space and time~\cite{gottesman2022opportunities}. Taking advantage of the temporal dimension, Floquet codes~\cite{HH_code,Haah_2022_boundaries_honeycomb,Gidney_honeycomb_2021,Gidney_Planar_honeycomb_2022,Paetznick_floquet_codes_majoranas,higgott_breuckmann_2023_hyperbolic,fahimniya2024_hyperbolic_floquet,Anyon_condensation_2024,gu2023fault,mclauchlan_2024_defectsfloquet,xyzrubycode_2024,hilaire2024enhanced,Davydova_2023_floquet_codes_without_parents,Dua_2024,alam2024dynamical}, or more generally dynamical codes~\cite{Davydova_2024_DA_codes,fu2024error,aitchison2024_automorphisms}, form a large class of quantum error-correcting codes, which can achieve competitive fault-tolerant performance~\cite{HH_code,Gidney_honeycomb_2021,Gidney_Planar_honeycomb_2022,Paetznick_floquet_codes_majoranas,Anyon_condensation_2024,fahimniya2024_hyperbolic_floquet,higgott_breuckmann_2023_hyperbolic,gu2023fault,xyzrubycode_2024,mclauchlan_2024_defectsfloquet,hilaire2024enhanced} while reducing the weights of check measurements performed for the error correction. Several of these codes~\cite{Haah_2022_boundaries_honeycomb,Gidney_Planar_honeycomb_2022,Davydova_2023_floquet_codes_without_parents,Anyon_condensation_2024} also benefit from being defined on a lattice with sparser connectivity than that for the surface code~\cite{kitaev2003fault,Fowler_2012_Surface}: each qubit is only connected to three other qubits. In architectures where two-qubit parity check measurements are native, Floquet codes could achieve higher thresholds~\cite{Paetznick_floquet_codes_majoranas,Gidney_honeycomb_2021,hilaire2024enhanced} and lower space-time overheads~\cite{Gidney_honeycomb_2021,Paetznick_floquet_codes_majoranas} than the surface code. Without requiring additional connectivity, such codes can be deformed around defective components caused by highly noisy qubits or gates~\cite{mclauchlan_2024_defectsfloquet,aasen2023_dead_qubits}. Moreover, dynamical codes allow for implementations of arbitrary Clifford, and even some non-Clifford gates, through low-weight parity check measurements~\cite{Davydova_2024_DA_codes}. In addition, dynamical measurement schedules can result in certain errors being self-corrected in the Floquet-Bacon-Shor code~\cite{alam2024dynamical} -- this code was recently demonstrated in superconducting qubit experiments~\cite{sun2025universal}.

Although Floquet codes (in particular, the honeycomb code) have been studied under various noise models~\cite{Gidney_Planar_honeycomb_2022,QEC_spin_qubits_2024_Wootton}, there have not been any Floquet codes that are specifically tailored for an improved performance under biased noise. A biased noise model is one in which a specific type of error, for example, phase errors, occurs more frequently than other errors, such as bit flip errors. This biased noise is typical to most quantum platforms, for example,  bosonic ``cat" qubits~\cite{Lescanne_2020_cat_qubits,Kerr_cat_qubit}, spin-optical~\cite{de2024spin,hilaire2024enhanced}, neutral atoms~\cite{evered2023high}, quantum-dot spin qubits~\cite{watson2018programmable,tanttu2024assessment} and Majorana qubits~\cite{Aasen2016milestones,McLauchlan_2024_Majorana_codes}. In these platforms, biased noise can arise due to different predominant error mechanisms. For example, by increasing the photon number of the resonators, the cat qubits~\cite{Lescanne_2020_cat_qubits,Kerr_cat_qubit} can be made exponentially protected against the bit-flip error at the expense of only a linear increase in the phase-flip error. For the spin qubits in both spin-optical~\cite{hilaire2024enhanced} and quantum dot architectures~\cite{watson2018programmable,tanttu2024assessment},  the noise is predominantly dephasing due to the short spin coherence time, $T_2$. Moreover, the two-qubit gates of quantum-dot spin qubits have also  been experimentally shown to exhibit phase-biased noise which is caused by the $T_2$ dephasing, non-Markovian error sources, AC-Stark shift, and calibration errors~\cite{tanttu2024assessment}. In neutral-atom qubits, detailed modeling of the experiment shows that the two-qubit gate noise is predominantly phase-biased due to the short Rydberg state $T_2^*$ lifetime caused by finite atomic temperature and light-shift fluctuations of the lasers~\cite{evered2023high}. Lastly, for Majorana qubits, the noise is expected to be biased due to residual coupling between the Majoranas and thermal fluctuations that excite the qubit to an above-gap quasiparticle state~\cite{aasen2025roadmap}.

For enhanced performance, quantum-error-correcting codes need to be designed such that they possess symmetries that can be utilised to simplify the error-syndrome decoding problem given the noise structure~\cite{Higgott2023Improved,Tuckett_bias_1,Tuckett_bias_2,Tuckett_bias_3,XZZX_surface_code,xu2022tailoredxzzxcodesbiased,san2023cellular,Clifford_deformed_SCs,Domain_wall_CC,forlivesi2024quantum}. While there have been a number of proposals for bias-tailored static codes~\cite{sarvepalli2009asymmetric,Tuckett_bias_1,Tuckett_bias_2,XZZX_surface_code,Domain_wall_CC,Srivastava_2022_XYZ2_code,Tuckett_bias_3,Roffe_2023_bias_LDPC,XZZX_surface_code,Clifford_deformed_SCs,xu2022tailoredxzzxcodesbiased,claes2023tailored,Eric2023Tailoring,san2023cellular,Shruti2023High,Sahay2023Tailoring,forlivesi2024logical,forlivesi2024quantum}, designing Floquet codes for high performance under biased noise is still an open problem. Owing to the experimental relevance of biased noise and given the  ease of implementation of Floquet codes, which require only two-qubit parity measurements that can be implemented even in architectures with sparse connectivity, it is therefore imperative to tailor Floquet codes for biased noise and study how the performance of such dynamical codes can be improved.

In this paper, we present the X$^3$Z$^3$ bias-tailored Floquet code, a Clifford-deformed~\cite{Domain_wall_CC} 
version of the Calderbank–Shor–Steane (CSS) Floquet code~\cite{Davydova_2023_floquet_codes_without_parents,Anyon_condensation_2024}. Despite not having a fixed stabiliser group (as static codes have), remarkably the X$^3$Z$^3$ Floquet code possesses a stabiliser-product symmetry under infinitely phase-biased noise, simplifying decoding in biased noise regimes.
We perform an in-depth study of this code, along with the CSS Floquet code~\cite{Davydova_2023_floquet_codes_without_parents,Anyon_condensation_2024}, and two types of honeycomb codes: one introduced by Hastings and Haah~\cite{HH_code} and  the other by Gidney \textit{et al.}~\cite{Gidney_honeycomb_2021}. We simulate all codes under biased-noise models, and find that the X$^3$Z$^3$ code has the best performance. Using a matching decoder, we find that, as the noise changes from fully depolarising to pure dephasing, the X$^3$Z$^3$ Floquet code threshold increases from $1.13 \%$ to $3.09\%$ under a code-capacity noise model and increases from  $0.76\%$ to  $1.08\%$ under a circuit-level noise model mimicking hardware with noisy direct entangling measurements.  Furthermore, we show that its sub-threshold performance is also substantially better under biased noise than other Floquet codes.

Compared to its static counterparts, the X$^3$Z$^3$ Floquet code has an advantage that it can be 
realised using only two-qubit parity check measurements. This makes it particularly suitable for devices with constrained connectivity, such as the honeycomb and heavy-hexagonal lattice (currently IBM's preferred superconducting-qubit architecture)~\cite{hetenyi2024heavy_hex,Chamberland_heavy_hex_2}.
Moreover, we demonstrate that the two-qubit parity measurements of the Floquet code can be performed in a bias-preserving way even in hardware without direct entangling measurements, thus enabling high performance implementation in such devices. Crucially, we show that even in the presence of mid-gate errors which degrade noise bias in the target qubits of conventional CNOT gates~\cite{Puri_2020_bias_preserving}, two of the three required parity-check circuits, built using these conventional gates, can be made fully $Z$-bias preserving, i.e., sustain the $Z$ noise bias on both data qubits, while the other one is partially phase-bias preserving, i.e., protects the $Z$ noise bias in only one of the data qubits.

We argue that other dynamical codes defined on the same architecture, and built from two-body measurements, would likely not have drastically improved performance compared to the X$^3$Z$^3$ Floquet code. To support this argument, we prove that decoding graphs of such dynamical codes under infinitely phase-biased noise have connectivities that are too high for the decoding problem to be reduced to a simple decoding of repetition codes, as is the case for static codes~\cite{XZZX_surface_code,Domain_wall_CC}. This can be understood as resulting from the fact that error syndromes of dynamical codes possess less symmetry than their static code counterparts.

The paper is laid out as follows. In the Secs.~\ref{subsec:dynamical_code}-~\ref{subsec:CSS}, we first review the basics of Floquet codes together with two widely studied examples: honeycomb and CSS Floquet codes. Readers who are already familiar with Floquet codes can skip directly to Sec.~\ref{subsec:X3Z3} in which we discuss our X$^3$Z$^3$ Floquet code. In Sec.~\ref{subsec:persist_sym}, we show that there exists a persistent symmetry in the code's error syndrome under pure dephasing noise that allows for a simplified decoding. Subsequently, in Sec.~\ref{subsec:circuits} we introduce ancilla-assisted bias-preserving parity measurement circuits that allow for high-performance code implementation in devices without native entangling measurements. In Secs.~\ref{subsec:simulation} and~\ref{subsec:subthreshold}, we provide the simulation results for memory experiments. In Sec.~\ref{subsec:theorems}, we give two theorems showing that the decoding hypergraphs of dynamical codes cannot be reduced to 1D graphs. Finally, we conclude and present future research directions in Sec.~\ref{sec:Discussion}. In Sec.~\ref{sec:Methods}, we provide the description of our noise models, details of numerical simulations and proofs of the no-go theorems. In the Supplementary Information, we provide a more thorough review of the basics of honeycomb and CSS Floquet codes, details of our parity check circuits, plots for determining thresholds, details of hyperedges in the honeycomb codes, and results for Floquet codes with elongated dimension and twisted periodic boundary conditions. 

\section{Results}\label{sec:Background}
\subsection{Dynamical and Floquet codes}\label{subsec:dynamical_code}
\begin{figure*}
    \centering
\includegraphics[width=0.8\linewidth]{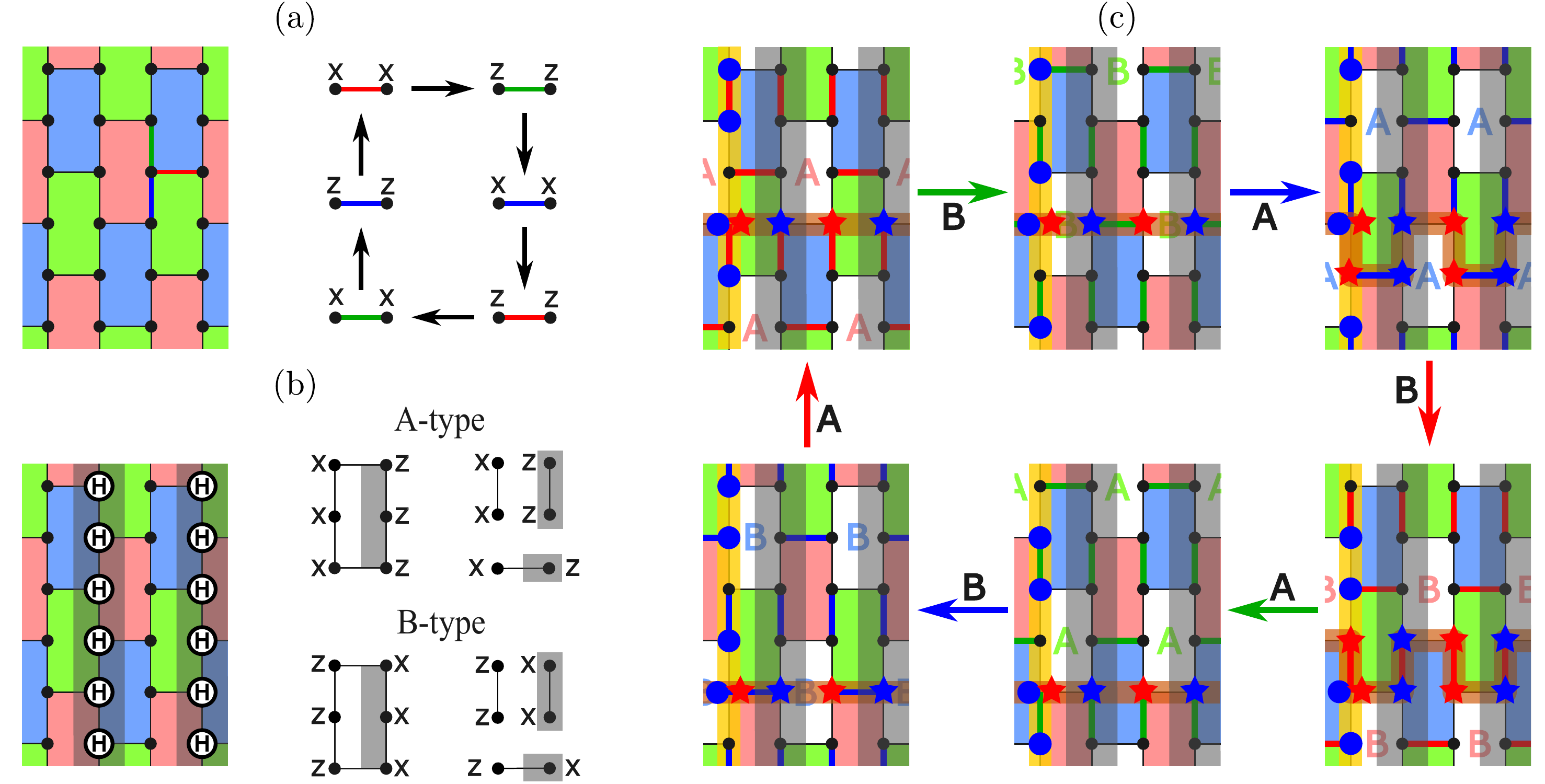}
\caption{\textbf{X$^3$Z$^3$ and CSS Floquet codes on the hexagonal lattice with their stabilisers, checks, and logical operators.} (a) CSS and (b,c) X$^3$Z$^3$ Floquet codes. (a,b) Left: Hexagonal lattice with qubits sitting on vertices and opposite boundaries identified. Plaquettes and edges are assigned one of three colours [red, green, blue; examples are shown in (a)] and one of two types (X- or Z-type for CSS and A- or B-type for X$^3$Z$^3$). 
(a) Right: CSS Floquet code measurement schedule. (b) Left: The X$^3$Z$^3$ Floquet code is obtained from the CSS code via Hadamard gates applied to shaded domains' qubits. Right: Plaquette and edge (check) operators are A- or B-type, depending on their support basis on the shaded/unshaded domains. (c) The X$^3$Z$^3$ Floquet code measurement schedule. Arrows indicate the type and colour of the edge operator measured at each step, where the edges just measured (members of the ISG) are highlighted in the lattice. Uncoloured plaquettes host only a single type of stabiliser, either A or B, indicated by the letters in the plaquettes, while coloured plaquettes host both A- and B-type stabilisers in the ISG. One set of anti-commuting logical operators is shown by yellow and brown strings, where their qubit supports are depicted using big circles and stars, respectively, with the $X$ ($Z$) bases highlighted by red (blue) colouring.  The other set (not shown) is similar to the set shown but offset by three measurement subrounds and with an $X \leftrightarrow Z$ interchange of the qubit supports' bases.}
\label{fig:X3Z3_figure}
\end{figure*}

We begin by defining dynamical and Floquet codes. Here we consider the Floquet codes to be defined on the lattice of a two-dimensional colour code, which is trivalent and three-colourable. 
A trivalent lattice has each vertex incident to three edges, and a three-colourable lattice has every face assigned one of three colours in such a way that there are no two adjacent faces of the same colour. Throughout this paper, we will use the honeycomb lattice as an example of such a lattice (see Fig.~\ref{fig:X3Z3_figure}).

We define qubits as residing on the vertices of the lattice and error-correction is performed by measuring two-qubit operators defined on edges of the lattice (i.e., acting on qubits incident to a given edge). Each edge is coloured the same as the plaquettes it connects [see e.g., the coloured edges in the middle right of the lattice in Fig.~\ref{fig:X3Z3_figure}(a)]. We perform error-correction by measuring edge operators in a particular sequence. After any given subround of edge measurements, the system will be in the $+1$-eigenstate of the operators in an ``instantaneous stabiliser group" (ISG), which will change at each time step. The ISG at time step $t+1$ is defined as
\begin{align}
    \text{ISG}_{t+1} =&\,\langle S, \pm M \; | \; M \in \mathcal{M}_{t+1}, \; S\in \text{ISG}_t, \nonumber\\
    &\quad\text{such that }[S, M] = 0, \; \forall M\in \mathcal{M}_{t+1}\rangle .
\end{align}
In the above, $\mathcal{M}_{t+1}$ is the set of ``check'' measurements performed at time step $t+1$. The ISG$_{t+1}$ also includes ``plaquette stabiliser operators'' $S\in \text{ISG}_t$ which commute with all $M\in \mathcal{M}_{t+1}$. 
The check measurement operators are chosen in such a way that those check operators at time $t$ that have overlapping qubit supports with check operators at $t+1$ anti-commute. For Floquet codes, the measurement sequence is periodic, such that $\mathcal{M}_{t+T} = \mathcal{M}_t$ for some integer $T$. For such a code, we will be performing quantum memory experiments with $mT$ time steps, for some integer $m$. We will refer to $m$ as the number of ``QEC rounds" in the experiment, while we will refer to $mT$ as the number of ``measurement subrounds" in the experiment.

The logical operators at time $t$ are given by $\mathcal{C}(\mathrm{ISG}_{t})/\text{ISG}_t$, where $\mathcal{C}(\text{ISG}_t)$ is the centraliser of ISG$_t$, i.e., the group of Pauli operators commuting with all $S\in\text{ISG}_t$. A (potentially trivial) logical operator ``representative" is some member of $\mathcal{C}(\text{ISG}_t)$. Each lowest-weight nontrivial logical operator representative for the codes considered will be a string-like Pauli operator at each time step [see Fig.~\ref{fig:X3Z3_figure}(c)]. To avoid anti-commuting with the next-subround edge measurements, certain check measurement results along a logical operator's path have to be multiplied into that logical operator. Hence, the logical operators will evolve from one time step to the next. As an illustration, take for example the vertical logical operator immediately after the $g$B checks are applied, i.e., the operator on the yellow string in the top-centre panel of Fig.~\ref{fig:X3Z3_figure}(c). This operator is obtained by multiplying the vertical green-B checks on the vertical logical operator's path [the two green edges on the yellow path shown in the top-centre panel of Fig.~\ref{fig:X3Z3_figure}(c)] into the previous time-step vertical logical operator [the operator shown by the yellow string in the top-left panel of Fig.~\ref{fig:X3Z3_figure}(c)]. Note that after the multiplication, the logical operator commutes with the next check measurements, which are the blue-A edge measurements.

We can detect errors if we can find sets of measurements, called detectors, that always multiply to the value $+1$ in the absence of noise, thus registering no error. Over some number of QEC rounds we will have extracted several detector outcomes. A detector (or decoding) hypergraph is formed by first defining a node for each (independent) detector in the code's history. Subsequently, for each potential fault (e.g., Pauli or measurement errors) that might have occured, a (hyper)edge is drawn between the detectors whose signs are flipped by this fault. Each (hyper)edge is assigned a weight based on the probability of the corresponding error occurring~\cite{Dennis_2002}. The codes we will be examining are amenable to minimum-weight perfect matching decoding~\cite{pymatching}, upon decomposing hyperedges into edges. Given a ``syndrome'' (a set of detectors whose measurements return $-1$ rather than $+1$), the decoder attempts to pair up the triggered detectors to determine a highest-probability correction operation. The decoder succeeds if the error combined with the correction is a trivial logical operator. 

Having discussed the general idea of Floquet codes, we now briefly review the commonly studied examples: two variants of the honeycomb code~\cite{HH_code,Gidney_honeycomb_2021} and the CSS Floquet code~\cite{Davydova_2023_floquet_codes_without_parents,Anyon_condensation_2024}. We will later modify the CSS Floquet code to achieve the bias-tailored X$^3$Z$^3$ Floquet code. More details of the honeycomb and CSS Floquet codes are presented in Supplementary Sec.~I of the Supplementary Information~\cite{Supplementary_material}. \\

\subsection{Honeycomb codes}\label{subsec:honeycomb_codes}
We begin by first discussing the honeycomb codes. The first variant we review is due to Gidney \textit{et al.}~\cite{Gidney_honeycomb_2021}, which we call the P$^6$ Floquet code, since its plaquette operators are six-body operators of the form $P^{\otimes 6}$ for $P=X,Y,Z$. We define edge operators of three types: on red edges we define an $XX$ operator, on green edges a $YY$ operator and on blue edges a $ZZ$ operator. We measure edge operators in the periodic sequence $r\rightarrow g\rightarrow b$. When this code is defined with periodic boundary conditions it stores two logical qubits (it is equivalent to the toric code concatenated with a two-qubit repetition code at each time step~\cite{HH_code}). The code's logical operators evolve through the measurement cycle (see Supplementary Sec.~IA of the Supplementary Information~\cite{Supplementary_material}). While the measurement sequence has period 3, the logical operators only return to their initial values (up to signs) with period 6. 

We define one stabiliser operator for each plaquette, where red, green, and blue plaquettes host the $X^{\otimes 6}$, $Y^{\otimes 6}$, and $Z^{\otimes 6}$ operators, respectively.  These plaquette operator eigenvalues are inferred from edge measurements in two consecutive subrounds. Detectors are formed  from consecutive plaquette operator measurements.

The second honeycomb code variant, which we call the XYZ$^2$ honeycomb code, is due to Hastings and Haah~\cite{HH_code}. It differs from the P$^6$ code by single-qubit Clifford rotations acting on the qubits. While the  XYZ$^2$ code's edges are still coloured the same as the P$^6$ edges, the XYZ$^2$ code's edge operators have their Pauli bases defined according to their orientations within each T junction, i.e., horizontal edges are of $Z$ type while the $X$ and $Y$ checks are respectively those edges which are 90$^{\circ}$ clockwise and counter-clockwise rotated from the horizontal edges (see Supplementary Sec.~IA of the Supplementary Information~\cite{Supplementary_material}). All plaquettes have the same stabiliser operator, i.e., the XYZ$^2$ operator. While the logical operators of the XYZ$^2$ code have the same qubit supports as those of the P$^6$ code, the qubit support bases of the XYZ$^2$ logical operators are not uniform throughout, but involve $X$, $Y$, and $Z$ Paulis~\cite{HH_code}. We note that even though these honeycomb code variants have been studied by several works  in the literature (e,g.,~Refs.~\cite{HH_code,Gidney_honeycomb_2021,Haah_2022_boundaries_honeycomb,Gidney_Planar_honeycomb_2022}), there have not been any studies comparing the performance of these two codes under biased noise. \\

\subsection{CSS Floquet code}\label{subsec:CSS}
Having discussed the honeycomb codes, we now give a brief review of another type of Floquet code, the CSS Floquet code, which was first presented in Refs.~\cite{Davydova_2023_floquet_codes_without_parents,Anyon_condensation_2024}. We refer the readers to Supplementary Sec.~IB of the Supplementary Information~\cite{Supplementary_material} for more details on the code. The CSS Floquet code is defined on the same honeycomb lattice. We show the code's measurement cycle in Fig.~\ref{fig:X3Z3_figure}(a). We measure operators defined on edges using the $r\rightarrow g\rightarrow b$ cycle, but alternate between measuring $XX$ and $ZZ$ operators on these edges. Hence, this code has a period-6 measurement cycle, and its logical operator evolutions also have period-6 (see Supplementary Fig.~2 of the Supplementary Information~\cite{Supplementary_material} for an illustration). Even though the honeycomb code's measurement schedule has only a period of 3, for consistency, we will define 1 QEC round to be 6 measurement subrounds for all codes studied in this paper. The check operators, stabilisers and logicals all are either $X$- or $Z$-type and, for this reason, the code is called the CSS Floquet code. 

Unlike the honeycomb codes, the CSS Floquet code has no persistent stabiliser operators. Instead, each plaquette stabiliser is repeatedly reinitialised and measured out, with subsequent measurement values compared to form a detector. In contrast to the honeycomb codes where the values of the plaquette stabilisers are inferred from measurements of check operators in two consecutive subrounds, here the plaquette operator measurements take only a single subround~\cite{Anyon_condensation_2024,Davydova_2023_floquet_codes_without_parents}. Because of these single-step measurements, at each measurement subround, there is always one type of plaquette operator that anti-commutes with the check measurements, and hence their values are undetermined (see Supplementary Sec.~IB of the Supplementary Information~\cite{Supplementary_material}). These ``inactive" plaquettes, however, will be reinitialised at the next measurement subround. As a result, at each time step, the ISG contains $X$-type and $Z$-type stabiliser operators defined on two of the three colours of plaquette and \textit{either} an $X$-type \textit{or} a $Z$-type operator defined on the other colour. For instance, after measuring the red-$X$ checks, the ISG contains $X^{\otimes 6}$ and $Z^{\otimes 6}$ operators on both blue and green plaquettes, but only $X^{\otimes 6}$ red plaquette operators, since $Z^{\otimes 6}$ red plaquette operators anti-commute with the red-$X$ checks (see Supplementary Sec.~IB of the Supplementary Information~\cite{Supplementary_material} for more details).

The CSS Floquet code is naturally suited to a minimum-weight perfect matching (MWPM) decoding, since single-qubit ($X$ or $Z$) Pauli and measurement errors all lead to graph-like syndromes~\cite{Anyon_condensation_2024}: they each trigger a pair of detectors. There are two decoding graphs formed from the $Z$-type and $X$-type detectors. Only $Y$ errors form hyperedges that need to be decomposed into edges in the two detector graphs.\\

\subsection{X\texorpdfstring{$^3$}{}Z\texorpdfstring{$^3$}{} bias-tailored Floquet code}\label{subsec:X3Z3}
We now describe the X$^3$Z$^3$ bias-tailored Floquet code, which is shown in Figs.~\ref{fig:X3Z3_figure}(b) and~\ref{fig:X3Z3_figure}(c). This code is related to the CSS Floquet code by Hadamard gates applied to the qubits in alternating strips [i.e., the grey strips in Fig.~\ref{fig:X3Z3_figure}(b)] along vertical non-trivial cycles of the lattice, and is a Floquetified version of the domain wall colour code~\cite{Domain_wall_CC}. As in the CSS Floquet code, the X$^3$Z$^3$ code also has two types of plaquettes and edges: one originates from the Pauli $X$ and the other from the Pauli $Z$ plaquettes and edges in the CSS code before the Hadamard deformation. We refer to these modified operators as A-type and B-type, respectively; these are shown in Fig.~\ref{fig:X3Z3_figure}(b). The measurement sequence is analogous to the CSS Floquet code sequence: $r\text{A}\rightarrow g\text{B} \rightarrow b\text{A} \rightarrow r\text{B} \rightarrow g\text{A} \rightarrow b\text{B}$, where $c\text{A}$ represents the measurement of A-type check operators along $c$-coloured edges, and similarly for $c\text{B}$. Just as with the CSS Floquet code~\cite{Anyon_condensation_2024}, this code can also be defined on a planar lattice with boundary. 

\begin{figure}
\centering
\includegraphics[width=\linewidth]{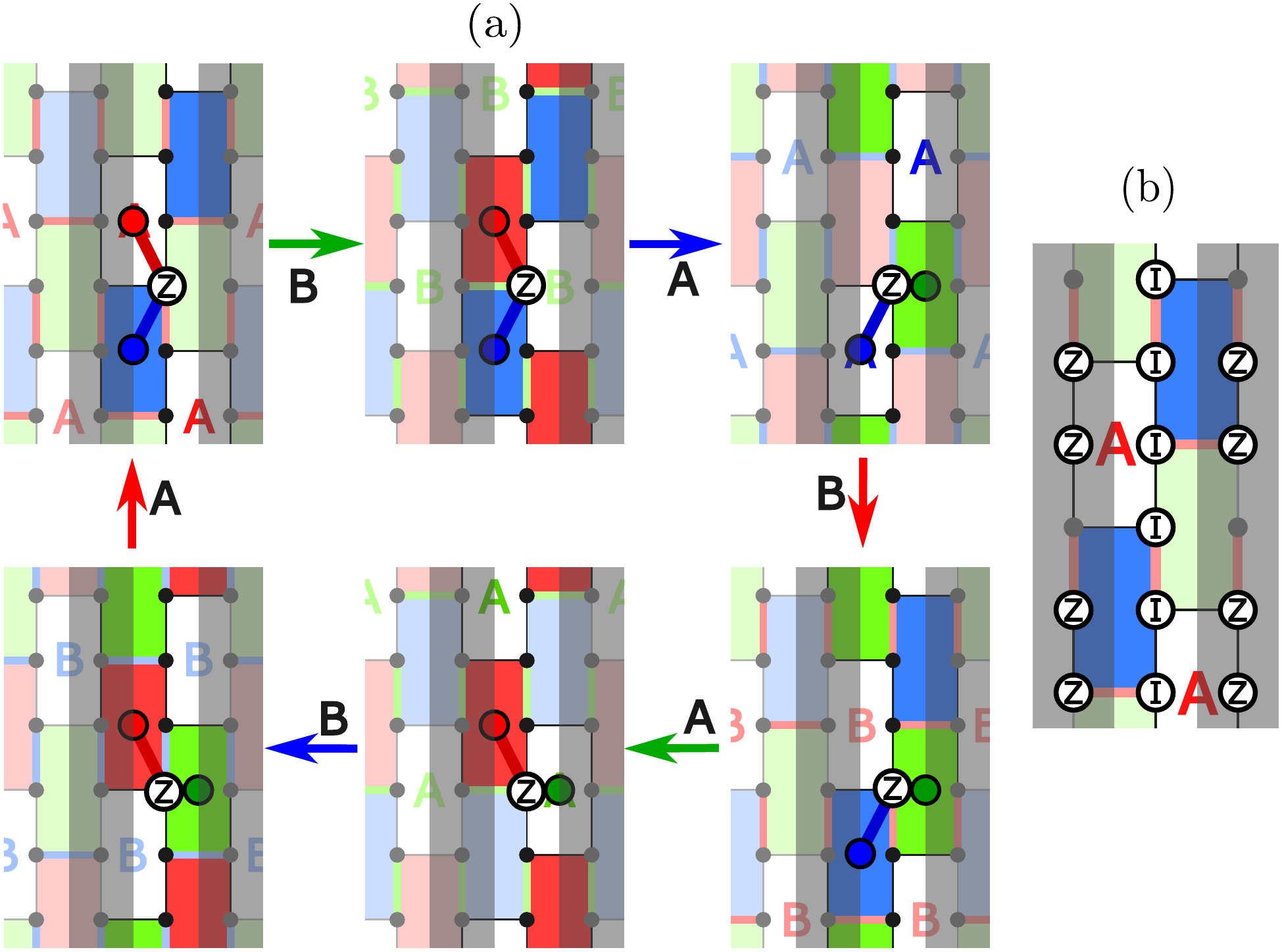}
\caption{\textbf{Persistent stabiliser-product symmetry of the X$^3$Z$^3$ Floquet code leads to improved performance under $Z$-biased noise.} (a) An example of such a symmetry along a single unshaded strip of the lattice, together with a Pauli-$Z$ error, is shown throughout the measurement cycle. Arrows indicate the type of check ($c\text{A}$ or $c$B for some colour $c$) measured at each subround. A-type plaquette operators whose product forms the symmetry at that subround are highlighted in darker colours. Plaquettes indicated by coloured dots are the locations of syndromes that would be triggered if the $Z$ error shown were to occur at that measurement subround.
(b) An example of a stabiliser-product symmetry just before $g$B checks are measured is the product of red and blue A-type plaquettes shown. Since this product is the identity along the vertical unshaded strip and commutes with $Z$ errors on all qubits, the syndromes/anyons [coloured dots in (a)] must appear in pairs along the vertical strips.}

\label{fig:X3Z3_Floquet_code_sequence_symmetry}
\end{figure}

\begin{figure}
    \centering
    \includegraphics[width=\linewidth]{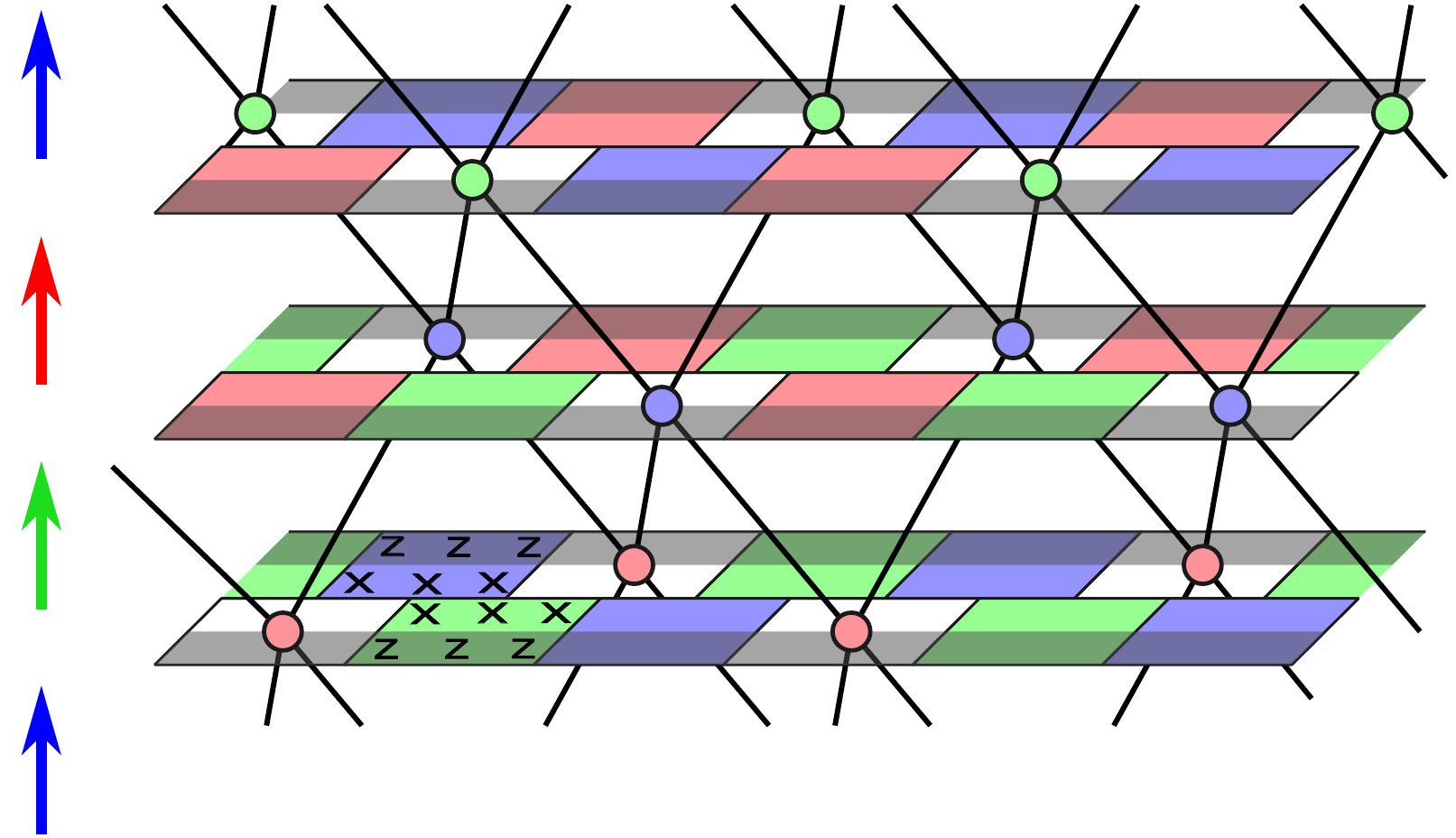}
    \caption{\textbf{The 2D A-type detector graph of the X$^3$Z$^3$ Floquet code under pure dephasing noise.} Only a single vertical strip [e.g., that supporting the stabiliser-product symmetry in Fig.~\ref{fig:X3Z3_Floquet_code_sequence_symmetry}(b)] is shown. The illustrated part of the graph is disconnected from those of  neighbouring strips. Arrows represent only the A-type measurements. The B-type subrounds are not shown, since they do not influence the detectors in this graph. Each node (representing a detector) is placed in the centre of a plaquette and at the subround in which the corresponding detector is formed. For example, the bottom layer corresponds to the blue A-check measurements, at which time the red A-type plaquettes are measured.}
    \label{fig:Detector_graph}
\end{figure}

The decoding of the X$^3$Z$^3$ Floquet code under $Z$-biased noise is simplified by the presence of a symmetry in the decoding graphs, which we will explain below. This is a space-time analogue of the symmetries present in bias-tailored static codes such as the XZZX code~\cite{XZZX_surface_code} and the domain wall colour code~\cite{Domain_wall_CC}. In such codes, under the infinitely phase-biased code-capacity noise model (with only single-qubit $Z$ errors, for example), syndromes are forced to come in pairs along one-dimensional (1D) strips of the lattice. This results from strips of stabilisers multiplying together to the identity along one domain and  commuting with all $Z$ errors on all qubits:
these are 1D symmetries of the stabiliser code under infinitely phase-biased noise. We will refer to (plaquette) stabilisers in the ISG that are flipped by an error as anyons. In bias-tailored static codes, anyons can propagate within strips but cannot move outside the strip without changing their Pauli type. We will show that there also exists a similar symmetry in the bias-tailored Floquet code.

As in the CSS Floquet code, there are two disjoint decoding graphs for the X$^3$Z$^3$ Floquet code: the A-type and B-type graphs, whose edges correspond to single-qubit Pauli errors. In even measurement subrounds, we perform B-type measurements and form detectors for the B-type graph while in odd measurement subrounds we only form detectors for the A-type graph. Note that under a more complicated noise model, such as one that includes measurement or two-qubit errors, edges can exist between the A- and B-type graphs and they are no longer disjoint.\\

\subsection{Persistent stabiliser-product symmetry and two-dimensional decoding of the X$^3$Z$^3$ Floquet code under biased noise}\label{subsec:persist_sym}
There is no constant stabiliser group when viewing the X$^3$Z$^3$ Floquet code as a subsystem code (in a subsystem code, the stabiliser group is defined as the centre of the gauge group, which, in our case, is generated by all edge operators)~\cite{Anyon_condensation_2024,Davydova_2023_floquet_codes_without_parents}. Surprisingly, despite this fact, at every time step there do exist operators that form a symmetry under pure dephasing noise (without measurement errors). An example of such a symmetry is shown in Fig.~\ref{fig:X3Z3_Floquet_code_sequence_symmetry}. It is a symmetry on one of the unshaded domains and it is formed by the product of A-type plaquettes.
A similar symmetry can also be found on the shaded strips which are formed by the product of B-type plaquettes. 

To see why such symmetries exist, consider one particular time step, e.g., just before the green B-type measurements, and one vertical unshaded domain in the lattice. This is shown in Fig.~\ref{fig:X3Z3_Floquet_code_sequence_symmetry}(b). As depicted, the product of red and blue A-type plaquettes in the ISG is the identity on all qubits in the domain, and yields Pauli $Z$ operators on some qubits in the neighbouring shaded vertical domains. To be more concrete, consider the plaquettes that can detect $Z$ errors along the unshaded vertical domain $\calD$ at time step $t$. These are the A-type plaquette operators with support in $\calD$. For a plaquette stabiliser operator $S$, which is measured in a time step $t'> t$, to form a detector (i.e., a node in the detector graph), it has to be in ISG$_t$ and ISG$_{t+1}$, since it must commute with at least the next measurement subround. We can define ISG$_{(t,t+1)}^{A,\calD}$ as the subgroup of ISG$_t\cap \text{ISG}_{t+1}$ formed from A-type plaquettes with non-trivial supports on domain $\calD$. We then have
\begin{align}
    \prod_{S\in \text{ISG}_{(t,t+1)}^{A,\calD}} S = \left( \bigotimes_{k\in \calD} I_k \right)\left( \bigotimes_{j\in \bar{\calD}_{Z,t}} Z_j\bigotimes_{j'\in\bar{\calD}_{I,t}} I_{j'}\right),
\end{align}
where $\bar{\calD}_{Z,t}$ and $\bar{\calD}_{I,t}$ are some subsets of qubits in the domains adjacent to domain $\calD$ on which the product of the A-type plaquette stabilisers yields Pauli $Z$ and the identity, respectively [for the example given in Fig.~\ref{fig:X3Z3_Floquet_code_sequence_symmetry}(b), the domain $\cal{D}$ is the unshaded vertical strip, the domains  $\bar{\calD}_{Z,t}$ and $\bar{\calD}_{I,t}$ comprise the qubits in the neighbouring shaded vertical strips with $Z$-circled label and no label, respectively]. We use a subscript $t$ in the notations $\bar{\calD}_{Z,t}$ and $\bar{\calD}_{I,t}$ to indicate that the qubits belonging to these two subsets depend on the time step $t$. Note that we suppress the identity factors outside of $\calD$ and its adjacent domains. In particular, by restricting this operator to domain $\calD$, we have the following persistent stabiliser-product symmetry:
\begin{align}
     \prod_{S\in \text{ISG}_{(t,t+1)}^{A,\calD}} S |_\calD &= \bigotimes_{k\in \calD} I_k,
\end{align}
which implies that the error syndrome obeys a conservation law in the domain $\calD$: the syndrome comes in pairs along this strip.
At each subround, there also exists a similar symmetry on the shaded domains, which is formed by the product of B-type plaquettes.
Therefore, for the pure $Z$ Pauli noise model, we can perform minimum-weight perfect matching decoding \textit{within} each domain.

To explain this more concretely, we use the anyon picture. These anyons, shown in Fig.~\ref{fig:X3Z3_Floquet_code_sequence_symmetry}(a), are interpreted as having locations given by the detectors that \emph{would} be flipped if the corresponding error were to occur at that time step. For example, if the $Z$ error shown in Fig.~\ref{fig:X3Z3_Floquet_code_sequence_symmetry}(a) occurs just before the green B-check measurements (top-left panel of the figure), it will trigger a red A detector after two measurement subrounds and a blue A detector after a further two subrounds.  As a result of each of the symmetries, Pauli-$Z$ errors create anyons in pairs along each vertical strip. An example of such anyons, which are formed due to a Pauli $Z$ error at a particular measurement subround, is shown in Fig.~\ref{fig:X3Z3_Floquet_code_sequence_symmetry}(a). While the plaquette locations of the anyons, triggered by the $Z$ error occurring on the same qubit, change every other subround, they always respect the symmetry, i.e., are aligned along the domain (see Fig.~\ref{fig:X3Z3_Floquet_code_sequence_symmetry}).

Unlike the symmetry in static stabiliser codes, this stabiliser-product symmetry does not allow for 1D decoding even when measurement errors are absent. While in static codes, only measurement errors produce ``time-like" edges, in Floquet codes, even Pauli errors produce time-like edges (between detectors formed at different times). To demonstrate this, we display in Fig.~\ref{fig:Detector_graph} a portion of the A-type detector graph under infinitely phase-biased single-qubit noise with no measurement errors. As can be seen, there is a disconnected subgraph defined along one unshaded domain of the code lattice. Even considering this simple noise model, the decoding graph in the infinite bias regime is two-dimensional (i.e., the graph is planar). This results from the fact that neighbouring plaquettes are measured at different times. We emphasise that although the above discussion is based on $Z$ errors, the same analysis also holds for $X$ errors because $X$ and $Z$ operators are interchangeable for the X$^3$Z$^3$ code. As a result, the performance of the X$^3$Z$^3$ code under $X$-biased noise is expected to be the same as that under the $Z$-biased noise model.\\

\subsection{Bias-preserving parity-check circuits}\label{subsec:circuits}
While two-body measurements are native to certain architectures, for example, Majorana~\cite{Paetznick_floquet_codes_majoranas,Karzig_2017_scalable,McLauchlan_2022_Majorana_Comp} and photonic~\cite{Paesani2023High,hilaire2024enhanced} platforms, most architectures require quantum circuits to carry out such measurements. To maintain the high-performance of bias-tailored codes in these hardware, the parity-check measurement circuits need to be constructed in a phase-bias preserving manner, such that they do not propagate frequently-occurring errors to rarely-occurring errors. To this end, we design two-qubit parity check circuits that preserve the $Z$ bias on data qubits. That is, the probability of $X$ and $Y$ errors on data qubits after these circuits remains small (proportional to single-qubit $X$ or $Y$ error probabilities). Our bias-preserving parity check measurement circuits are constructed by using an ancilla circuit connecting two data qubits, which can be realised even in devices with minimal connectivity, such as the heavy-hexagonal layout~\cite{Chamberland_heavy_hex_2,hetenyi2024heavy_hex}.

\begin{figure}
    \centering
    \includegraphics[width=\linewidth]{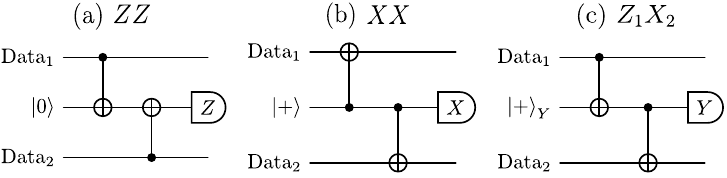}
    \caption{\textbf{Depth-2 parity check measurement circuits.} For the circuits to be bias-preserving, the CNOT gates in the circuits need to be also bias-preserving. Shown are the circuits for (a) $Z_1Z_2$, (b) $X_1 X_2$, and (c) $Z_1 X_2$ checks circuits, where the ancilla qubit is reset and later measured in the $Z$, $X$ and $Y$ bases, respectively.}
\label{fig:depth_2_cts}
\end{figure}

For a circuit to be bias preserving, it can be constructed using only gates that do not change the error type under conjugation. For example, one way to construct bias-preserving circuits is to use only CNOT gates in the measurement circuits as they only propagate errors to others with the same Pauli type. For the X$^3$Z$^3$ code, we need to construct bias-preserving circuits for three different kinds of  parity checks: $ZZ$, $XX$ and $ZX$. All these three circuits can be constructed using only CNOT gates as shown in Fig.~\ref{fig:depth_2_cts}. As depicted, these circuits also include resets and measurements of the ancilla qubit in the three Pauli bases, i.e., $Z$, $X$ and $Y$ bases for $ZZ$, $XX$ and $ZX$ checks, respectively. Note that the circuits are still phase-bias preserving even if the resets/measurements in the $X$ and $Y$ bases are compiled in terms of resets/measurements in the $Z$ basis with additional single-qubit Clifford gates. This is because a $Z$ error on any qubit at any time step in the compiled circuits still propagates to data qubits as a $Z$ error, or as a check operator about to be measured.

The parity-check circuits shown in Fig.~\ref{fig:depth_2_cts}, however, may not preserve the noise bias if errors happen during the CNOT gates. This is because a $Z$ error occurring on the target qubit during the application of a conventional (non-bias preserving) CNOT gate may result in a combination of $Z$ and $X$ errors on the target qubit after the CNOT gate~\cite{Puri_2020_bias_preserving}. One way to remedy this problem is to use bias-preserving CNOT gates~\cite{Puri_2020_bias_preserving}. These gates, however, are available only in specific platforms such as cat qubits. Here, we propose a more general approach to constructing $Z$-bias preserving parity check circuits without requiring the bias-preserving gates. While our proposed circuits can preserve only a specific type of noise bias, i.e., the $Z$ bias, these circuits are built using conventional (which may not be phase-bias preserving) two-qubit gates and thus can be implemented in  many different architectures. 

\begin{figure}
    \centering
    \includegraphics[width=\linewidth]{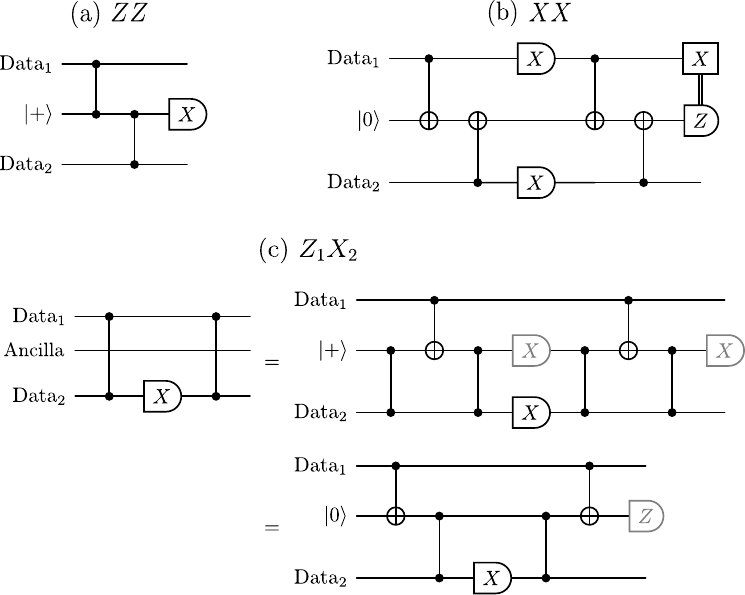}
    \caption{\textbf{Parity-check measurement circuits built from   conventional two-qubit gates and constructed in ways that the $Z$-noise bias on the data qubits is preserved under mid-gate errors.} Shown are  (a) $ZZ$, (b) $XX$, and (c) $Z_1X_2$ check circuits. The circuits in (a) and (c) preserve the $Z$ noise bias on data qubits even in the presence of mid-gate errors. The circuit in (b) includes a classically controlled $X$ gate targeting one of the data qubits, which is non-$Z$ bias preserving. Data qubits are labelled by Data$_1$ and Data$_2$, while the middle qubit in each circuit is a measurement ancilla. In (c), the first equality can be checked by noting that the depth-3 circuit before and after the mid-circuit measurement performs a next-nearest-neighbour CZ gate, if the measurement ancilla is initially in the $\ket{+}$ state. The second equality can be checked by commuting all gates past the measurement. The grey measurements can be optionally included to provide flag information.}    \label{fig:Measurement_circuits}
\end{figure}

Our proposed phase-bias preserving parity-check circuits are depicted in Fig.~\ref{fig:Measurement_circuits}. For the $ZZ$ and $ZX$ check measurement circuits, they are constructed using either CZ gates or CNOT gates targeting ancilla qubits, which propagate Pauli errors on the ancilla qubits only as $Z$ errors on the data qubits. In other words, these circuits do not leave any residual Pauli $X$ or $Y$ errors occurring with probability $\mathcal{O}(p_Z)$ and therefore do not degrade the $Z$ noise bias on the data qubits.  This is true even if we consider mid-gate Pauli errors (see Supplementary Sec.~II in the Supplementary Information~\cite{Supplementary_material} for details). The $XX$ measurement circuit shown in Fig.~\ref{fig:Measurement_circuits}(b), however, is not fully $Z$-bias preserving in the presence of mid-gate errors. This circuit degrades the $Z$ bias on one of the data qubits, i.e., the data qubit that is the target of the classically-controlled $X$ gate. (Note that instead of implementing the classically controlled $X$ gate, we may alternatively perform a CNOT gate, with the control being the ancilla qubit and target being data qubit 1, immediately preceding the ancilla measurement.) However, since the $XX$ parity-check circuit shown in Fig.~\ref{fig:Measurement_circuits}(b) degrades the $Z$-bias on only one data qubit, it has a better $Z$-bias-preserving performance compared to the $XX$ check circuit in Fig.~\ref{fig:depth_2_cts}(b), which degrades the $Z$-bias on both data qubits due to the fact that both data qubits are used as target qubits of the CNOT gates in the circuit.

We now elaborate on how the parity-check circuits work. We begin by noting that while the $ZZ$ parity check measurements [Fig.~\ref{fig:Measurement_circuits}(a)] are performed by reading out the ancilla qubits like in standard syndrome extraction circuits, the  $XX$ and $ZX$ parity values [Figs.~\ref{fig:Measurement_circuits}(b) and~\ref{fig:Measurement_circuits}(c)] are obtained by measuring the data qubits after some two-qubit gates are applied on them. These two-qubit gates are required for the parity measurements so as not to reveal each individual data qubit's state.  For the $XX$ parity-check circuit, the two mid-circuit $X$-measurements in Fig.~\ref{fig:Measurement_circuits}(b), are equivalent to reading out the $X_1X_a$ and $X_2X_a$ parity values, where the Pauli operator $X_a$ acts on the ancilla qubit which is prepared in the $\ket{0}$ state. Hence, the product of these two measurement outcomes is the required $X_1X_2$ parity check outcome. We can disentangle the ancilla qubit by performing a final $Z_a$ measurement (which requires correcting a potential bit flip on one of the data qubits with the classically controlled $X$ gate), or by using a CNOT gate (not shown) before the $Z_a$ measurement (see Supplementary Sec.~II of the Supplementary Information~\cite{Supplementary_material} for details). 

The mixed-type $Z_1 X_2$ parity circuit can be constructed using CZ gates between the data qubits, as shown in Fig.~\ref{fig:Measurement_circuits}(c). If the connections between the data qubits are not available in the device such as  in hardware with a heavy-hexagonal lattice, these CZ gates have to be implemented using ancilla qubits in-between the data qubits. We show such an implementation of next-nearest-neighbour CZ gates in the first equality of Fig.~\ref{fig:Measurement_circuits}(c) (this is adapted from Ref.~\cite{hetenyi2024heavy_hex}).
This circuit also includes optional measurements, coloured in grey, which provide ``flag information" for detecting $Z$ errors on the ancilla qubit that may have propagated to the data qubits. Such flag measurements can assist in decoding~\cite{Chamberland_heavy_hex_2}. In the second equality of Fig.~\ref{fig:Measurement_circuits}(c), we reset the ancilla qubit in the $\ket{0}$ state instead of the $\ket{+}$ state, allowing for a shorter-depth circuit to implement the $Z_1X_2$ measurement. This comes at the cost of removing one flag measurement. To see that this circuit works as intended, we can commute the two-qubit gates past the $X_2$ measurement, resulting in a $Z_1Z_aX_2$ parity measurement. From the result of this measurement, together with the fact that $Z_a$ value is set to be $+1$ at the ancilla reset, we can then infer the value of $Z_1X_2$.

We emphasise that, even if the $X$ measurements in the parity-check circuits above are implemented using $Z$ measurements sandwiched by Hadamard gates, this does not degrade the bias, since a $Z$ error occurring between those Hadamard gates is harmless. Specifically, a $Z$ error happening before the measurement is immediately absorbed by the $Z$ measurement without flipping its outcome, and a $Z$ error after the measurement is a stabiliser of the state (up to a sign), so does nothing. 
Moreover, the circuit will not change the noise bias significantly even if the Hadamard gates are noisy, since in many architectures, the single-qubit gate errors are not the predominant error source and are usually much smaller than the two-qubit gate errors~\cite{krantz2019quantum,bruzewicz2019trapped,burkard2023semiconductor,wintersperger2023neutral}. 

In Supplementary Sec.~II of the Supplementary Information~\cite{Supplementary_material}, we show in more detail how the parity check circuits preserve the noise bias. We also show that the phase-bias preserving parity-check circuits presented in this paper have optimal circuit depth.\\

\subsection{Memory experiment simulation}\label{subsec:simulation}
To study the performance of our proposed X$^3$Z$^3$ code, we perform quantum memory experiment simulations under two biased noise models. These noise models are the generalisations of the standard
code-capacity and standard depolarising entangling measurement (SDEM3) noise models (see the Methods section for details of the noise models). We simulate the X$^3$Z$^3$ code along with three other codes: CSS, P$^6$ and XYZ$^2$ Floquet codes, with varying degrees of noise bias $\eta$ (see the Methods section for details of our numerical simulations). The noise bias $\eta = p_Z/(p_X+p_Y)$ is defined as the ratio of the $Z$-error $p_Z$ to other errors where the total physical error rate is $p = p_X + p_Y+p_Z$. As the noise asymmetry increases, the noise bias $\eta$ increases from $\eta = 0.5$, which corresponds to fully depolarising noise ($p_X = p_Y = p_Z = p/3$), to $\eta = \infty$, corresponding to a pure $Z$-biased noise ($p_Z = p$, and $p_X = p_Y =0$).
\\ 

\begin{figure}[t]
\includegraphics[width=\linewidth]{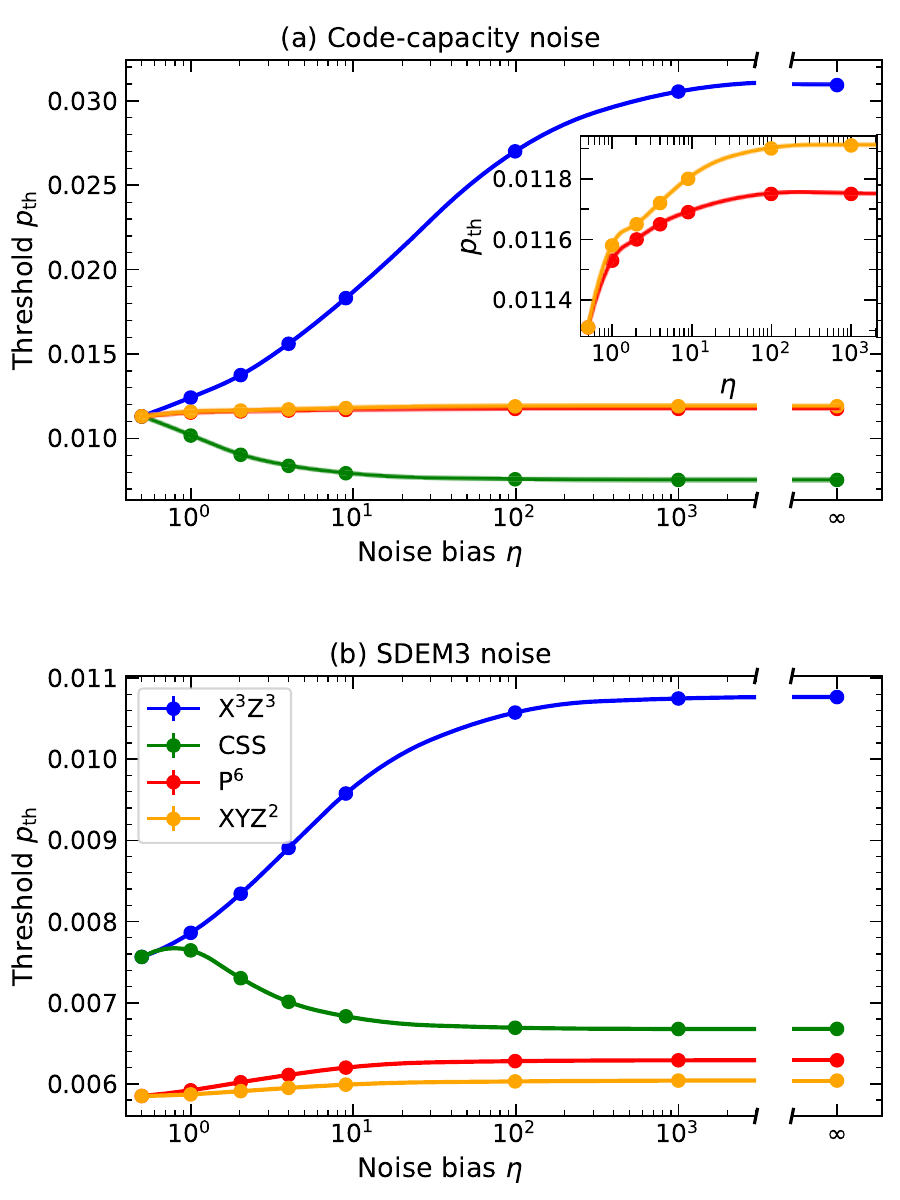}
	\caption{ \textbf{Thresholds ($p_{\mathrm{th}}$) of different Floquet codes as a function of noise bias $\eta$.} Codes studied are X$^3$Z$^3$ (blue), CSS (green), P$^6$ (red) and XYZ$^2$ (orange). Results are computed for two different noise models: (a) code-capacity and (b) SDEM3. (a) Inset: Zoom-in threshold plots for the P$^6$ (red) and XYZ$^2$ (orange) codes. 
  For better visualisation, we fit all curves with quadratic splines.  }\label{fig:Threshold}
 \end{figure}

\begin{figure}[t]
\includegraphics[width=\linewidth]{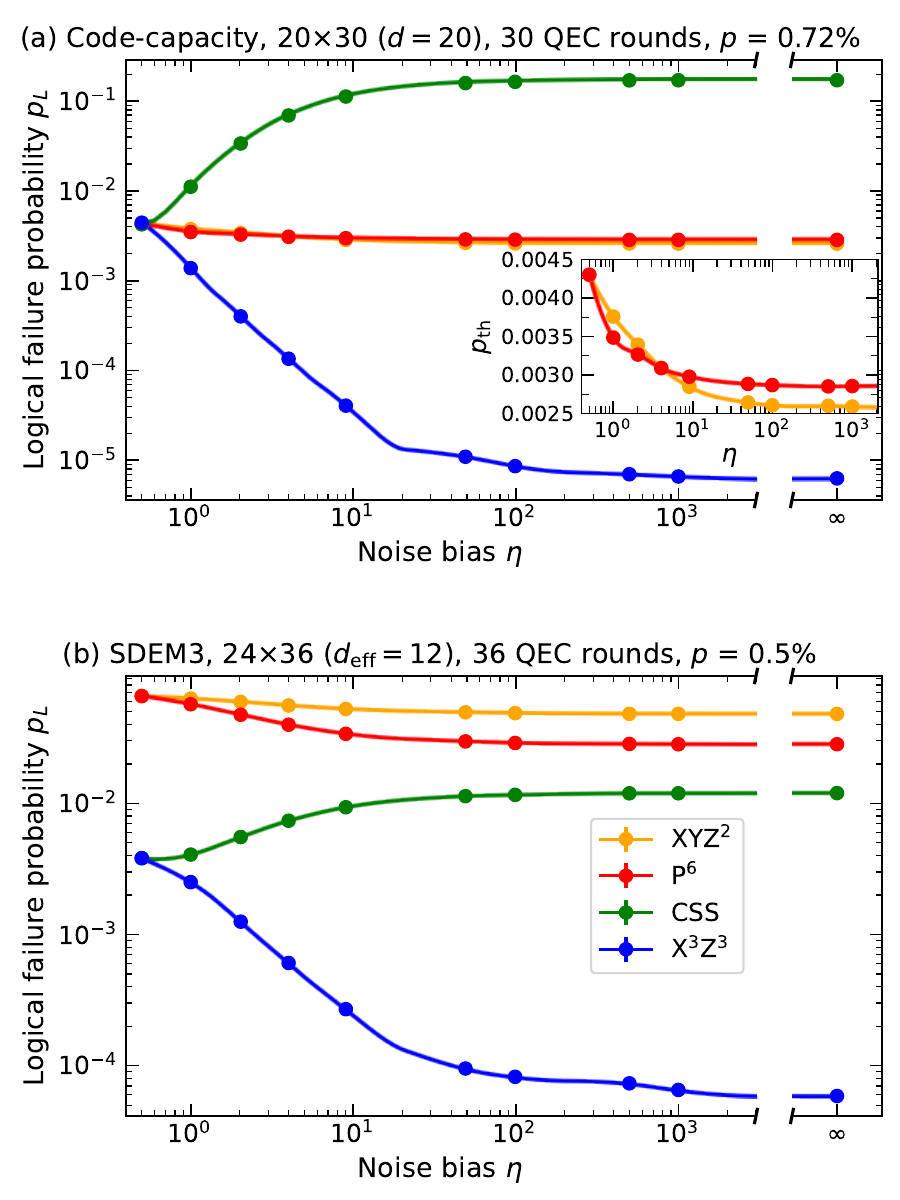}
	\caption{\textbf{Sub-threshold logical failure probability $p_L$ of different Floquet codes as a function of noise bias $\eta$.} Codes studied are X$^3$Z$^3$ (blue), CSS (green), P$^6$ (red) and XYZ$^2$ (orange). Results are computed for two different noise models: (a) code-capacity and (b) SDEM3. Inset: Zoom-in subthreshold performance plots of the P$^6$ (red) and XYZ$^2$ (orange) codes. Results are calculated using (a) lattice size $20 \times 30$ ($d = 20$) for 30 QEC rounds with $p = 0.72\%$ and (b) lattice size $24 \times 36$ ($\deff = 12$) for 36 QEC rounds with $p = 0.5\%$. Each data point is averaged over $10^6-10^8$ Monte Carlo shots.  For better visualisation, we fit all curves with quadratic splines. For each of the noise models, we choose the largest code distance and the smallest subthreshold physical error rate that can be simulated given our computational resources.  }\label{fig:subthreshold_performance}
 \end{figure}

\noindent
\textbf{Thresholds}\\
Figure~\ref{fig:Threshold} shows the thresholds of all codes for various levels of noise bias $\eta$ calculated for (a) code-capacity and (b) SDEM3 noise models. Each threshold is obtained from the intersection of the logical failure probabilities $p_L$ vs physical error rate $p$ curves of different code distances $\deff \equiv d/2 = 6,8,10$ and 12. 
Here, the effective distance $\deff$ is defined as the minimum number of faults under SDEM3 depolarising noise that produce a logical error. This is half of the distance $d$ of the code-capacity noise model. For the methods and plots used to determine the thresholds, see the Methods section and Supplementary Figs.~5-8 of the Supplementary Information~\cite{Supplementary_material}. We also provide the threshold data in Ref.~\cite{StimCircuitsZenodo}.

Since the code-capacity noise model, which considers only single-qubit noise, is a more benign model than the SDEM3 noise model, which also includes two-qubit and measurement errors, the code performance calculated under the code-capacity noise is better than that of the SDEM3 noise. For both noise models, as shown in Figs.~\ref{fig:Threshold}(a) and~\ref{fig:Threshold}(b), the performance of the X$^3$Z$^3$ code becomes increasingly better than those of all other tested codes as the noise bias increases. In particular, as the noise changes from fully depolarising to a pure dephasing type, the X$^3$Z$^3$ Floquet code's threshold increases from $\approx 1.13 \%$ to $\approx 3.09\%$ for the code-capacity noise and from $\approx 0.76\%$ to $\approx 1.08\%$ for the SDEM3 noise model. The threshold therefore increases by a factor of 2.7 and 1.4 for the code capacity and SDEM3 noise models, respectively.

It is interesting to note that for the fully depolarising code-capacity noise, all codes have the same threshold. This is because at every measurement subround under this noise model, all the above codes have two types of errors at each fault location (occurring with a total probability $2p/3$) that give rise to edges in the detector graph and one type (occurring with probability $p/3$) that produces hyperedges. For the CSS and X$^3$Z$^3$ codes, these hyperedges result from $Y$ Pauli errors. On the other hand, the hyperedge errors in the P$^6$ and XYZ$^2$ codes are those that anti-commute with the check measurements that occur just before and immediately after the errors~\cite{fahimniya2024_hyperbolic_floquet,fu2024error}. Therefore, the Pauli type of the hyperedge errors for the honeycomb codes varies between measurement subrounds. For example, $Z$ errors create hyperedges only when they occur between the $XX$ and $YY$ checks of the honeycomb codes. In summary, all the above codes have the same performance for fully depolarising code-capacity noise because their (weighted) detector hypergraphs are all equivalent.

As shown in Figs.~\ref{fig:Threshold}(a) and~\ref{fig:Threshold}(b), while the CSS Floquet code has the same threshold as the X$^3$Z$^3$ code when the noise is fully depolarising, its threshold decreases with increasing noise bias, where its value is $0.752\%$ and $0.668\%$ at infinite bias for the code-capacity and SDEM3 noise, respectively. This decrease is due to the fact that the CSS Floquet code has pure $X$ and pure $Z$ detectors where in the presence of biased noise, half of the detectors, i.e., those with the same type of the dominant error, will become less useful in detecting the biased errors. 

Figures~\ref{fig:Threshold}(a) and~\ref{fig:Threshold}(b) show that the thresholds for the honeycomb codes have only minor improvements as the noise bias increases, where the thresholds for both honeycomb codes increase only by $\leq 6\times 10^{-4}$. Specifically, the thresholds of the honeycomb codes at $\eta = \infty$ are only about $1.03-1.08$ times larger than  their thresholds at $\eta = 0.5$ (where their fully-depolarising noise thresholds are $1.13\%$ and $0.585\%$ for the code-capacity and SDEM3  models, respectively). We note that for the SDEM3 noise, the thresholds of both honeycomb codes are lower than those of the X$^3$Z$^3$ and CSS Floquet codes at all noise biases. This is partly explained by noting that the SDEM3 noise model contains measurement errors which give rise to hyperedges in the decoding hypergraphs of the honeycomb codes~\cite{fahimniya2024_hyperbolic_floquet} but only graph-like edges for the CSS and X$^3$Z$^3$ Floquet codes~\cite{Anyon_condensation_2024}. As we explain below, these hyperedges degrade the MWPM decoder performance. 

We find that there are differences, albeit modest ones, between the performance of the XYZ$^2$ and P$^6$ honeycomb codes under biased noise; these differences have not been pointed out before in the literature. While  the XYZ$^2$ honeycomb code has a better performance than the P$^6$ honeycomb code for the code-capacity noise model, we find that its performance is inferior to that of the P$^6$ Floquet code for the SDEM3 noise model. This is surprising, considering the similarity of the former to the bias-tailored XYZ$^2$ static code~\cite{Srivastava_2022_XYZ2_code}. However, owing to stabilisers being measured in two measurement subrounds, the XYZ$^2$ 
Floquet code no longer possesses all the symmetries of the static version making the dynamical code not able to inherit all the benefits of its static counterpart.  

\begin{figure*}[t]
\includegraphics[width=0.95\linewidth]{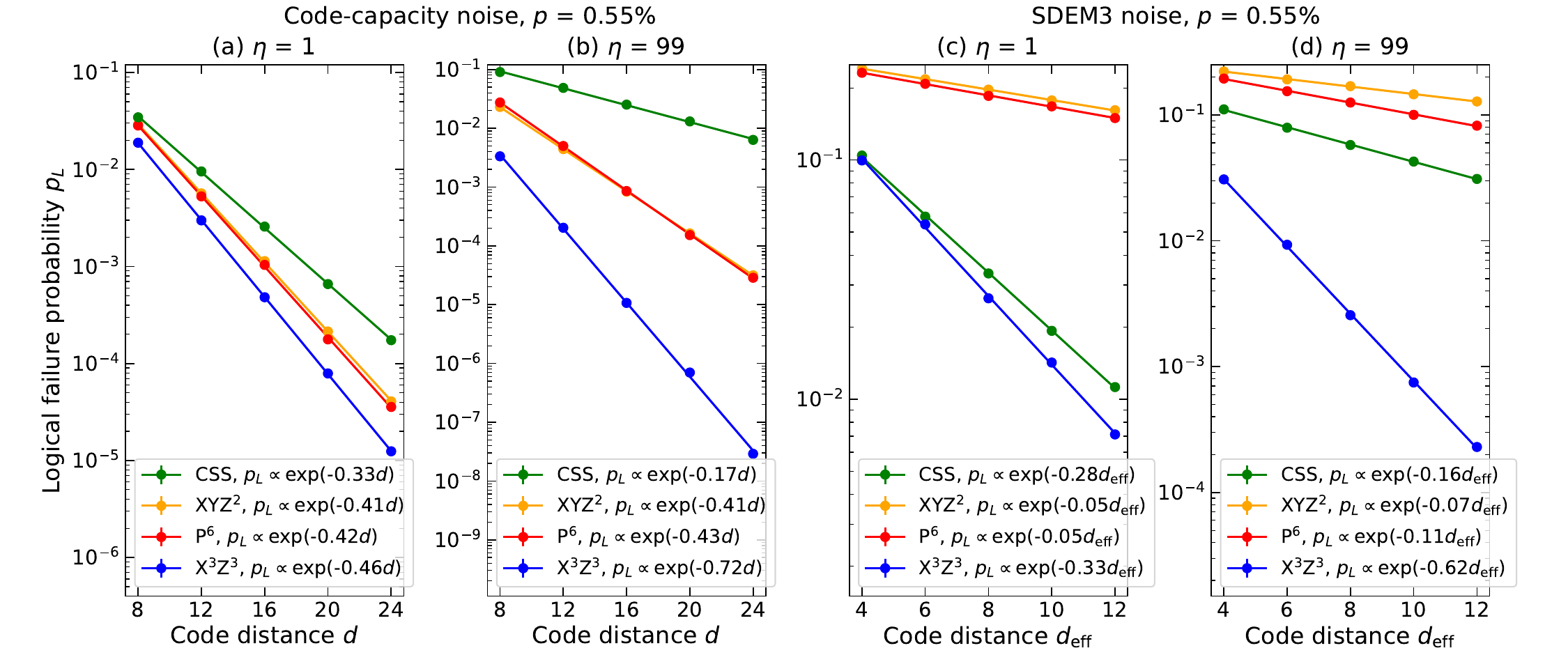}
	\caption{\textbf{Exponential suppression of the sub-threshold logical failure probabilities $p_L$ with respect to code distance.} We use code distances $d$ or $\deff$ depending on the noise model and compare different Floquet codes: CSS (green), XYZ$^2$ (orange), P$^6$ (red), and X$^3$Z$^3$ (blue). Results are calculated for different noise models: (a,b) code-capacity and (c,d) SDEM3 noise models, with a subthreshold physical error rate $p = 0.55\%$, which is small enough, yet can still be simulated using our computational resources up to code distance $d = 24$ or $\deff = 12$, for all the different codes and parameters corresponding to different curves in the plots. The same physical error rates are chosen for all curves such that the subthreshold performance under different noise bias strengths and noise models can be compared. Plots are  computed using two different bias strengths; one representing noise near the depolarizing regime: (a,c) $\eta = 1$ and the other representing noise in the strongly dephasing regime: (b,d) $\eta = 99$. All curves can be fitted to an exponential decay function $f \propto \mathrm{exp}(-\gamma d)$ or $f \propto \mathrm{exp}(-\gamma \deff)$ where $\gamma$ depends on the bias strength $\eta$ and is an increasing function of $(p_{\mathrm{th}} - p)$. Each  data point is averaged over $10^5-10^9$ shots. }\label{fig:Error_distance}
 \end{figure*}

The performance difference between the  XYZ$^2$ and P$^6$ honeycomb codes can be understood from the distribution of hyperedge-like syndromes (namely, those with four triggered detectors) in the detector hypergraphs under the biased SDEM3 noise model. These hyperedges generally degrade the code's performance when using a matching decoder since they must be decomposed into edges (see Ref.~\cite{fahimniya2024_hyperbolic_floquet} for a detailed description of the decoding in a hyperbolic version of the honeycomb code). An infinitely phase-biased SDEM3 noise model that includes Pauli noise but no measurement errors can already form hyperedges in the honeycomb codes' detector hypergraphs. In particular, the P$^6$ code can have hyperedges resulting from $Z$ errors occurring only after the red check measurement subrounds, while the hyperedges in XYZ$^2$ code can result from certain $Z$ errors happening in any subround. While there is no difference in the overall number of hyperedges resulting from single-qubit $Z$ errors in the two hypergraphs, the different arrangement of the hyperedges due to the single-qubit errors in these two codes gives rise to different kinds of syndromes for the two-qubit $ZZ$ errors. Specifically, $ZZ$ errors can lead to hyperedge syndromes in the XYZ$^2$ Floquet code but only edge-like syndromes for the P$^6$ code (see Supplementary Sec.~IV of the Supplementary Information~\cite{Supplementary_material} for details). As a result, in an infinitely phase-biased regime, where Pauli errors after MPP gates are evenly distributed between $Z_1$, $Z_2$ and $Z_1Z_2$, there are more hyperedge syndromes triggered in the XYZ$^2$ code than in the P$^6$ code. This is the reason underlying the difference in the thresholds,  as shown in Fig.~\ref{fig:Threshold}(b), of the two honeycomb codes under the SDEM3 noise. Since honeycomb codes are prone to hyperedge errors, better performance of the honeycomb codes might be expected using a correlated MWPM decoder such as the one used in Ref.~\cite{Gidney_honeycomb_2021}.\\

\subsection{Subthreshold performance}\label{subsec:subthreshold}
As a result of the shift of the thresholds with the noise bias, the subthreshold performance of the codes changes accordingly. As shown in  Figs.~\ref{fig:subthreshold_performance}(a) and~\ref{fig:subthreshold_performance}(b), the subthreshold logical failure probability improves significantly for the X$^3$Z$^3$ code, while it becomes only slightly better for the honeycomb codes and deteriorates for the CSS code. In general, one expects the subthreshold performance of the codes to improve by increasing the code distance along the direction where the biased errors most likely form logical strings. For example, for the $Z$ biased noise considered here, it would be by increasing the vertical code distance $d_V$. 
To this end, we study the subthreshold performance of elongated CSS and X$^3$Z$^3$ Floquet codes with $d_V > d_H$, where $d_V$ and $d_H$ are the vertical and horizontal code distances, respectively. The details are given in Supplementary Sec.~V of the Supplementary Information~\cite{Supplementary_material}.

Besides thresholds, another quantity of importance is the scaling of the code sub-threshold performance with the code distance. We plot the sub-threshold logical failure probability as a function of $d$ or $\deff$ for varying levels of noise bias in Fig.~\ref{fig:Error_distance}. As shown, the logical failure probabilities for all codes decrease exponentially with the code distance, i.e., $p_L \propto \mathrm{exp}(-\gamma d)$ or $p_L \propto \mathrm{exp}(-\gamma \deff)$. Among all codes presented, the X$^3$Z$^3$ code has the largest logical error suppression rate $\gamma$.
At higher noise bias, this error suppression rate becomes significantly larger for the X$^3$Z$^3$ code, moderately increases for the two honeycomb codes and decreases for the CSS Floquet code. The reason is that as the noise bias increases, a fixed subthreshold physical error rate moves relatively with respect to the shifting threshold, so that it becomes much further below the threshold for the X$^3$Z$^3$ Floquet code, moves moderately away from the threshold for the two honeycomb codes and becomes closer to the threshold for the CSS code [see Figs.~\ref{fig:Threshold}(a) and~\ref{fig:Threshold}(b)]. 

We also numerically simulate the performance of the X$^3$Z$^3$ Floquet code with a twisted periodic boundary condition. This twisted code has a pure $Z$-type logical operator with a length that scales quadratically with the code distance of the untwisted code. As a result, we find better thresholds and more favorable subthreshold performance for the twisted code over the untwisted one. While these performance benefits hold for the \textit{infinite-bias} code-capacity noise, they do not apply to other noise biases and models tested (See Supplementary Sec.~VI of the Supplementary Information~\cite{Supplementary_material} for details). 

While in this paper, we focus on Floquet codes with periodic boundary conditions, we expect similar qualitative performance of the Floquet codes under open boundary conditions. As shown in Ref.~\cite{Gidney_Planar_honeycomb_2022} for the honeycomb code with depolarizing noise, the boundary conditions (whether periodic or open) do not significantly impact the threshold and subthreshold performance of the code. This should be expected even more so for the CSS and X$^3$Z$^3$ Floquet codes, because the boundary condition does not affect the measurement schedule for the codes. We leave the simulations of the codes under open boundary conditions to future works.

The error suppression rate $\gamma$ is related to the error suppression factor $\Lambda$ introduced in Ref.~\cite{google2021exponential} which is defined as the reduction factor in the logical failure probability as the code distance increases by 2. Mathematically, it is given by 
\begin{equation}
\Lambda \equiv \frac{p_L(d)}{p_L(d+2)} = e^{2\gamma }
\end{equation}
for the code-capacity noise and similarly for the SDEM3 noise model, but with $d$ replaced by $\deff$. For a noise model where the error is characterised by a single error rate $p$ and threshold value $\pth$, when $p \ll \pth$, QEC suppresses the logical error 
exponentially as the code distance increases, i.e.,
$p_L \approx (p/\pth)^{d/2}$. Since $\Lambda \equiv p_L(d)/p_L(d+2)$, we then have $\Lambda \approx \pth/p$. At the thresholds, we have $\Lambda(\pth) \approx 1$. The value of $\Lambda$ increases as the physical error rate $p$ moves further below the threshold $\pth$.  For physical error rate below the threshold, $\Lambda > 1$ and  larger $\Lambda$ means greater error correction.
Since the threshold depends on the noise bias $\eta$, $\Lambda$ depends on both $\eta$ and $p$.
 In Table~\ref{tab:lambda_factor} 
 we list the values of $\Lambda$ corresponding to the $\gamma$ values shown  in Fig.~\ref{fig:Error_distance}.\\

\begin{table}
        \centering
         \begin{tabular}{|l|c|c |c|c|}
\hline
 & \multicolumn{2}{c|}{Code-capacity} & \multicolumn{2}{c|}{SDEM3} \\

\cline{2-5}
 & $\eta = 1$   & $\eta = 99$  & $\eta=1$   & $\eta = 99$   \\
\hline
CSS      & 1.94 & 1.39 & 1.74 & 1.37   \\
\hline
XYZ$^2$       & 2.27 & 2.28  & 1.11 & 1.15   \\
\hline
P$^6$ & 2.31 & 2.37   & 1.12 & 1.24  \\
\hline
X$^3$Z$^3$   & 2.49 & 4.26 & 1.94 & 3.42 \\
\hline
\end{tabular}
\caption{\textbf{Error suppression factors $\Lambda$ for different Floquet codes.} These are calculated using two different noise models, code-capacity and SDEM3, with varying levels of noise bias $\eta$. Results are calculated with a physical error rate  $p = 0.55\%$. The lambda factor is related to the error suppresion rate $\gamma$ shown in Fig.~\ref{fig:Error_distance}.}
    \label{tab:lambda_factor}
    \end{table}

\subsection{Performance optimality of X$^3$Z$^3$ Floquet code and no-go theorems}\label{subsec:theorems}
As seen above, although the performance of the X$^3$Z$^3$ Floquet code is significantly better compared to other Floquet codes, its threshold under infinitely phase-biased code-capacity noise does not reach the 50$\%$ threshold of its bias-tailored static version~\cite{Domain_wall_CC}. To explain this, we show that there are fewer symmetries in the decoding graphs of  dynamical codes, which restricts the code performance. The key feature resulting in the high performance of bias-tailored static codes (i.e., their decoding can be understood as decoding a series of repetition codes~\cite{Tuckett_bias_1,Tuckett_bias_2,XZZX_surface_code,Domain_wall_CC}) is not possible for dynamical codes built from two-qubit parity measurements, as we will show below. 

\begin{figure}
    \centering
    \includegraphics[width=0.99\linewidth]{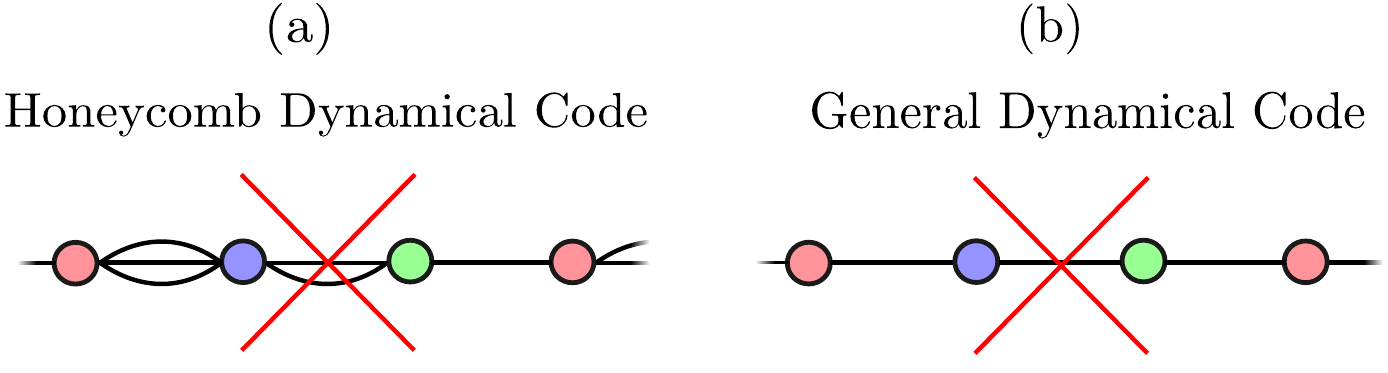}
    \caption{\textbf{Illustrations of the no-go theorems.} We display examples of the decoding graphs (under an infinitely biased code-capacity noise model; edges correspond to $Z$ errors and vertices to detectors) that are forbidden by (a) Theorem~\ref{thm:Degree_graph} for dynamical codes on the honeycomb lattice and (b) Theorem~\ref{thm:GDC_graphs} for general dynamical codes.}
    \label{fig:theorems}
\end{figure}

We argue that due to the above limiting constraint, the symmetry of the X$^3$Z$^3$ Floquet code has likely rendered its performance close to optimal under a matching decoder, despite the fact that even for the most favorable case of infinitely phase-biased code-capacity noise, its decoding can only be reduced to at most a series of disjoint 2D (planar) graphs. We formalise this by proving two no-go theorems for the decoding graphs of dynamical codes. We begin with the following informal statement of our first theorem:
\begin{theorem}
    (Informal) A dynamical code (not necessarily a Floquet code) on the honeycomb lattice operated over a sufficiently large number of time steps cannot have a 1D decoding graph under an infinitely biased code-capacity noise model, so long as it obeys the following properties:
    \begin{enumerate}[label=(\alph*)]
        \item At each time step, two-body Pauli operators on edges of a given colour are measured. Measurements in consecutive time steps occur on edges of different colours and anti-commute.
        \item Detectors are supported on plaquettes.
        \item All non-trivial errors are detectable and produce syndromes of weight $>1$.
    \end{enumerate}
\end{theorem}
In principle, a 1D graph [such as the one shown in Fig.~\ref{fig:theorems}(a)] could still result in better code performance than a general planar decoding graph. The fact that this type of graph is not possible suggests that planar graph decoding is already optimal for dynamical codes on the honeycomb lattice. 

To generalise this result beyond codes defined on the honeycomb lattice, we provide the following theorem for general dynamical codes built from two-qubit parity measurements. 
\begin{theorem}
    (Informal) A dynamical code (not necessarily a Floquet code) operated over a sufficiently long time scale cannot have a decoding graph, under an infinitely biased code-capacity noise model, equivalent to that of a collection of repetition codes [see Fig.~\ref{fig:theorems}(b)], so long as: all measurements are two-qubit parity measurements, each qubit is in the support of one measurement in each time step, overlapping measurements in consecutive time steps anti-commute, and errors produce syndromes of weight $>1$. 
\end{theorem}
This theorem shows that (given reasonable and typical assumptions) the decoding graphs of general dynamical codes cannot be decomposed into a collection of 1D repetition-code decoding graphs, which could otherwise allow them to achieve higher thresholds.  

We provide the formal statements of the above two theorems and their proofs in the Methods section. We display the decoding graphs prohibited by the above two theorems in Fig.~\ref{fig:theorems}.

\section{Discussion}\label{sec:Discussion}
In this paper, we introduce the X$^3$Z$^3$ Floquet code, the first bias-tailored dynamical code based on two-qubit parity check measurements. We show that, despite having no constant stabilisers, the code possesses a persistent stabiliser-product symmetry under pure dephasing (or pure bit-flip) noise which allows for a simplified decoding. This results in a substantially improved threshold and sub-threshold performance under biased noise, when compared to other Floquet codes. We demonstrate the enhanced performance through our simulation results obtained from using a fast matching decoder and two error models: a simplistic code-capacity noise model and a noise model approaching realistic circuit-level errors. Besides the superior performance of the X$^3$Z$^3$ Floquet code, our results also show that there are differences, albeit modest ones, in the threshold and subthreshold performance between the XYZ$^2$ and P$^6$ honeycomb codes under biased noise.

To explain why the X$^3$Z$^3$ Floquet code does not reach the same high performance as the bias-tailored static codes,  we prove that a dynamical code on the honeycomb lattice (obeying certain assumptions common to standard Floquet codes) cannot have a 1D \textit{decoding} graph, the crucial requirement for the high performance of static codes. 
Despite this limitation, the bias-tailored X$^3$Z$^3$ Floquet code has the advantage over its static counterpart in that it requires only lower-weight measurements. Specifically for devices without native two-qubit parity measurements, we devise phase-bias preserving parity check measurement circuits for any qubit architecture, which allow for high-performance implementation of the code.  Our work therefore demonstrates that the X$^3$Z$^3$ Floquet code is a leading quantum error correction code especially for devices with limited connectivity such as the hexagonal and heavy-hexagonal architectures.
 
We now give several directions for future work. Since the decoding of any general Floquet codes with two-qubit parity measurements can be reduced, at most, to a 2D decoding problem, it would be interesting to investigate the performance improvements of the X$^3$Z$^3$ Floquet code using a more accurate decoder (such as a tensor network decoder) and to consider ways to analytically derive the best achievable thresholds for the X$^3$Z$^3$ Floquet code. Future studies may also investigate fault-tolerant logic in the X$^3$Z$^3$ code, via lattice surgery~\cite{Haah_2022_boundaries_honeycomb}, aperiodic measurement sequences~\cite{Davydova_2024_DA_codes}, twist braiding~\cite{ellison2023floquetcodestwist}, or transversal gates~\cite{Aliferis_2007}. Finally, it will be of interest to study the performance of memory experiments and fault-tolerant logical gate implementations in hardware where parity checks are implemented using bias-preserving circuits such as the circuits presented in this paper and compare them to the performance in hardware with native direct multi-Pauli product measurements.

\section{Methods}\label{sec:Methods}
\subsection{Noise models}
In this paper, we consider two noise models: code-capacity and entangling measurement (SDEM3) noise models. In the code-capacity noise model, we apply single-qubit Pauli noise on all the data qubits independently at every measurement subround. The single-qubit Pauli noise channel is given by 
\begin{equation}\label{eq:singlequbit_channel}
\mathcal{E}_{1q}(\rho) = (1-p)\rho + p_X \hat{X}\rho \hat{X} + p_Y \hat{Y}\rho \hat{Y} + p_Z \hat{Z}\rho \hat{Z}.
\end{equation}
Here, $p \equiv p_X + p_Y+ p_Z$ is the total error probability. As in the literature, we define the noise bias as $\eta = p_Z/(p_X + p_Y)$ and assume $p_X = p_Y$. Several values of $\eta$ are worth listing: \begin{enumerate}\label{enum_bias_points}
\item $\eta = 0 \rightarrow p_Z = 0$, and $p_X = p_Y = p/2$,
\item $\eta = 0.5 \rightarrow p_X = p_Y = p_Z = p/3$,
\item $\eta = \infty \rightarrow p_Z = p$, and $p_X = p_Y = 0$.
\end{enumerate}

 We note that, since for Floquet codes one QEC round consists of several measurement subrounds, the single-qubit noise channel in the code-capacity model is applied several times in one QEC round instead of just once as is the case with static codes. The code-capacity noise model is often used as a preferred initial noise model to study before going to a more involved model as its simplicity often offers insight into understanding the code performance. 

\begin{table}
        \centering
\begin{tabular*}{\linewidth}{|p{0.5\linewidth}|  >{\centering\arraybackslash}p{0.463\linewidth}|}
\hline
  \textbf{Noise model} & \textbf{Generalised SDEM3} \\
  \hline  
 &$M_{PP}(p,\eta)$\\
  Noisy gate set &Init($p$,$\eta$) \\
  &$M(p)$ \\
 \hline
 Measurement ancilla & None\\
 \hline
 Measurement subrounds per one QEC round & 6\\
\hline
\end{tabular*}\caption{\textbf{Description of the generalised SDEM3 noise model.} The definition of the noisy gates is given in Table~\ref{tab:noisygateset}.}
    \label{tab:SDEM3noise}
    \end{table}

\begin{table*}
        \centering
\begin{tabular}
{|p{0.1\linewidth}|p{0.9\linewidth}|}
\hline
  \textbf{Noisy gate set} & \textbf{Description}  \\
  \hline  
 {$M_{PP}(p,\eta)$}&
  Measurement of Pauli Product $PP$ on a pair of qubits:
\begin{itemize}
  \item independently register incorrect measurement result with probability $p$, 
  \item apply a two-qubit $Z$-biased noise channel with total error probability $p$ and noise bias $\eta$ after measurement, i.e., apply:  
  \begin{itemize}
  \item the trivial identity operator ($II$) with a probability $1-p$,
  \item $ZI$, $IZ$, and $ZZ$ operators, each with probability $\zeta p/3$,
  \item the remaining Pauli operators, each with probability $(1-\zeta)p/12$, where $\zeta = \frac{3}{5}\left(\frac{\eta}{1+\eta}\right)^2 + \frac{2}{5}\left(\frac{\eta}{1+\eta}\right)$.
\end{itemize}

\end{itemize}
 \\
  \hline
  {Init($p$,$\eta$)}& Qubit initialization in some Pauli basis, followed by applying a single-qubit $Z$-biased noise channel with total error probability $p$ and noise bias $\eta$ after the reset, i.e., apply:
    \begin{itemize}
  \item the trivial identity operator ($I$) with a probability $1-p$,
  \item Pauli $Z$ operator with probability $p\eta/(\eta+1)$,
  \item Pauli $X$ and $Y$ operators, each with probability $p/[2(\eta+1)]$. 
\end{itemize}\\ 
\hline
  {$M(p)$}& Measure the qubit in some Pauli basis and register the incorrect measurement result with probability $p$.\\
 \hline
\end{tabular}
\caption{\textbf{Description of noisy gates in Table~\ref{tab:SDEM3noise}.}}
    \label{tab:noisygateset}
    \end{table*}

    On the other hand, the SDEM3 model is a more elaborate error model involving single-qubit noise channels after every single-qubit gate, measurement and reset, a two-qubit noise channel after every multi-Pauli product (MPP) parity measurement gate, and a classical flip after each measurement. As in Ref.~\cite{Gidney_Planar_honeycomb_2022}, we assume that each of these error channels occurs with a total probability $p$. The SDEM3 noise model is therefore close to standard circuit-level noise and would be a more accurate description of a realistic noise channel, particularly in hardware with native two-qubit measurements, for example,
Majorana~\cite{Karzig_2017_scalable,Paetznick_floquet_codes_majoranas,McLauchlan_2022_Majorana_Comp} and photonic~\cite{Paesani2023High,hilaire2024enhanced} architectures. 

To take into account biased noise, we generalise the SDEM3 model for depolarising noise used in  Refs.~\cite{Gidney_Planar_honeycomb_2022,mclauchlan_2024_defectsfloquet}. Here the single-qubit noise channel has the same form as given in Eq.~\eqref{eq:singlequbit_channel} for the code-capacity noise model. On top of the single-qubit errors, this model also has a two-qubit noise channel applied after each of the MPP gates, which is given by 
\begin{equation}
\mathcal{E}_{2q}(\rho) = (1-p)\rho + \sum_{O \in \{I,X,Y,Z\}^{\otimes 2}\backslash \{I\otimes I\} } p_O \hat{O}\rho \hat{O}.
\end{equation} To conform with how the bias is defined for the single-qubit noise channel, we also use $\eta$ to characterize the bias of the two-qubit noise channel. Specifically, we define the bias $\eta$ for the two-qubit noise channel such that
\begin{enumerate}
\item $\eta =0 \rightarrow p_{ZZ} = p_{IZ} = p_{ZI} = 0$ and each of the other probabilities is $p/12$,
\item $\eta =0.5 \rightarrow$ each of the Pauli errors occurs with $p/15$, and
\item $\eta =\infty \rightarrow p_{ZZ} = p_{IZ} = p_{ZI} = p/3$, and the other probabilities are 0.
\end{enumerate}
Just as for the single-qubit noise, the above definition of $\eta$ ensures that the two-qubit noise at the three special points, $\eta =0$, $\eta =0.5$ and $\eta =\infty$, are $Z$-error free, depolarising, and pure dephasing, respectively. Given that the two-qubit error probabilities at these $\eta$ values must satisfy the conditions above, we define $\zeta \in [0, 1]$ and write the two-qubit Pauli error probabilities as 
\begin{align}\label{eq:pZZ}
p_{ZZ} &= p_{ZI} = p_{IZ} = \zeta p/3, \nonumber\\
p_{O} &= (1-\zeta)p/12,\nonumber\\ &\qquad\mathrm{for}\,  O \in \{I,X,Y,Z\}^{\otimes 2}\backslash \{II,ZI,IZ,ZZ\}, 
\end{align}
with $\eta$ and $\zeta$ related via: 
\begin{equation}
\zeta = \frac{3}{5}\left(\frac{\eta}{1+\eta}\right)^2 + \frac{2}{5}\left(\frac{\eta}{1+\eta}\right).
\end{equation}
Note that Eq.~\eqref{eq:pZZ} is defined such that the total probability of all the Pauli errors $\sum_{O\in \{I,X,Y,Z\}^{\otimes 2}\backslash \{I\otimes I\} } p_O = p$ for all noise biases $\eta$. 
Apart from the single- and two-qubit Pauli noise channels, we also apply a classical flip of the measurement results with probability $p$ after each of the single and two-qubit measurements. These measurement flips are uncorrelated with the single and two-qubit noise channels. We summarise the description of the SDEM3 noise model and its noisy gate set in Tables~\ref{tab:SDEM3noise} and~\ref{tab:noisygateset}, respectively.\\

\subsection{Stabiliser circuits}
We simulate the circuits and generate the error syndromes using Stim~\cite{gidney2021stim}; example Stim circuits are provided at Ref.~\cite{StimCircuitsZenodo}. 
We construct Stim circuits from various quantum operations (resets, measurement gates, etc.) with noise channels associated with each quantum operation (i.e., with error probabilities set by the noise models introduced above), detectors and logical observable updates. 
From these circuits, Stim can generate the detector error models which list the error mechanisms, the associated syndromes, and the logical observables flipped by the errors. A detector error model is fed to the decoder which then predicts the most likely error based on a given syndrome. We refer the reader to Ref.~\cite{gidney2021stim} for more details on Stim.\\

\subsection{MWPM decoder}
To decode the error syndromes, we apply a minimum-weight
perfect matching (MWPM) decoder~\cite{dennis2002topological}, which is implemented using PyMatching~\cite{pymatching}. The syndrome decoding is mapped by the MWPM decoder onto a graph problem, which is subsequently solved by utilizing Edmonds’ algorithm~\cite{edmonds1965maximum,edmonds1965paths} for finding a perfect matching which has minimal weight, i.e., finding a minimum-total-weight set of edges, for which every vertex is incident to exactly one edge~\cite{fowler2013minimum}.\\

\subsection{Details of numerical simulations}
Our simulations are performed for four different Floquet codes using different values of  physical error rates $p$ and various strengths of noise bias $\eta$. Moreover, we consider two different noise models: code-capacity and SDEM3 noise models. For each code, the simulations are run with effective distances $\deff \equiv d/2 = 4, 6, 8, 10, 12$, where the effective distance $\deff$  is defined as the minimum number of faults under SDEM3 depolarising noise that produce a logical error. This is half of the distance $d$ of the code-capacity noise model. The calculated effective distance differs between noise models since in the SDEM3 noise model, two-qubit errors occurring after an MPP parity gate count as a single fault, while under the code-capacity noise model they would count as two faults. 

We choose lattices with periodic boundary conditions. For most simulations, we choose lattice sizes $L \times 3L/2$, where $L = d = 2\deff$ is the lattice length which varies with respect to the code distance. We choose lattice sizes $L \times 3L/2$ such that the horizontal and vertical code distances are the same, as can be inferred from the minimum weight of the horizontal and vertical logical operators (see e.g., Fig.~\ref{fig:X3Z3_figure}(c) for lattice length $L = 4$).  We simulate the memory experiments for $3\deff$ QEC rounds  or $18\deff$ measurement subrounds (1 QEC round consists of 6 measurement subrounds).  For Supplementary Sec.~V of the Supplementary Information~\cite{Supplementary_material} where we study elongated codes, we use lattice sizes of $d_H \times 3 d_V/2$ where $d_V \geq d_H$. Here, $d_H$ and $d_V$ are the minimum weights of the logical operators in the horizontal and vertical directions, respectively. We simulate the horizontal and vertical logical memory experiments for $3d_H/2$ and $3d_V/2$ rounds, respectively.\\

We select a single logical qubit for each Floquet code (they all encode two logical qubits on a lattice with periodic boundary conditions) to test for the logical failure probability. Depending on the codes and parameter regimes, we run different numbers of Monte Carlo shots ranging from $10^5-10^{10}$ (larger number of shots for longer code distances and smaller physical error rates) for each of the horizontal and vertical logical observables, which give us the horizontal $p_H$ and vertical $p_V$ logical failure probabilities.  We then report the combined logical error probability: 
\begin{equation}\label{eq:combinedlogical}
p_L = 1 - (1-p_H)(1-p_V),
\end{equation}
which is an estimate of the probability that \textit{either} a vertical or a horizontal logical error occurs. Equation~\eqref{eq:combinedlogical} assumes that the horizontal and vertical logical errors occur independently. For the infinitely phase-biased code-capacity noise model, one of the logical errors which is of pure $Z$-type or pure $X$-type has zero logical error probability, because the pure biased noise cannot form the other logical operator which is of mixed-type and hence cannot flip the pure-type logical operator. As a result, for the infinitely phase-biased code-capacity noise, the maximum combined logical error probability is $p_L = 0.5$ [Eq.~\eqref{eq:combinedlogical}]. This is in contrast to other noise biases and models where the maximum combined logical error probability is $p_L = 0.75$ since $p_V, p_H \leq 0.5$.\\

\subsection{Determining the thresholds}
To extract thresholds of the codes, we perform a finite-size  collapse~\cite{wang2003confinement,Harrington2004Thesis} of the logical failure probability $p_L$ data taken for various physical error rates $p$ and code distances. This is done by fitting the data to the curve $A + Bx + Cx^2$  where $x = (p-\pth) d^{1/\nu}$ or $x = (p-\pth) \deff^{1/\nu}$ for the code-capacity and SDEM3 noise models, respectively. Here, $\nu$ is the critical exponent, $\pth$ is the Pauli threshold, $A,B,C$ are the fit parameters, $d$ and $\deff$ are code distances under the code-capacity and SDEM3 noise models, respectively. This data is presented in Supplementary Figs.~5-8 of the Supplementary Information~\cite{Supplementary_material}.\\

\subsection{No-go theorems for 1D decoding graphs of dynamical codes}
Here, we provide formal statements and proofs for the two no-go theorems presented in the Results section. 
To this end, we begin by providing several definitions.
\begin{definition}
    Let $V$ be a set of detectors for a dynamical code. For every qubit and every time step $(q,t)$, i.e., every fault location in the code, define a hyperedge $e_{(q,t)} = (v_1,v_2,\ldots)$, where the $v_i\in V$ are detectors that return $-1$ outcomes if a single $Z$ error occurs at the fault location $(q,t)$. The \textbf{$\mathbf{Z}$-detector hypergraph} (ZDH) is defined as $G = (V, E)$, where $E = \bigcup_{q, t} e_{(q,t)}$ is the set of all hyperedges. A $Z$-detector graph (ZDG) is a $Z$-detector hypergraph in which all hyperedges are edges (their size is 2). 
\end{definition}

\begin{definition}
    A \textbf{1D-decodable} $Z$-detector graph is one in which all vertices have neighbourhoods of size no greater than 2.
\end{definition}

\noindent

Finally, we define dynamical codes on the honeycomb lattice in the following way:
\begin{definition}\label{def:HDC}
    A honeycomb dynamical code (HDC), with ``duration" $T$ and constant ``initial(final)-time boundary offsets" $t_i$ ($t_f$), is a finite-depth measurement circuit acting on qubits of the honeycomb lattice (without boundary). The sets of measurements in the circuit, $\mathcal{M}_t$ ($t\in \lbrace 1,\ldots, T\rbrace$), are composed of two-body Pauli measurements along coloured edges (either $r$, $g$, or $b$, for each time step) of the lattice, performed sequentially on a state stabilised by some group $\mathcal{S}$. The HDC obeys the following properties:
    \begin{enumerate}[label=(\alph*)]
        \item Overlapping measurements in consecutive time steps anti-commute and are supported on different edges.
        \item Detectors (for time steps $t>t_i$) are associated with plaquettes in the lattice (they have support only on qubits around their associated plaquette). 
        \item If a single-qubit or two-qubit error occurring in time steps $t_i<t<T-t_f$ anti-commutes with future measurements, it is detectable, unless the two-qubit error is the same as an edge operator just measured or to be measured in the next time step. 
        \item All single-qubit errors in time steps $t_i < t < T- t_f$ have syndromes of weight $>1$. 
    \end{enumerate}
\end{definition}

Let us first comment on our definition of an HDC.  
The chosen properties are very natural. The anti-commutation of consecutive measurement subrounds ensures ``local reversibility" of the code, so as to preserve the quantum information and its locality ~\cite{aasen2023measurementquantumcellularautomata} (although note that we do not require the code to have logical qubits in our definition). The requirement that consecutive sets of measurements are supported on different edges is not very restrictive: if two (anti-commuting) measurements act on the same edge in consecutive time steps, we can replace the second with a Clifford gate and commute that to the end of the circuit. This merely changes the bases of subsequent measurements without changing their (anti-)commutation. Since the Clifford gates are unimportant, we can ignore them and hence we end up with a shorter duration $T$. We restrict our attention to errors occurring in the temporal ``bulk" of the code, $t_i < t < T-t_f$. This avoids complications due to errors occurring close to the initial and final-time boundaries where detectors here may be formed in different ways than the bulk detectors. Specifically, the detectors associated with errors at the initial and final time boundaries are obtained by comparing the edge or plaquette measurement values with the initial state and final read-out of the physical qubits, respectively. The exclusion of these final time boundary detectors is also required here since we do not include final read-out measurements in our definition of an HDC. In summary, we consider errors occurring in all ``detection cells"~\cite{Anyon_condensation_2024} that are completed before time step $T$ and begin at time step $t\geq 1$, where a detection cell consists of all the spacetime points at which a detector can identify errors.

We will show that an HDC cannot even have a 1D-decodable ZDG, which implies that its decoding graph cannot be equivalent to a collection of the much simpler repetition code's graphs. Note that a repetition code graph is not only 1D-decodable but also has a maximum degree of 2, which means there are no double edges between neighbouring vertices. Here, however, we allow for these double edges in our definition of 1D-decodability, and show that this more general property is also impossible. Such double edges naturally arise in Floquet codes, where they correspond to two-qubit undetectable errors (i.e., edge operators just measured or about to be measured).

We begin by first proving the following lemma which is going to be used in the proof of Theorem~\ref{thm:Degree_graph}.
\begin{lemma}\label{lem:nbring_plaqs}
    In a honeycomb dynamical code, at each time step $t>1$, no detectors can be formed on neighbouring plaquettes.
\end{lemma}
\begin{proof}
    Without loss of generality, suppose neighbouring plaquettes $F_1$ and $F_2$ are coloured red and blue, respectively. Suppose, for the sake of contradiction, that they both host detectors at a time step $t>1$. 
    Let $\mathcal{D}_1,\mathcal{D}_2 \subset \mathcal{M}_t$ be the sets of check measurements from $\mathcal{M}_t$ whose products around the plaquettes $F_1$ and $F_2$, respectively, form the detectors at the corresponding plaquettes at time $t$.
    That is, the measurements in $\mathcal{D}_j$ act only on qubits around $F_j$.  
    Since at every time step, only edges of one particular colour are being measured and green edges are shared by blue and red plaquettes, 
    the measurement set $\mathcal{M}_t$ at time $t$ can contain only measurements on green edges. Otherwise, either $\mathcal{D}_1$ or $\mathcal{D}_2$ would have to be empty or they would include measurements on qubits not adjacent to $F_1$ or $F_2$, respectively, which is not allowed by the definition of $\mathcal{D}_1$ and $\mathcal{D}_2$. 
    
    To form detectors at both plaquettes, we require that, for all $M\in\mathcal{M}_{t-1}$, $[M,\prod_{D\in\mathcal{D}_j} D] = 0$, for $j=1,2$. Otherwise the plaquette operator outcomes would be indeterministic. However, from property (a) of Definition~\ref{def:HDC} of an HDC, for each $D\in\mathcal{D}_1$, there is a measurement in $\mathcal{M}_{t-1}$ that anti-commutes with it. Therefore, $\prod_{D\in\mathcal{D}_j}  D$ must have even overlap with all edge measurements in  $\mathcal{M}_{t-1}$, for $j=1,2$. This condition can only be satisfied for both $j=1,2$ if all $M\in \mathcal{M}_{t-1}$ are also on green edges (if they are supported on red edges, there will be one measurement with odd overlap with each $D\in\mathcal{D}_1$, and similarly for $D\in\mathcal{D}_2$ if they are supported on blue edges). But now both $\mathcal{M}_{t-1}$ and $\mathcal{M}_t$ are made up of measurements on green edges, contradicting property (a) of Definition~\ref{def:HDC}. 
\end{proof}

Using the properties of the HDC as defined above, we now present the first main theorem of our paper:
\setcounter{theorem}{0}
\begin{theorem}\label{thm:Degree_graph}
    A honeycomb dynamical code with duration more than $3t_f+t_i+3$ cannot have a 1D-decodable $Z$-detector graph, where  $t_i$ and $t_f$ are the initial and final  time-boundary offsets.
\end{theorem}
Since the proof of the theorem is quite long, here we begin by providing the sketch of the proof. We first show that the $Z$-detector graph of an HDC contains many more edges than vertices. This means the graph must contain many cycles. We will then find a \textit{subgraph} of the ZDG in which no two vertices are connected by more than one edge, where for such a subgraph, the size of the neighbourhood of each vertex is equal to the number of edges incident to it. In particular, we will show that there exists at least one such subgraph that \textit{still} possesses many more edges than vertices, meaning it is not possible for all vertices in the subgraph to have neighbourhoods of size $\leq 2$ -- if that were the case, then there would be at least as many vertices as edges. This in turn implies that the full ZDG must contain some vertices with neighbourhoods of size $> 2$.

The full proof of Theorem~\ref{thm:Degree_graph} is as follows.
\begin{proof}
We take an error occurring in time step $t$ to mean that it occurs immediately \textit{after} the measurements in $\mathcal{M}_t$. Suppose there are $T$ time steps and $n$ qubits in the HDC.  %
Let us consider part of the ZDG that contains vertices corresponding to all detectors formed in time steps $t>t_i$, but only those edges corresponding to \textit{detectable} errors, i.e., those occurring in time steps $t_i<t<T-t_f$. We will call this subgraph $\ZDGbulk$.
The number of vertices in this subgraph, corresponding to the number of detectors formed, is at most $n(T-t_i)/6$, since at each time step no detectors can be formed on neighbouring plaquettes (see Lemma~\ref{lem:nbring_plaqs}), and there are $n/6$ such non-neighbouring plaquettes, which correspond to all plaquettes of a single colour, in the honeycomb lattice without boundary. Meanwhile, $Z$ errors generate $n$ edges between each time step, resulting in a total of $n(T-1-t_i-t_f)$ edges in the $\ZDGbulk$.

Let us now create a bipartition of the qubits into sets $C$ and $D$ such that no qubits in $C$ are adjacent to one another, and similarly for qubits in $D$. This is possible since the honeycomb lattice is bipartite. 
Let us consider the subgraph of $\ZDGbulk$, denoted by $\ZDGbulkC$, that contains only edges corresponding to $Z$ errors on qubits in $C$. The subgraph $\ZDGbulkC$ therefore contains $n(T-1-t_i-t_f)/2$ edges. Suppose two vertices are connected in this graph by more than one edge. In such a scenario, we can choose any two from the collection of edges between these vertices. These two edges correspond to a weight-2 error, i.e, an error with support only on two qubits, in $C$, which is undetectable. This error is undetectable because the combination of the two errors, where each flips the values of the same two detectors, triggers no detector.

Consider one such undetectable weight-2 error, $Z_{i,t} Z_{j,t'}$, where $Z_{i,t}$ denotes an error that acts on qubit $i$ at time step $t$ and similarly for $Z_{j,t'}$. We will show that for $Z_{i,t} Z_{j,t'}$ \textit{not} to be detectable, the two errors must act on the same qubit ($i= j$) but at adjacent time steps ($t = t'\pm 1$). To this end, we begin by ruling out other possibilities. First, we show that if the qubits $i,j \in C$ are different then the error $Z_{i,t} Z_{j,t'}$ will anti-commute with some future measurements. This, along with the fact that $Z_{i,t}Z_{j,t}$ is not an operator just measured or about to be measured (since $i$ and $j$ are not connected by an edge), means that the error is detectable by property (c) of Definition~\ref{def:HDC} of an HDC. We begin by analysing the case where the two errors occur at the same time step ($t = t'$). Suppose that this error commutes with all $M\in \mathcal{M}_{t+1}$. Since every measurement in $\mathcal{M}_{t+1}$ must have support only on one qubit in $C$, then the two check measurements in $\mathcal{M}_{t+1}$, one with support on qubit $i$ and the other on qubit $j$, must have Pauli $Z$ operator supports on those respective qubits.  Now consider the measurement $M' \in \mathcal{M}_{t+2}$ with support on qubit $i$. By property (a) of Definition~\ref{def:HDC}, one measurement from the set $\mathcal{M}_{t+2}$ that has a support on qubit $i$ cannot act as a Pauli $Z$ operator on that qubit, since it has to anti-commute with the previous edge measurement from $\mathcal{M}_{t+1}$ that also has a $Z$ support on the same qubit. Therefore, $M'$ will anti-commute with $Z_{i,t} Z_{j,t}$ which, according to property (c) of Definition~\ref{def:HDC}, means that the error $Z_{i,t} Z_{j,t}$ is detectable. Similarly, if the errors occur on different qubits but at different time steps, they must also be detectable. To show this, we can argue using the reasons mentioned above that $Z_{i,t}$ must anti-commute with a future measurement in set $\mathcal{M}_{t+1}$ or $\mathcal{M}_{t+2}$. Since qubits $i$ and $j$ are not connected by an edge in the honeycomb lattice, the same measurement, however, cannot anti-commute with $Z_{j,t'}$.

 Using the above reasonings, we can infer that the only possibility for the weight-2 error to be undetectable is for the two errors to occur on the same qubit,  but at different time steps, i.e., $Z_{i,t} Z_{i,t'}$ (note that if the two errors on qubit $i$ occur at the same time step then they cancel each other out, which is equivalent to there being no error at all). Suppose, without loss of generality, that $t< t'$. For the weight-2 error to be undetectable, the measurements in between them must commute with $Z_i$. Since consecutive measurements with support on the same qubit anti-commute, the two errors acting on qubit $i$ must be separated only by one measurement time step where the in-between  measurement  commutes with the two errors. 

The above analysis implies that vertices in $\ZDGbulkC$ can be connected by only at most two edges. All such edges would have to correspond to errors on the same qubit and if there were more than two, we could find a pair of edges that correspond to an undetectable error $Z_{i,t} Z_{i,t'}$ with $t' > t+1$, which we have shown above to be impossible. Therefore, at most half of the edges in $\ZDGbulkC$ can be removed without changing which vertices are connected. This means that there is a spanning subgraph of $\ZDGbulkC$ with at least $\lceil n(T-1-t_i-t_f)/4 \rceil$ edges with no vertices in the subgraph connected by more than one edge.

All things considered, this implies that the full ZDG has a subgraph with $\lceil n(T-1-t_i-t_f)/4 \rceil$ edges and a number of vertices $\leq n(T-t_i)/6$, such that no two vertices are connected by more than one edge. For $T > 3t_f + t_i+3$, there are more edges than vertices in this subgraph, which means that it cannot have all vertices with neighbourhoods of size $\leq 2$. Therefore, some vertices in the full ZDG also must have neighbourhoods with size greater than 2.
\end{proof}

The above theorem implies that no matter how an HDC is bias-tailored, it likely can never achieve as high a threshold as the repetition code or the bias-tailored stabiliser codes. Indeed, for large $T$ ($T \gg t_i, t_f$), there will be at least approximately $6$ times as many edges as vertices in the ZDG, and there is a subgraph with $3/2$ as many edges as vertices without any pair of vertices sharing more than one edge. This means the graph must have many cycles with length at least three edges.

It is clear that the codes examined in this paper obey the properties of an HDC. 
In particular, any Floquet codes related to the CSS Floquet code by Hadamard gates applied to a subset of the qubits (as is the case for the X$^3$Z$^3$ Floquet code) do not have 1D-decodable ZDGs. The reason we have focused on dynamical codes that have ZDGs, e.g., Hadamard-deformed CSS codes, instead of codes with hyperedges in their $Z$-detector hypergraphs, is because their decoding graphs are immediate generalisations of that of the CSS code. The fact that 1D-decodable ZDGs are not possible for an HDC suggests that there are not enough independent symmetries to reduce its decoding graph to the extent possible for stabiliser codes. It is interesting to note that even with hyperedges in the $Z$-detector hypergraph, a stabiliser code can still be bias-tailored~\cite{Tuckett_bias_1} such that it possesses large enough symmetries to render its thresholds close to the hashing bound~\cite{LLoyd1997Capacity,devetak2005private}. While such high-thresholds are possible for static codes, we do not believe this would be the case for HDCs.

Finally, we consider more general codes defined on an arbitrary graph $G$. 
\begin{definition}
    A general dynamical code (GDC) is a finite-depth measurement circuit acting on qubits associated with vertices of a graph $G$. It has a duration $T$ and initial (final) time boundary offsets $t_i$ ($t_f$), as defined above. The sets of measurements $\mathcal{M}_t$ are two-body Pauli measurements associated with edges of $G$ (each qubit is in the support of one measurement in each time step), performed sequentially on a state stabilised by some group $\mathcal{S}$. A GDC obeys properties (a) and (d) of an HDC.
\end{definition}

Having defined GDCs, we now provide the second main theorem of our paper:
\begin{theorem}\label{thm:GDC_graphs}
A GDC with a duration at least $2t_f + t_i +2$ (i.e., a constant), where $t_i$ and $t_f$ are the initial and final time-boundary offsets, respectively, cannot have a $Z$-detector graph equivalent to a collection of disjoint repetition codes.
\end{theorem}
\begin{proof}
    Since there are $n/2$ measurements in a given time step, there can be at most $n/2$ independent detectors formed from such measurements.
    However, there will in fact be far fewer for a GDC, because each individual measurement anti-commutes with a measurement from the previous time step resulting in a random  value for each edge measurement. Therefore, a detector with a deterministic outcome must be formed from at least a product of two non-deterministic measurements within a time step. 
    Suppose the GDC has a $Z$-detector graph (edges in the ZDG correspond to particular $Z$ errors). Again, we consider the subgraph corresponding to $Z$ errors occurring in time steps $t_i<t<T-t_f$, where $T$ is the GDC's duration, and detectors formed in time steps $t> t_i$. There are $< n(T-t_i)/2$ vertices and $n(T-1-t_i-t_f)$ edges in this subgraph. Since the number of vertices of a collection of repetition code decoding graphs cannot be less than the number of edges, this means that the ZDG of a GDC cannot be a collection of repetition code graphs for $T\geq  2t_f + t_i +2$.
\end{proof}

 Owing to this theorem, we do not expect to see thresholds in dynamical codes as high as 50\%. The reason that static codes perform better than dynamical codes under biased noise is that they obey two separate symmetries under an infinitely phase-biased code-capacity noise model. There is a spatial symmetry that forces syndromes to appear in pairs along 1D strips of the lattice, but there is also a temporal symmetry: syndromes must appear in pairs within a given time slice. These two symmetries simplify the detector graph into a collection of disjoint repetition code decoding graphs, leading to a high performance (up to 50\% threshold) of the static codes under biased noise. On the other hand, even for the case where measurement errors are not considered, GDCs necessarily break the temporal symmetry (for HDCs, this is a consequence of Lemma~\ref{lem:nbring_plaqs}), leading to edges between vertices in different time slices of the detector graph  (see Fig.~\ref{fig:Detector_graph}). We thus expect that such codes can only possess a single symmetry under an infinitely phase-biased code-capacity noise model, as the X$^3$Z$^3$ Floquet code does.\\

\acknowledgments
    We would like to thank our colleagues at Riverlane, especially Alexandra Moylett, Ophelia Crawford, and Gy\"{o}rgy Geh\'{e}r for helpful discussions. We are also grateful to Ophelia Crawford and Hari Krovi for their comments on the manuscript. CM would like to thank Benjamin Brown for pointing out a useful circuit for performing next-nearest-neighbour entangling gates. CM acknowledges support from the Intelligence Advanced Research Projects Activity (IARPA), under the Entangled Logical Qubits program through Cooperative Agreement Number W911NF-23-2-0223.\\

\noindent
\textbf{Author contributions}\\
FS developed the X$^3$Z$^3$ Floquet code and biased noise models. FS also performed the numerical simulations for all Floquet codes investigated. CM characterised the symmetry and decoding features of the codes tested. In addition, CM developed the bias-tailored circuits and the proofs of the no-go theorems. Both authors contributed to the writing of the manuscript.\\

\noindent
\textbf{Data availability}\\
The threshold data is available in \href{https://doi.org/10.5281/zenodo.14258878}{https://doi.org/10.5281/zenodo.14258878} (Ref.~\cite{StimCircuitsZenodo}). Other numerical data generated in this work is available from the authors upon reasonable request. The video talk on this manuscript is available in \href{https://www.youtube.com/watch?v=nqQT-5IRC9w}{https://www.youtube.com/watch?v=nqQT-5IRC9w}.\\

\noindent
\textbf{Code availability}\\
Circuits used in the simulations are available in \href{https://doi.org/10.5281/zenodo.14258878}{https://doi.org/10.5281/zenodo.14258878} (Ref.~\cite{StimCircuitsZenodo}).\\

\noindent
\textbf{Competing interests}\\
The authors declare no competing interests.\\


\bibliography{References_NPJQI}

\end{document}


\title{Supplementary Information for ``Tailoring Dynamical Codes for Biased Noise: The X\texorpdfstring{$^3$}{}Z\texorpdfstring{$^3$}{} Floquet Code"}
\author{F. Setiawan}
\email{setiawan.wenming@gmail.com}
\affiliation{Riverlane Research Inc., Cambridge, Massachusetts 02142, USA}
\author{Campbell McLauchlan}\email{campbell.mclauchlan@gmail.com}
\affiliation{Centre for Engineered Quantum Systems, School of Physics, The University of Sydney, Sydney, NSW 2006, Australia}
\affiliation{Riverlane, Cambridge, CB2 3BZ, UK}

\maketitle

\section{Details of honeycomb and CSS Floquet codes}\label{app:Code_details}

\subsection{Honeycomb code}\label{app:honeycomb}

We here provide details of the two honeycomb code variants discussed in the main text. The codes' logical operators evolve through their measurement cycles. In Supplementary Fig.~\ref{fig:honeycomb_code}(a), we show one set of logical operators evolving through two measurement cycles. While the measurement sequence has a period of 3, the logical operators only return to their initial values (up to signs) with a period of 6. 

In Supplementary Fig.~\ref{fig:honeycomb_code}(b), we show the plaquette and edge operators for both honeycomb code variants. The P$^6$ code~\cite{Gidney_honeycomb_2021} has edge operators with Pauli bases defined by the edge colour, while the XYZ$^2$ code~\cite{HH_code} has edge operators with bases defined by the edge's orientation in the T junction, i.e., horizontal for the $Z$ edges, and rotated 90$^\circ$ clockwise and counter-clockwise from the horizontal edges, for the $X$ and $Y$ edges, respectively [see  Supplementary Fig.~\ref{fig:honeycomb_code}(b)]. The plaquette operators of the P$^6$ code also differ between plaquettes of different colours, while the plaquette operators of the XYZ$^2$ code are all the same. Each plaquette operator is a product of the edge operators along the plaquette boundary. For example, a red plaquette operator is the product of the green and blue edge operators  around its boundary. The plaquette operators are measured in two subrounds. For instance, after measuring the $g$ and $b$ checks sequentially, we can learn the value of each red plaquette by multiplying the measurement values of the green and blue checks around its boundary. Unlike the CSS and X$^3$Z$^3$ Floquet codes, the honeycomb code has constant plaquette stabilisers.

\begin{figure}
    \centering
\includegraphics[width=\linewidth]{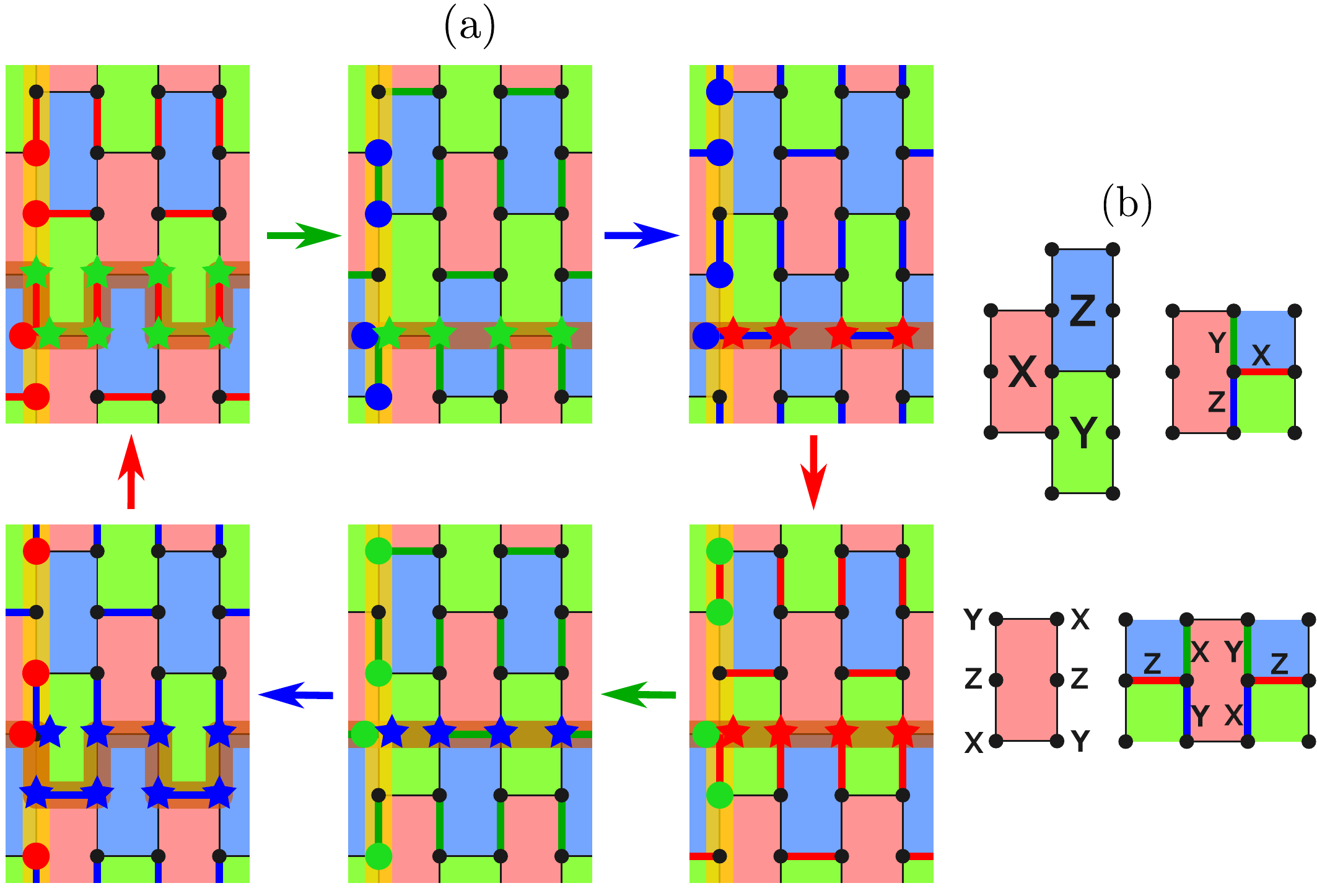}
\caption{\textbf{The honeycomb code measurement cycle, logical operators, plaquettes and edges.} 
(a) A distance-4 (under code-capacity noise) P$^6$ code~\cite{Gidney_honeycomb_2021} is shown throughout its evolution. The type of check measured at each measurement subround is indicated by the arrows and the highlighted/coloured edges in the lattice. One set of anti-commuting logical operators is shown in yellow and brown-coloured paths, where their qubit supports are depicted using big circles and stars, respectively. Each logical is a product of Pauli operators on the coloured qubits along its cycle, where red, green and blue qubits support $X$, $Y$, and $Z$ Pauli bases, respectively. The other set of logical operators is offset by three measurement subrounds. (b) Top: Red, green and blue edges/plaquettes of the P$^6$ honeycomb code support $X$, $Y$ and $Z$ operators, respectively. Bottom: All plaquettes of the XYZ$^2$ honeycomb code support the same XYZ$^2$ operator, as shown on the left. Edge operators are defined based on their orientation in the T junction (rather than the edge colour), as shown on the right.}
    \label{fig:honeycomb_code}
\end{figure}

\subsection{CSS Floquet code}\label{app:CSS}

In Supplementary Fig.~\ref{fig:CSS_code_mmt_cycle}, we illustrate the measurement cycle of the CSS Floquet code~\cite{Davydova_2023_floquet_codes_without_parents,Anyon_condensation_2024} and the resulting evolution of (one set of) its logical operators. Its logical operators are either of $Z$-type or $X$-type for all time, as shown. We show a $Z$-type logical operator defined on a vertical non-trivial cycle of the lattice. However, there exists another $X$-type logical operator (not shown) defined on that same vertical cycle. One can find this operator (and its anti-commuting partner) by noting the symmetry in the code obtained by proceeding forward by three measurement subrounds ($t\mapsto t+3$) and applying a Hadamard gate to all qubits ($X\leftrightarrow Z$). 

The plaquette operator members of the ISG at each step of the measurement cycle are indicated in Supplementary  Fig.~\ref{fig:CSS_code_mmt_cycle}. The value of each plaquette operator is inferred, in one time step, from the product of the measurement values of three edges  surrounding the plaquette. %
At each measurement subround, only one kind [which is of one type ($X$ or $Z$) and one colour] of plaquette is reinitialised and one other kind is re-measured to form a detector. The reinitialised and re-measured plaquettes have the same ($X$ or $Z$) type as the checks performed at that particular subround, but are of different colours from each other and also from the checks~\cite{Davydova_2023_floquet_codes_without_parents,Anyon_condensation_2024}. The colours of the reinitiliased and re-measured plaquettes are respectively the same as the colours of the checks performed just before and immediately after the current subround.

To see how detectors are formed in this code, let us follow a single blue $X$ plaquette through the measurement sequence. After red $X$ checks (top left of Supplementary  Fig.~\ref{fig:CSS_code_mmt_cycle}), the blue $X$ plaquette operator is measured (we infer its outcome by multiplying together the check measurement outcomes around its boundary). This initialises the plaquette. It is only re-measured after the green $X$ check measurements, four subrounds later, at which point a detector is formed. After this, blue $Z$ checks are measured, which anti-commute with the blue $X$ plaquette operator. Hence, its measurement outcome after this subround is random. 

\begin{figure}
    \centering
    \includegraphics[width=\linewidth]{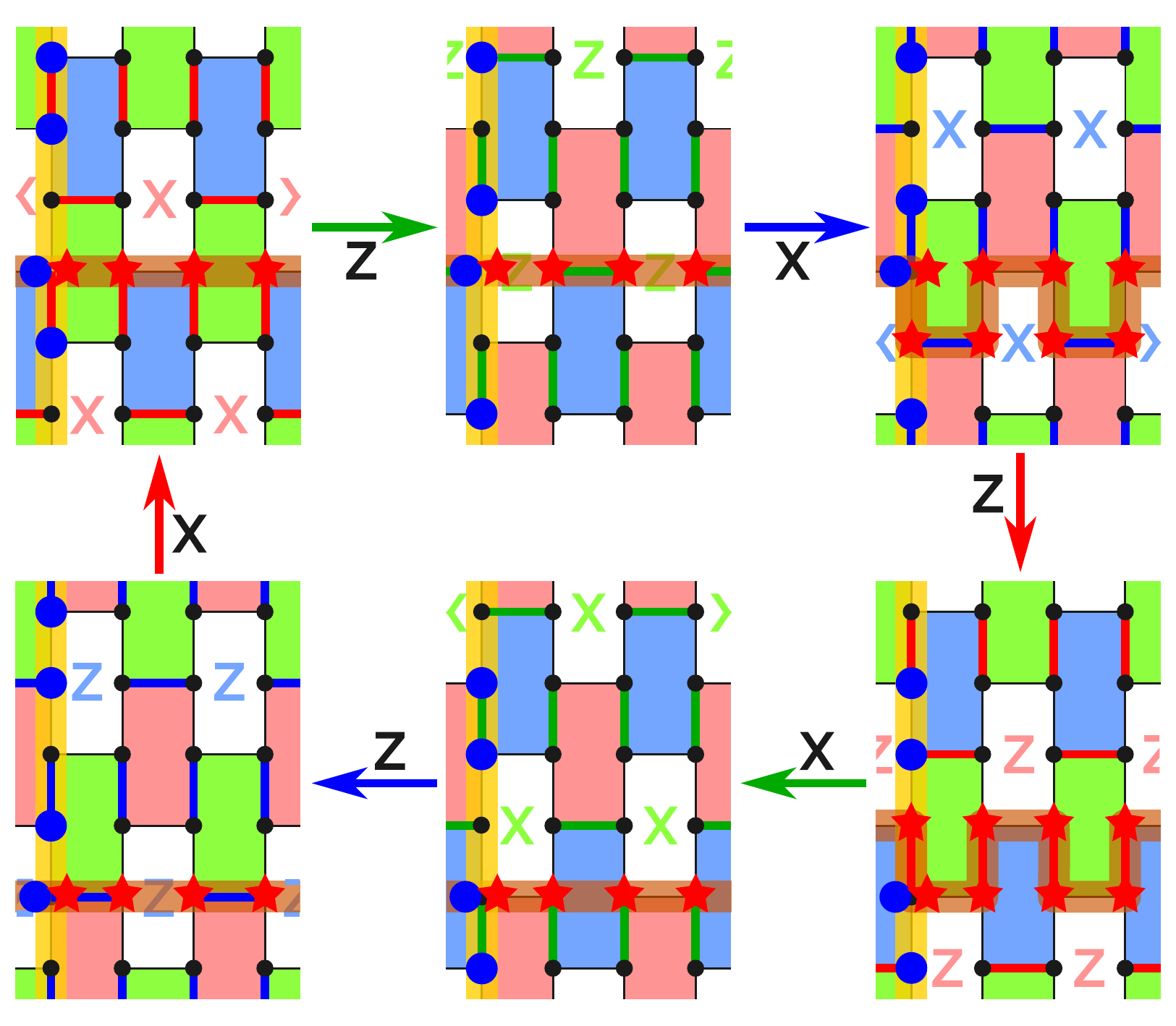}
    \caption{\textbf{Measurement cycle and logical operator evolution for the CSS Floquet code.} $X$ or $Z$ edge operators are measured alternately, with the colour of edge and Pauli type indicated by the labelled arrows. Plaquettes labelled with $X$ or $Z$ labels host only a stabiliser operator of the corresponding type, while other plaquettes host both $X$- and $Z$-type stabilisers. One set of anti-commuting logical operators is shown by yellow and brown paths, where their qubit supports are depicted using big circles and stars, respectively. Each logical is a product of only one type of Pauli operators, either $X$ or $Z$, represented by the red or blue coloured qubits.}
    \label{fig:CSS_code_mmt_cycle}
\end{figure}

\section{Parity-check measurement circuits}

\subsection{Error propagation}
\label{subsec:bias_preservation_and_degradation}
\begin{figure}
    \centering
    \includegraphics[width=\linewidth]{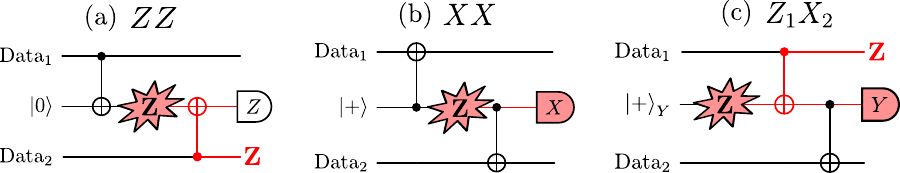}
    \caption{\textbf{Error propagation in the depth-2 parity-check circuits of Fig.~4 in the main text.} Shown are the examples of $\mathbf{Z}$ errors occurring at certain space-time points of (a) $ZZ$, (b) $XX$, and (c) $Z_1X_2$ parity check circuits. These errors can spread to at most single-qubit $Z$ errors on the data qubits and flip certain measurement outcomes (shown in red).}
    \label{fig:circuits_with_noise_1}
\end{figure}

\begin{figure}
    \centering
    \includegraphics[width=\linewidth]{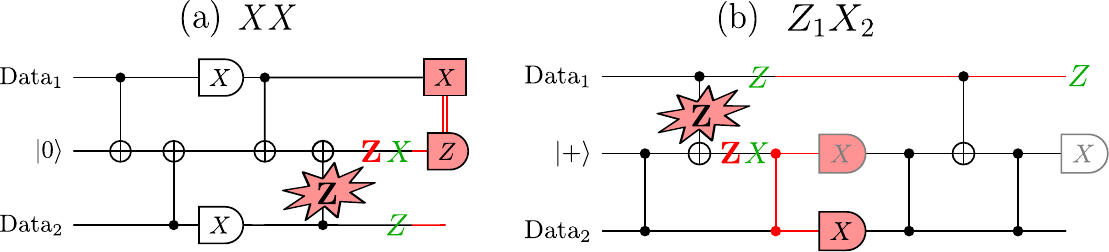}
    \caption{\textbf{Error propagation in the parity check circuits of Fig.~5 in the main text.} Shown are examples of $\mathbf{Z}$ errors occurring on the target qubits during the application of a CNOT gate in two different parity-check circuits: (a) $XX$ and (b) $Z_1X_2$. Green Paulis represent Pauli errors occurring in a specific branch of the state vector, while red Paulis occur in all branches. Red lines track the spread of Pauli errors and red measurements may be flipped by the errors.}
    \label{fig:circuits_with_noise_2}
\end{figure}

\begin{figure*}[t!]
\includegraphics[width=\linewidth]
{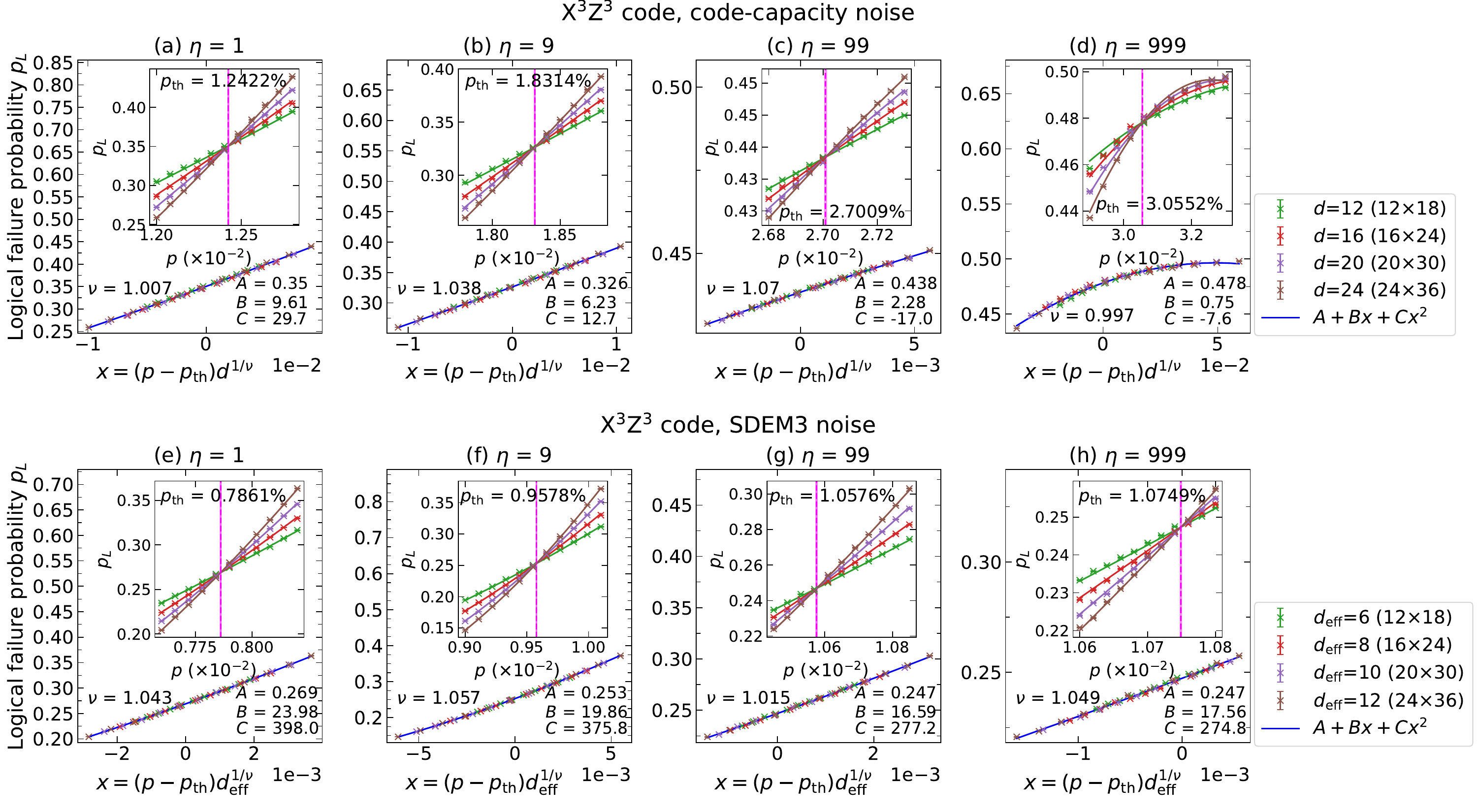}
\caption{\textbf{Logical failure probability $p_L$ of the X$^3$Z$^3$ Floquet code as a function of $x = (p-\pth)d^{1/\nu}$ or $x = (p-\pth)\deff^{1/\nu}$.} They are calculated for two different noise models: Code capacity (upper panels) and SDEM3 (lower panels), with various bias strengths: (a,e) $\eta = 1$, (b,f) $\eta = 9$,
(c,g) $\eta = 99$, and (d,h) $\eta = 999$. We fit the results to the function $A+Bx+Cx^2$ (blue line). Insets: Logical failure probability $p_L$ vs physical error rate $p$; thresholds $\pth$ are shown by the magenta vertical lines. Results for each distance $d$ and physical error rate $p$ are averaged over $10^6-10^7$ number of shots.}
\label{fig:Harr_fit_X3Z3}
\end{figure*}

\begin{figure*}[t!]
\includegraphics[width=\linewidth]
{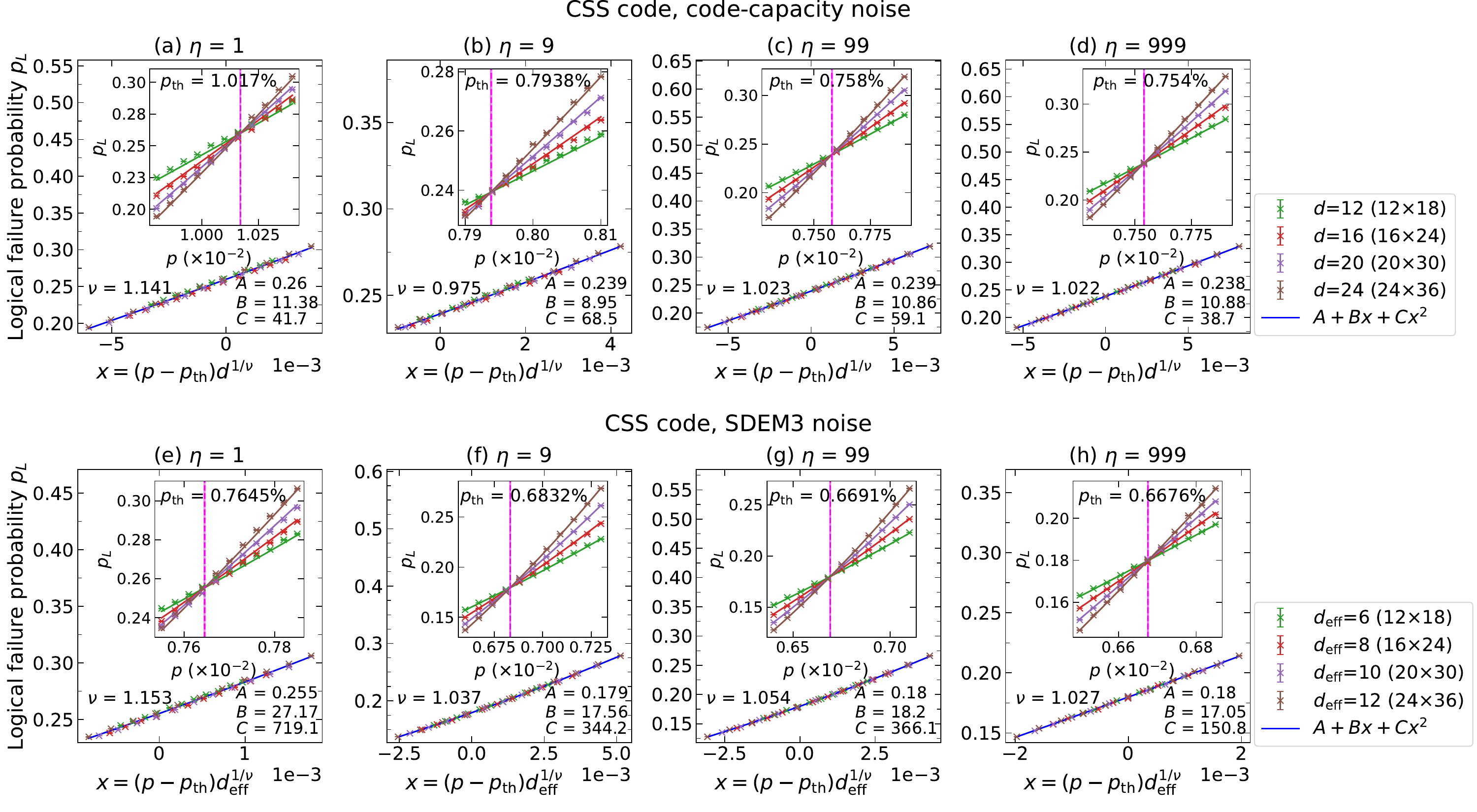}
\caption{\textbf{Logical failure probability $p_L$ of the CSS Floquet code as a function of $x = (p-\pth)d^{1/\nu}$ or $x = (p-\pth)\deff^{1/\nu}$.} They are calculated for two different noise models: Code capacity (upper panels) and SDEM3 (lower panels), with various bias strengths: (a,e) $\eta = 1$, (b,f) $\eta = 9$,
(c,g) $\eta = 99$, and (d,h) $\eta = 999$. We fit the results to the function $A+Bx+Cx^2$ (blue line). Insets: Logical failure probability $p_L$ vs physical error rate $p$; thresholds $\pth$ are shown by the magenta vertical lines. Results for each distance $d$ and physical error rate $p$ are averaged over $10^6$ number of shots.}
\label{fig:Harr_fit_CSS}
\end{figure*}

\begin{figure*}[t!]
\includegraphics[width=\linewidth]
{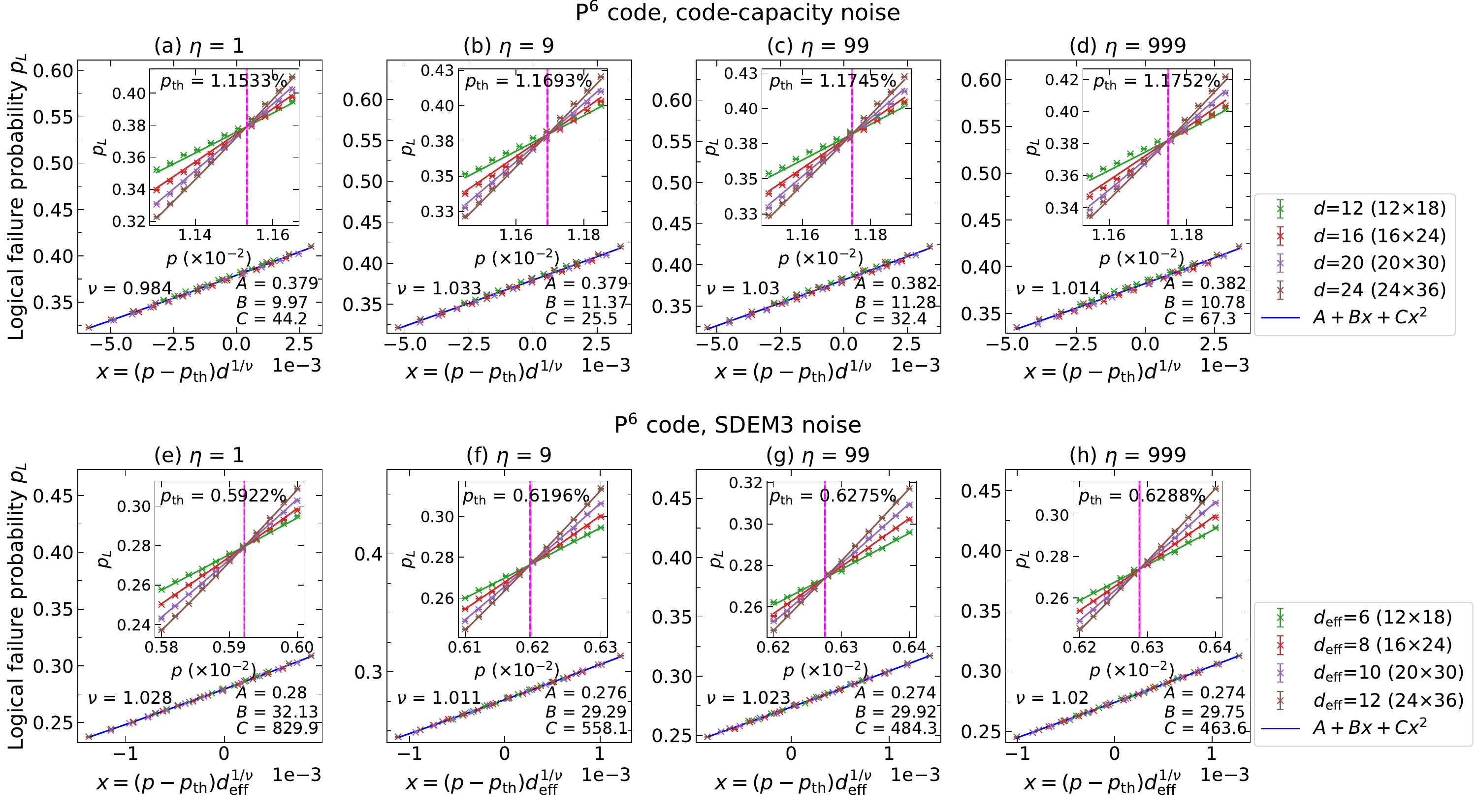}
\caption{\textbf{Logical failure probability $p_L$ of the P$^6$ honeycomb code as a function of $x = (p-\pth)d^{1/\nu}$ or $x = (p-\pth)\deff^{1/\nu}$.} They are calculated for two different noise models: Code capacity (upper panels) and SDEM3 (lower panels), with various bias strengths: (a,e) $\eta = 1$, (b,f) $\eta = 9$,
(c,g) $\eta = 99$, and (d,h) $\eta = 999$. We fit the results to the function $A+Bx+Cx^2$ (blue line). Insets: Logical failure probability $p_L$ vs physical error rate $p$; thresholds $\pth$ are shown by the magenta vertical lines. Results for each distance $d$ and physical error rate $p$ are averaged over $10^6$ number of shots.}
\label{fig:Harr_fit_Gidney}
\end{figure*}
    
\begin{figure*}[t]
\includegraphics[width=\linewidth]
{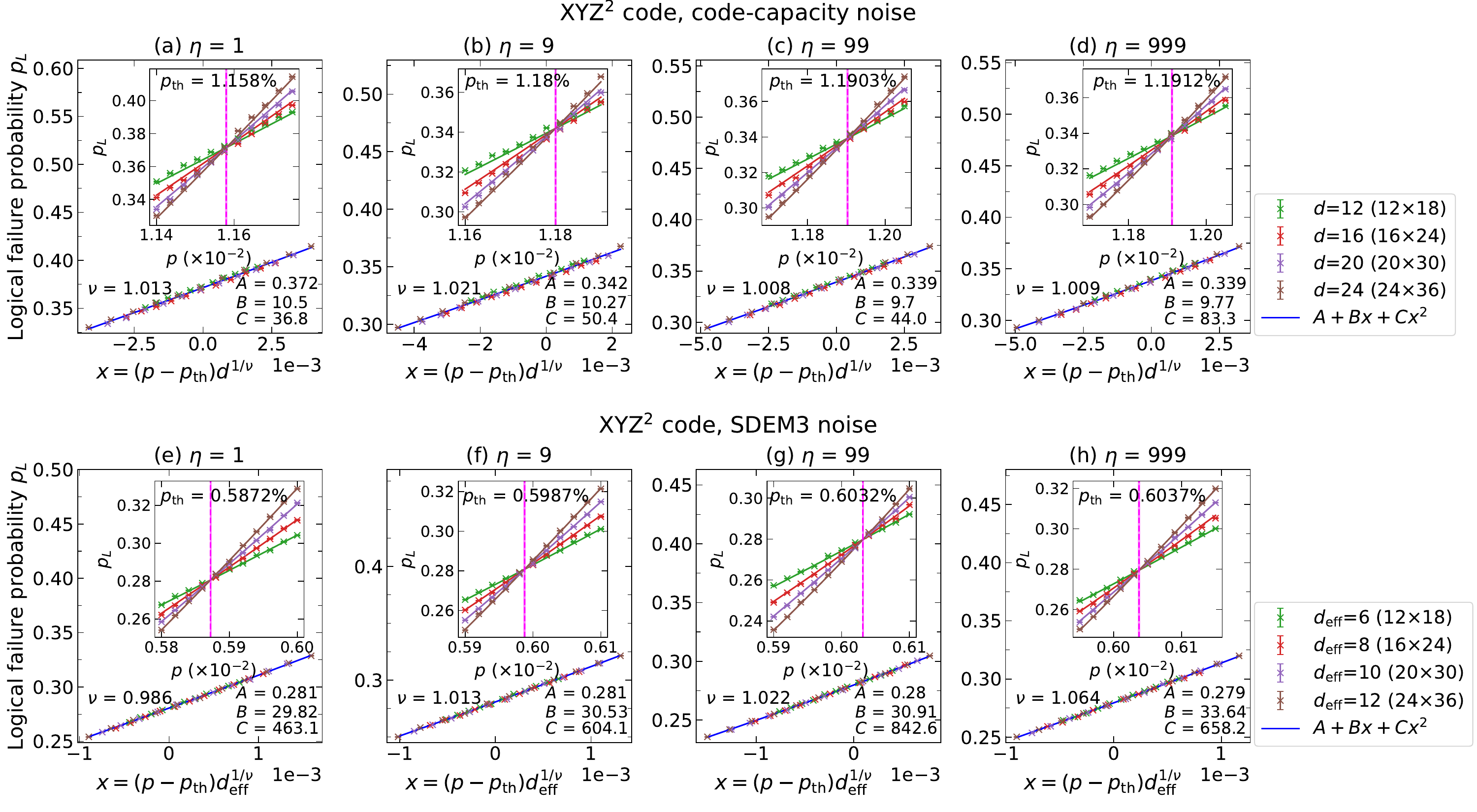}
\caption{\textbf{Logical failure probability $p_L$ of the XYZ$^2$ honeycomb code as a function of $x = (p-\pth)d^{1/\nu}$ or $x = (p-\pth)\deff^{1/\nu}$.} They are calculated for two different noise models: Code capacity (upper panels) and SDEM3 (lower panels), with various bias strengths: (a,e) $\eta = 1$, (b,f) $\eta = 9$,
(c,g) $\eta = 99$, and (d,h) $\eta = 999$. We fit the results to the function $A+Bx+Cx^2$ (blue line). Insets: Logical failure probability $p_L$ vs physical error rate $p$; thresholds $\pth$ are shown by the magenta vertical lines. Results for each distance $d$ and physical error rate $p$ are averaged over $10^6$ number of shots.}
\label{fig:Harr_fit_HH}
\end{figure*}

Here we provide more details on the propagation of errors, including mid-gate errors, in the parity-check circuits shown in Figs.~4 and~5 of the main text. We first assume there are no mid-gate errors, and demonstrate that the parity check circuits in Fig.~4 of the main text is bias-preserving. To this end, in Supplementary Fig.~\ref{fig:circuits_with_noise_1} we provide examples of $\mathbf{Z}$ errors on ancilla qubits and show that these errors can only spread to the data qubit as a $\mathbf{Z}$ error; thus, the strength of the noise bias is not degraded.

When considering mid-gate errors, the analysis is more involved. To see how the mid-gate errors propagate, we consider  a CNOT gate implemented via the following Hamiltonian~\cite{Puri_2020_bias_preserving}:
\begin{align}\label{eq:HCX}
    H_\text{CX} = V \frac{1}{2}[(\mathds{1}_c+Z_c)\otimes \mathds{1}_t + (\mathds{1}_c-Z_c)\otimes X_t],
\end{align}
where $V$ is the interaction strength between the control $c$ and target qubit $t$. This Hamiltonian gives rise to a unitary gate $U(T) = \exp(-iTH_\text{CX})$ at time $T$ where for $VT = \pi/2$, the resulting unitary is a CNOT gate. Note that a phase-flip error $Z_c$, occurring on the control qubit $c$ during the application of the CNOT gate, always propagate as a $Z_c$ error since it commutes with the Hamiltonian $H_{\text{CX}}$ [Supplementary Eq.~\eqref{eq:HCX}]. On the other hand, a mid-gate phase flip error $Z_t$ on the target qubit can propagate as a phase flip on the control qubit and a combination of phase flip and bit flip error on the same target qubit, as shown below. When the mid-gate error $Z_t$ happens at time $0<\tau < T$, the resulting unitary rotation of the two qubits is:
\begin{align}\label{eqn:unitary_error}
   & U(T- \tau) Z_t U(\tau) = Z_t e^{iV(T-\tau)(\mathds{1}_c-Z_c)X_t}U(T)\nonumber \\
   &=  Z_t \left(\cos^2\theta + i\frac{1}{2}\sin2\theta (X_t - Z_cX_t) + \sin^2\theta Z_c \right) U(T),
\end{align}
where $\theta \equiv V(T-\tau)$.
The unitary error $\exp[iV(T-\tau)(\mathds{1}_c-Z_c)X_t]$ in Supplementary Eq.~\eqref{eqn:unitary_error} is an $X$ rotation on the target qubit $t$ controlled by qubit $c$. As shown by the terms in the parentheses in the last line of Supplementary Eq.~\eqref{eqn:unitary_error}, the resulting unitary error is as a superposition of different Pauli errors: $\mathds{1}$, $X_t$, $Z_c$ and $Z_cX_t$.

We now consider specifically the error shown in Supplementary  Fig.~\ref{fig:circuits_with_noise_2}(a), occurring during the final CNOT gate. The green Paulis represent errors occurring in a particular branch of the wavefunction. The red $\mathbf{Z}$ operator is the Pauli error that exists for all branches. As can be seen, the $X_a$ error on ancilla qubit $a$ flips the $Z_a$ measurement on the ancilla. Since the ancilla is used as the control qubit of the subsequent classically controlled X gate on data qubit 1, this $Z_a$ measurement flip results in an incorrectly applied $X_1$ gate on data qubit 1. Thus, a mid-gate $Z_a$ error can propagate to an $X_1$ error. In contrast, consider the mid-gate error shown in Supplementary Fig.~\ref{fig:circuits_with_noise_2}(b). The error may flip some measurement outcomes (shown in red), but the resulting Pauli errors after the circuit is simply a superposition of $\mathds{1}$ and $Z$ errors on one of the data qubits. Hence, the noise bias is not degraded.

\subsection{Bias preservation of $XX$ parity-check circuit}

The $XX$ parity-check circuit needs to contain at least one two-qubit gate which does not preserve the $Z$ bias on the data qubit. This seems to be a necessary feature of an $XX$ measurement circuit. To see this, note that (as described in the Results section of the main text) the value of $X_1X_2$ can be obtained through a single measurement (whose result is the outcome of the $X_1X_2$ measurement) or by measuring the $X_1P_a$ and $X_2P_a$ operators, where $P_a$ is the Pauli operator on some ancillary degree of freedom. In the former case, we require CNOT gates targetting data qubits [see Fig.~4(b) of the main text for the case where we measure the ancilla qubit and for the case where data qubit is measured, we need long-range CNOT gates between the data qubits analogous to Fig.~5(c) of the main text], which degrade the $Z$ bias on the target data qubits. In the latter case,  we need to use a non-bias preserving classically conditioned $X$ gate on one of the data qubits after we perform measurement on the data qubits to obtain the $X_1P_a$ and $X_2P_a$ parity values. In order for the $X_1P_a$ and $X_2P_a$ measurements not to reveal the values of $X_1$ and $X_2$ separately, the ancilla is prepared and later measured in the eigenstate of $\bar{P}_a$, some Pauli basis which anticommutes with $P_a$. In Fig.~5(b) of the main text, we choose $P_a = X_a$ and $\bar{P}_a = Z_a$. The ancilla measurement result is then used to perform the classically controlled X gate on one of the data qubits.

To see why a classically controlled gate is required in the $XX$ parity check circuit in Fig.~5(b) of the main text, we can follow the evolution of the wavefunction of the data qubits and ancilla after the $X_1X_a$ and $X_2X_a$ measurements. We consider the ancilla to be prepared in the state $|0\rangle$ and the two data qubits are initialised in the state 
\begin{equation}
\psiket = \alpha |00\rangle_{1,2} + \beta |01\rangle_{1,2} + \gamma|10\rangle_{1,2} + \delta |11\rangle_{1,2}.
\end{equation}
After the $X_1X_a$ and $X_2X_a$ measurements, the wavefunction evolves as
\begin{align}\label{eq:XX_wavefunction}
&\psiket |0\rangle \nonumber\\
&\xrightarrow[]{X_1X_a} \frac{1}{\sqrt{2}} \left( \psiket|0\rangle_{a} + (-1)^{m_{X_1X_a}} X_1 \psiket|1\rangle_a \right)\nonumber\\
&\xrightarrow[]{X_2X_a} \frac{1}{2} \left\{\left( \psiket + (-1)^{m_{X_1X_a}+m_{X_2X_a}}X_1X_2\psiket
\right)|0\rangle_a\right.
\nonumber\\
&\hspace{1.6cm}\left.+\left[(-1)^{m_{X_1X_a}} X_1 \psiket + (-1)^{m_{X_2X_a}} X_2 \psiket \right] |1\rangle_a\right\} \nonumber\\
& \hspace{0.9cm}= \frac{1}{\sqrt{2}} \left( \psikettilde |0\rangle_a + (-1)^{m_{X_1Xa}}X_1\psikettilde |1\rangle_a\right),
\end{align}
where
\begin{equation}
\psikettilde \equiv \frac{1}{\sqrt{2}}\left[\psiket + (-1)^{m_{X_1X_2}}X_1X_2\psiket \right],
\end{equation}
is the eigenvector of $X_1X_2$ with eigenvalue $m_{X_1X_2} = 0,1$. In evaluating the last line of Supplementary  Eq.~\eqref{eq:XX_wavefunction}, we have used the relation $(-1)^{m_{X_1X_2}} = (-1)^{m_{X_1X_a} +m_{X_2X_a}}$ since $(X_1 X_a)\cdot (X_2X_a) = X_1X_2$, where
$m_{X_1X_a},m_{X_2X_a} = 0, 1$ are the $X_1X_a$ and $X_2X_a$ measurement results, respectively. 

As can be seen from last line of Supplementary Eq.~\eqref{eq:XX_wavefunction},  the state of the data qubits after the $X_1X_a$ and $X_2X_a$ measurements is entangled with the ancilla qubit's state. Depending on the ancilla measurement result, we will obtain two different data-qubit states where these two states differ by a Pauli $X$ operator acting on one of the data qubits and a trivial global phase. As a result, we need an $X$ gate acting on one of the data qubits classically conditioned on the ancilla measurement to cancel this residual $X_1$ operator. This gate, however, does not preserve the noise bias on the target data qubit [see Supplementary Fig.~\ref{fig:circuits_with_noise_2}(a)]. In conclusion, this observation leads us to believe that constructing fully bias-preserving $XX$ check circuit using conventional two-qubit gates is impossible.

\subsection{Circuit-depth optimality}
Finally, we discuss the circuit-depth optimality of these circuits. If we assume that bias-preserving two-qubit gates~\cite{Shruti2023High} are available, it is clear that the depth-2 parity check circuits in Fig.~4 of the main text are optimal for \textit{weight-two} parity checks to be bias-preserving. Next, for the case where the bias-preserving parity check circuits are constructed using conventional two-qubit gates, we will argue that the parity check circuits shown in Fig.~5 of the main text are also already optimal in circuit depth, where further reduction in the depth of the circuits may not be possible without degrading the noise-bias preservation of the circuit. First of all, the $ZZ$ circuit of Fig.~5(a) in the main text and $Z_1X_2$ circuit on the left-hand side of Fig.~5(c) of the main text (the latter of which assumes the availability of direct connections between data qubits) have a circuit depth of 2 which is optimal for \textit{weight-two} parity checks. For the case where  long-range connectivity between data qubits is not available, we now argue that the depth-4 circuit shown in the last equality of Fig.~5(c) of the main text is a shortest-depth implementation of the circuit on the left-hand side of Fig.~5(c) of the main text. The argument is as follows. We first note that we cannot have a \textit{bias-preserving} $Z_1X_2$ measurement by measuring the ancilla qubit connecting the two data qubits.  
This is because such a circuit would require a CNOT gate targetting data qubit 2 [as in Fig.~4(c) of the main text], which does not preserve the noise bias. As a result, the bias-preserving $Z_1X_2$ measurement needs to be constructed as in Fig.~5(c) of the main text, where we measure one of the data qubits after  applying nearest-neighbour two-qubit gates between the ancilla and data-qubits in a bias-preserving way, i.e., CNOT gates targetting ancilla qubits or CZ gates. Since at least 2 nearest-neighbour gates are required to implement one next-nearest neighbour two-qubit gate and the $Z_1X_2$ circuit requires two such next-nearest neighbour gates, the depth-4 $Z_1 X_2$ circuit on the right-hand side of the last equality in Fig.~5(c) of the main text is therefore optimal.

We now comment on the tradeoff between using the parity-check circuits in Fig.~4 and Fig.~5 of the main text, for the case where bias-preserving two-qubit gates are unavailable. Compared to the parity check circuits in Fig.~4 of the main text, the circuits presented in Figs.~5(b) and~5(c) of the main text have better mid-gate error bias-preserving quality at the expense of additional gate depth. This trade-off will only be beneficial if the adverse effect of the extra noise due to longer gate depth is outweighed by the beneficial bias-preserving quality of the circuit, which is expected to be true when the noise bias is sufficiently large. To justify this, one  would therefore need to compare the performance of the bias-preserving and non-bias preserving parity check circuits by implementing them experimentally or simulating them using noise parameters specific to the device considered. We leave this study to future works.

\section{Threshold calculations}\label{app:thresholds}
Supplementary Figures~\ref{fig:Harr_fit_X3Z3}-\ref{fig:Harr_fit_HH} show the resulting data collapse and close-to-threshold plots of the logical failure probabilities  for the X$^3$Z$^3$, CSS, P$^6$ and XYZ$^2$ Floquet codes, respectively. The plots for each code are calculated for various bias strengths. As described in the Methods section of the main text, the thresholds are obtained by fitting the logical failure probability $p_L$ data taken for various physical error rates $p$ and code distances to the function $A + Bx + Cx^2$,  where $x = (p-\pth) d^{1/\nu}$ or $x = (p-\pth) \deff^{1/\nu}$ for the code-capacity and SDEM3 noise models, respectively. 
Here, $A,B,C$ are the fit parameters, $\nu$ is the critical exponent, $\pth$ is the threshold,  $d$ and $\deff$ are code distances for the code-capacity and SDEM3 noise models, respectively.

\section{Hyperedge errors in honeycomb codes under infinite noise bias}\label{app:hyperedges}

\begin{figure}
    \centering
\includegraphics[width=\linewidth]{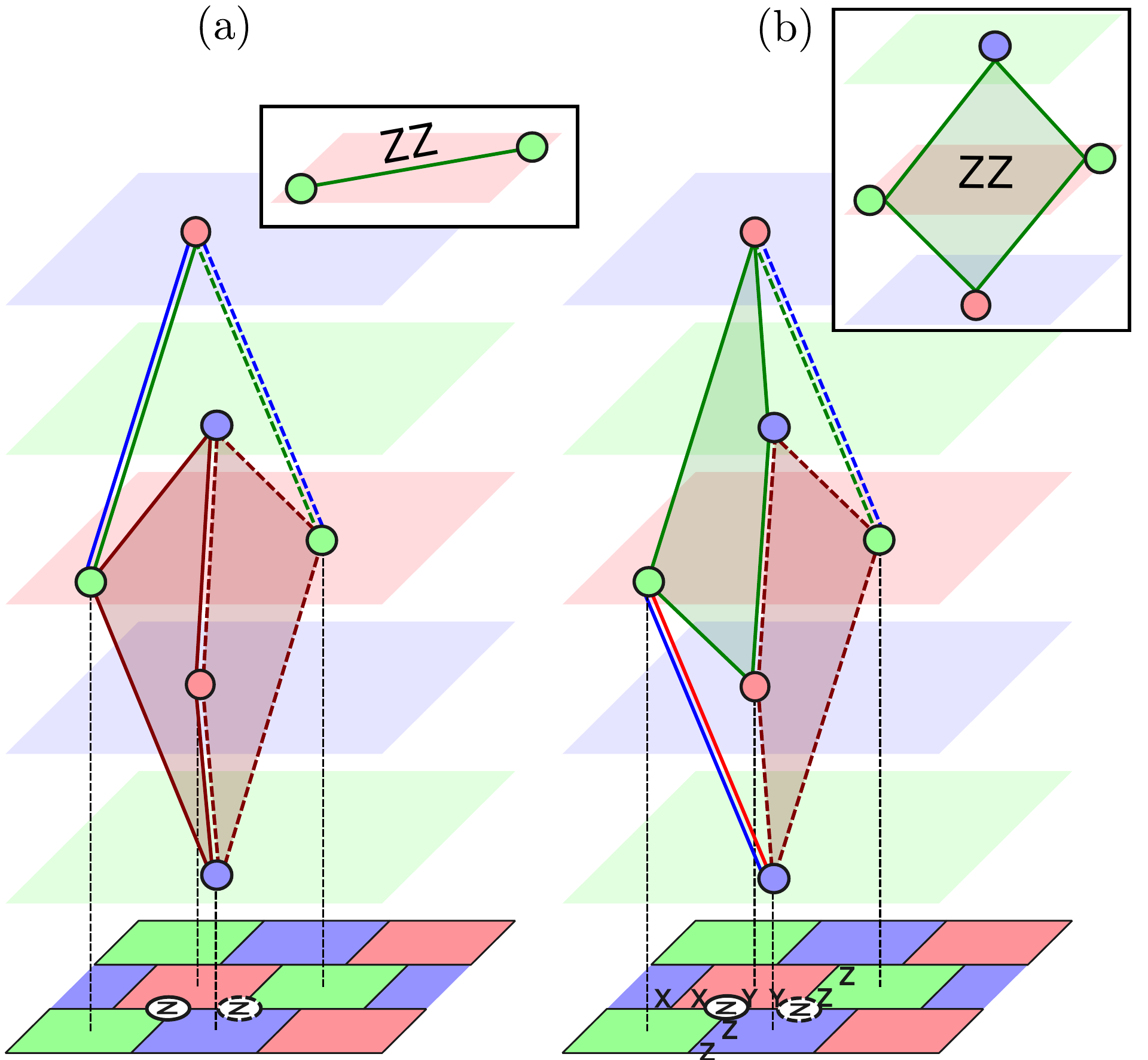}
\caption{\textbf{Detector hyperedges induced by $Z$ errors in honeycomb codes.} Shown are examples of detector hypergraphs of (a) P$^6$ and (b) XYZ$^2$ honeycomb codes. Coloured horizontal layers indicate type of measurements performed. Coloured nodes are detectors. Hyperedges and edges between detectors correspond to the syndromes of the single-qubit $Z$ errors shown by the uncoloured circles. Solid (dashed) edges correspond to syndromes due to the $Z$ error on the left (right) qubit. The colour of the (hyper)edge denotes the time step during which the $Z$ error occurs, e.g., a red hyperedge is the syndrome for a $Z$ error occurring immediately after red check measurements. Insets show the syndromes for the two-qubit $ZZ$ error occurring on the two qubits indicated (along a green edge) immediately after green check measurements. This syndrome is edge-like (hyperedge-like) for the P$^6$ (XYZ$^2$) code.}
\label{fig:hyperedge_detector_graph}
\end{figure}

In this section we explain the difference in performance between the XYZ$^2$ and P$^6$ codes. Specifically, we focus on the hyperedges formed in the detector hypergraphs under an infinitely phase-biased SDEM3 noise model. We will show that the frequency of hyperedges due to two-qubit errors is greater for the XYZ$^2$ code, resulting in an inferior performance under the biased SDEM3 noise, compared to the P$^6$ code.

Unlike the CSS and X$^3$Z$^3$ Floquet codes, the single-qubit $X$ and $Z$ Pauli errors in both of the honeycomb code variants  can produce hyperedge syndromes~\cite{fahimniya2024_hyperbolic_floquet,fu2024error} if they occur at specific measurement subrounds. These hyperedges degrade the performance of a matching decoder as they must be decomposed into edges, thereby losing some information about the correlations between detectors. We show in Supplementary  Fig.~\ref{fig:hyperedge_detector_graph} an example of such hyperedges which are formed from single-qubit $Z$ errors. Note that for a single-qubit $Z$ error to produce a hyperedge-like syndrome, it must occur between the measurements of the $XX$ and $YY$ edges of the $Z$-error affected qubit. 

First, let us consider the P$^6$ honeycomb code. Since in the P$^6$ code, the edge type is determined by its colour, a single-qubit $Z$ error here generates a hyperedge syndrome (four flipped detectors) if it occurs between red ($XX$) and green ($YY$) check measurements [see the hyperedge highlighted in red in Supplementary Fig.~\ref{fig:hyperedge_detector_graph}(a)] and an edge-like syndrome (two flipped detectors) if it occur at other measurement subrounds [shown by blue/green edges in Supplementary Fig.~\ref{fig:hyperedge_detector_graph}(a)]. Notice that the product of the two $Z$ errors occurring simultaneously always results in a graph-like syndrome whenever the error occurs [see inset of Supplementary Fig.~\ref{fig:hyperedge_detector_graph}(a)]. This is the same for all neighbouring $ZZ$ errors occurring along edges in the lattice.

In the XYZ$^2$ honeycomb code, by contrast, hyperedges can be formed from $Z$ errors occurring at  every measurement subround since edges of each colour consist of all check types ($XX$, $YY$ and $ZZ$). The hyperedge-forming $Z$ errors, however, need to occur at specific locations such that they happen between the measurements of the $XX$ and $YY$ edges of the noisy qubits. That is, a $Z$ error with coordinates $(x,y,t)$ could form a hyperedge syndrome for any $t$, given suitably chosen $x$ and $y$. Two such $Z$ errors and their syndromes for different time steps are shown in Supplementary Fig.~\ref{fig:hyperedge_detector_graph}(b). As can be seen, one $Z$ error (the error on the right) produces a hyperedge if it occurs after red check measurements, while the other (the error on the left) does so only after the green measurement subround. As such, a neighbouring $ZZ$ error occurring at a given time always produces a hyperedge syndrome. For example, if the $ZZ$ error shown (supported by a green edge) in Supplementary Fig.~\ref{fig:hyperedge_detector_graph}(b) occurs after the green MPP measurements, the resulting syndrome will include four detectors [see inset of Supplementary Fig.~\ref{fig:hyperedge_detector_graph}(b)].

In the SDEM3 noise model, two-qubit $ZZ$ errors along an edge can  occur only after the measurement of the corresponding edge operator. For example, a $ZZ$ error along a green edge can only occur after the green measurement subround. We have shown that such errors always produce edge-like syndromes for the P$^6$ code and hyperedge-like syndromes for the XYZ$^2$ code. On the other hand, the number of possible hyperedge-like syndromes from single-qubit $Z$ errors and measurement errors is the same for both codes. As a result, the total number of hyperedges is smaller for the detector hypergraph of the P$^6$ code, resulting in a better code performance.

\section{Performance of elongated CSS and X$^3$Z$^3$ Floquet codes}

\begin{figure*}[t]
\includegraphics[width=\linewidth]
{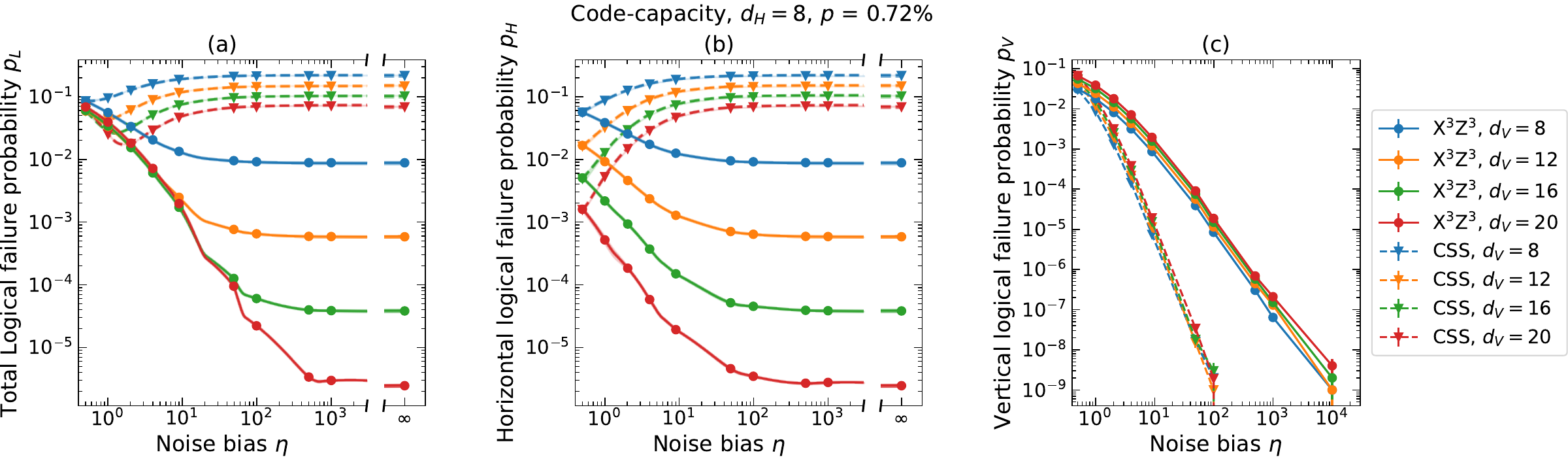}
\caption{\textbf{Subthreshold performance of elongated CSS and X$^3$Z$^3$ Floquet codes.} They are plotted as a function of noise bias $\eta$ for different vertical distances $d_V$. (a) Total logical failure probability $p_L$ [Eq.~(9) of the main text] calculated from (b) Horizontal logical failure probability $p_H$, and (c) Vertical logical failure probability $p_V$. Results are simulated under the code-capacity noise model for a fixed physical error rate $p = 0.72\%$ and horizontal distance $d_H = 8$. Each data point is averaged over $10^6-10^{10}$ number of shots.}
\label{fig:rectangular_X3Z3}
\end{figure*}

\begin{figure*}[t]
\includegraphics[width=0.75\linewidth]
{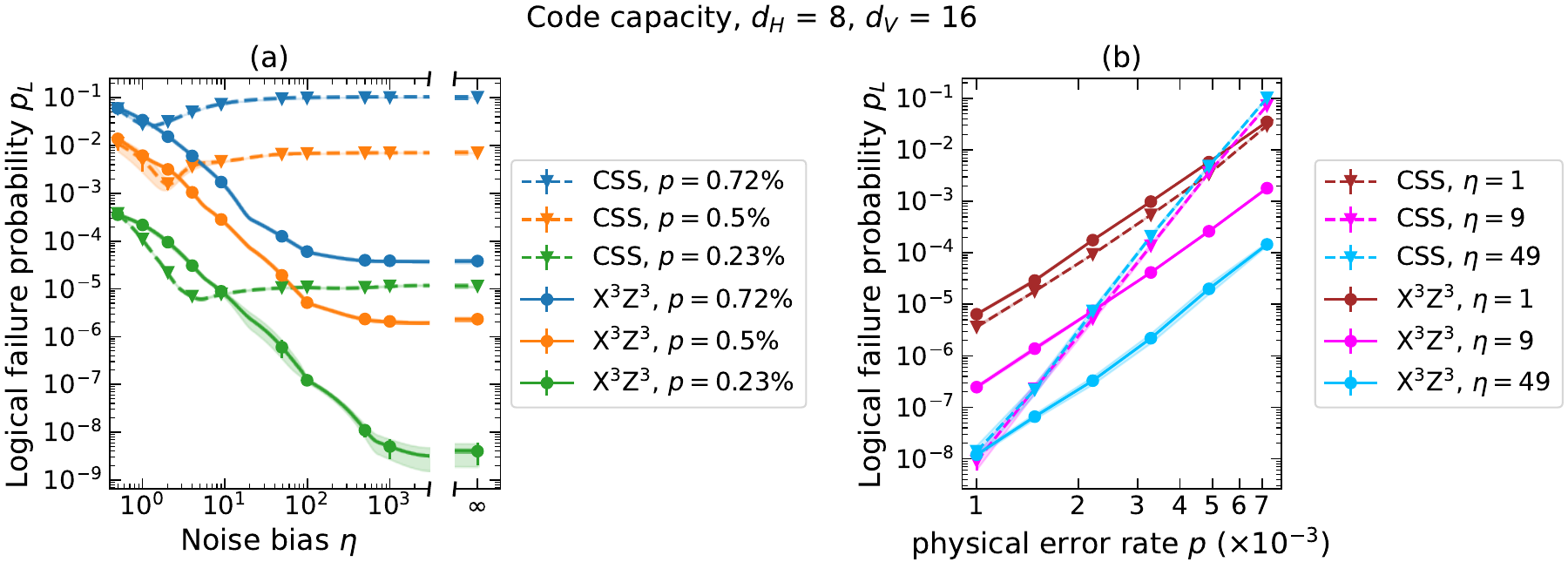}
\caption{\textbf{Logical failure probability of elongated CSS and X$^3$Z$^3$ Floquet codes.} They are plotted as a function of (a) noise bias $\eta$, and (b) physical error rate $p$. Results are calculated for the code-capacity noise model with code distances $d_H = 8$ and $d_V = 16$. Each data point is averaged over $10^8-10^{9}$ number of shots.}
\label{fig:rectangular_X3Z3_vs_p}
\end{figure*}

Supplementary Figure~\ref{fig:rectangular_X3Z3} shows the subthreshold performance of elongated CSS and X$^3$Z$^3$ Floquet codes with different aspect ratios $d_V/d_H$. The code distances $d_H$ and $d_V$ are defined as the minimum weight of the code's logical operators in the horizontal and vertical directions, respectively. The total logical failure probabilities plotted in Supplementary Fig.~\ref{fig:rectangular_X3Z3}(a) are calculated using Eq.~(9) in the main text, from the horizontal and vertical logical error probabilities shown in  Supplementary Figs.~\ref{fig:rectangular_X3Z3}(b) and (c), respectively.  As shown in Supplementary  Fig.~\ref{fig:rectangular_X3Z3}(a), both the CSS and X$^3$Z$^3$ Floquet codes have their logical errors improve with increasing $d_V$. Even though for a fixed lattice size, the CSS and X$^3$Z$^3$ Floquet codes have the same the horizontal and vertical code distances, these two codes have different performance under biased noise. In the noise regime near depolarising noise (and particularly for lower physical rate; see Supplementary Fig.~\ref{fig:rectangular_X3Z3_vs_p}), the logical error is slightly lower for the CSS code. On the other hand, as the noise becomes biased towards dephasing error, the X$^3$Z$^3$ code's logical error becomes increasingly better and can reach several orders of magnitude lower than those of the CSS Floquet code. 

The performance difference of these two codes in the two different noise regimes is due to the fact that for highly rectangular patches in the highly dephasing (depolarising) noise regime, the total logical error probability is dominated by the horizontal (vertical) logical failure probability, $p_H$ ($p_V$). For large $\eta$, $p_H$ is lower for the X$^3$Z$^3$ code, while for small $\eta$, $p_V$ is similar between the two codes [see Supplementary Figs.~\ref{fig:rectangular_X3Z3}(b) and (c)]. Indeed, Supplementary Fig.~\ref{fig:rectangular_X3Z3}(b) shows that as the noise bias increases, the horizontal logical error probability of the CSS code increases while that of the X$^3$Z$^3$ code decreases. The reason for this is that all the vertical logical operators in the CSS code are of pure Pauli $Z$s while half of the vertical logical strings for the X$^3$Z$^3$ code are of pure $X$-type. Therefore, as the $Z$ noise bias increases, the probability of the errors forming vertical logical strings that flip the horizontal logical observables increases for the CSS code while it decreases for the X$^3$Z$^3$ code. Furthermore, the X$^3$Z$^3$ code has a much simpler decoding problem, as explained in the main text.  

Supplementary Figure~\ref{fig:rectangular_X3Z3}(c) shows that as the $Z$ noise bias increases, the vertical logical error probability of the CSS code decreases much faster than that of the X$^3$Z$^3$ code. This is because the CSS code's horizontal logical operators, which flip the vertical logical observable, consist of only $X$ Paulis, and these pure $X$-horizontal operators become exceedingly less likely to form as the $Z$ noise bias increases. Meanwhile, each of the X$^3$Z$^3$ code's horizontal logicals comprise both $X$ and $Z$ Paulis. As a result, under $Z$-biased noise the probability of errors forming the horizontal logical strings that flip the $Z$-logical observables is lower for the CSS code. However, owing to $p_H\gg p_V$ for large $\eta$, the combined logical failure probability $p_L$ for the X$^3$Z$^3$ code is much lower than the CSS code, as can be seen in Supplementary Fig.~\ref{fig:rectangular_X3Z3}(a).

While our results are simulated for the periodic boundary case, the above conclusion is expected to hold also for the open boundary case. This is because the X$^3$Z$^3$ Floquet code with open boundary conditions has the same domains aligned in the same direction (i.e., the vertical shaded and unshaded strips) as in the periodic boundary condition case. As a result, pure $Z$-type logical operators are aligned along the elongated direction while mixed $X$/$Z$ operators lie in the transverse direction, just as is the case with the unrotated XZZX code [see Fig.~1(h) in Supplementary Ref.~\cite{XZZX_surface_code}]. Since changing the boundaries does not change the Pauli supports of the logical operators in the X$^3$Z$^3$ Floquet code, the likelihood of the errors forming logical string operators for the open boundary case will be similar to that of the periodic boundary case. This is in contrast to the rotated XZZX code where domains lie diagonally across the patch. As shown in Supplementary Ref.~\cite{forlivesi2024logical}, when a rectangular lattice is applied to a rotated XZZX code, it has a shorter Z-code distance which leads to a performance degradation.

\section{X\texorpdfstring{$^3$}{}Z\texorpdfstring{$^3$}{} Floquet code with twisted boundary conditions}\label{app:twisted_BCs}

\begin{figure}
    \centering
    \includegraphics[width=0.98\linewidth]{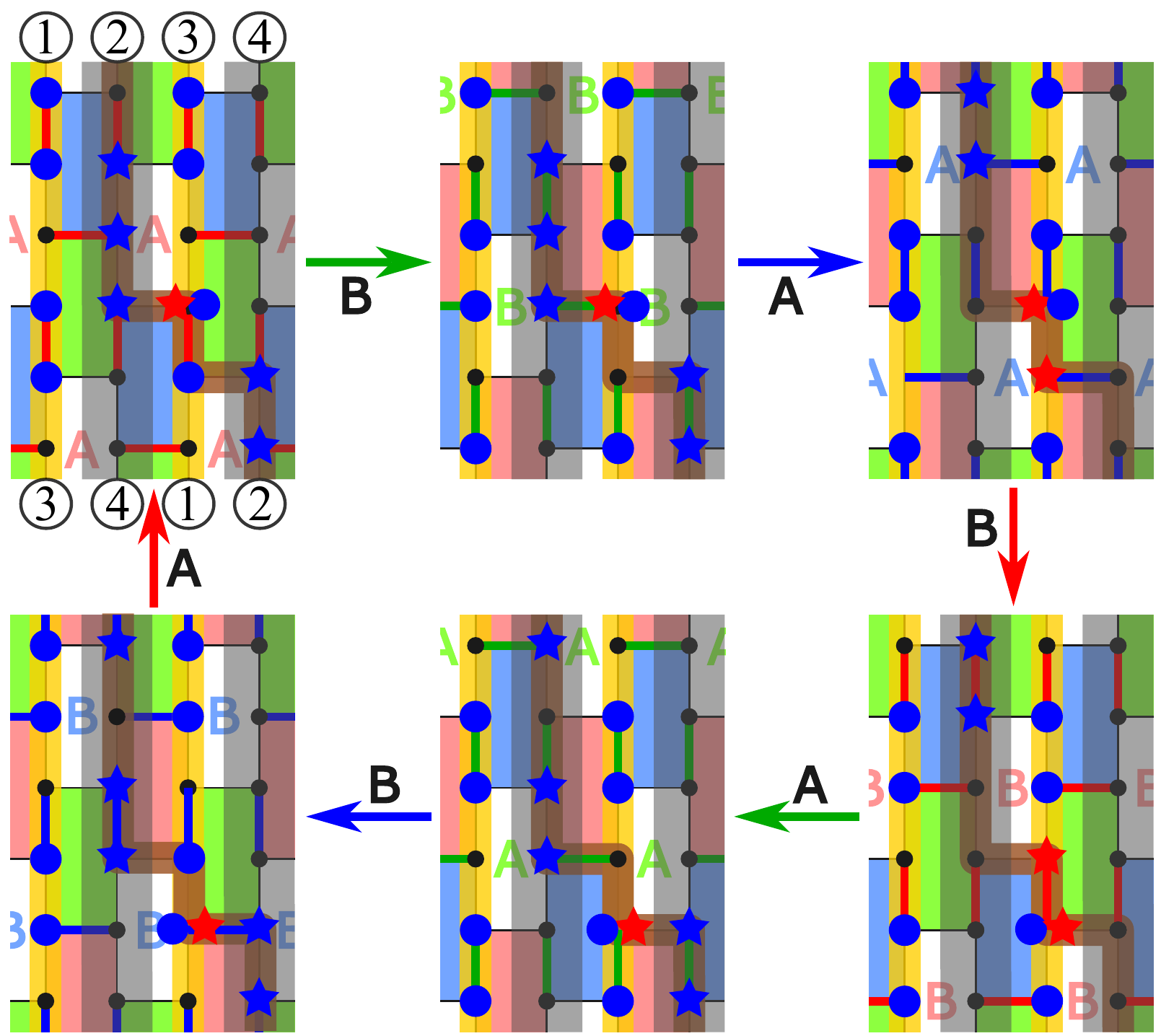}
    \caption{\textbf{Evolution of logical operators in the X$^3$Z$^3$ Floquet code with a twisted boundary condition.} Number-labelled edges  at the top and bottom boundaries indicate the twisted boundary condition. The logical operators are shown in yellow and brown non-trivial cycles, where their qubit supports are depicted using big circles and stars, respectively. Each logical is a product of Pauli operators on the coloured qubits along its cycle, where red and blue qubits support $X$, and $Z$ Pauli bases, respectively.}
    \label{fig:twisted_code}
\end{figure}

\begin{figure}
\includegraphics[width=\linewidth]{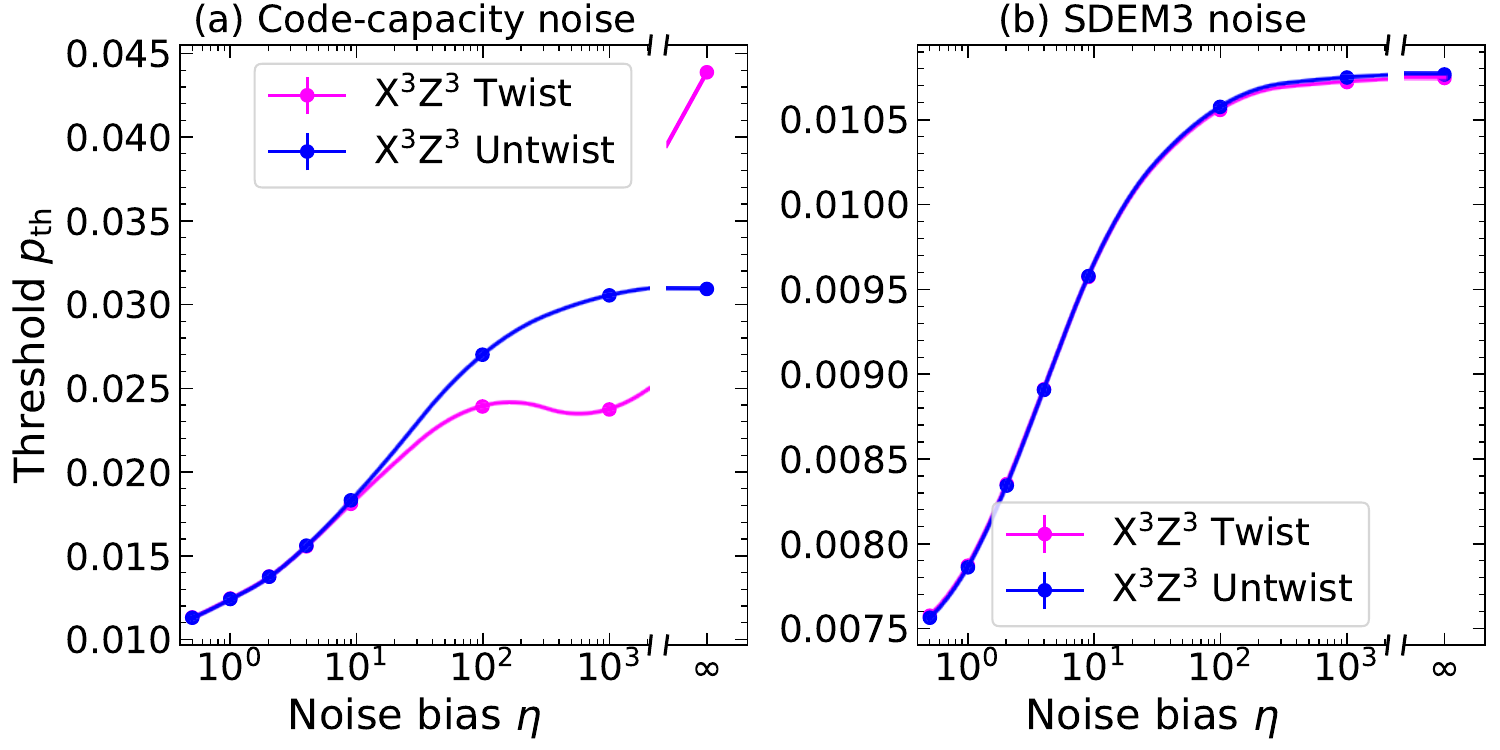}
	\caption{\textbf{Thresholds ($p_{\mathrm{th}}$) of the X$^3$Z$^3$ Floquet codes with different periodic boundary conditions.} Thresholds for twisted and untwisted periodic boundary conditions are shown in magenta and blue, respectively. They are plotted as a function of noise bias $\eta$ for two different noise models: (a) code-capacity and (b) SDEM3. We show some of the data points used in obtaining the thresholds in Fig.~\ref{fig:Harr_fit_twist_X3Z3}. 
   For better visualisation, we fit all curves with quadratic splines.  }\label{fig:Threshold_twist_X3Z3}
 \end{figure}

\begin{figure*}
\includegraphics[width=\linewidth]
{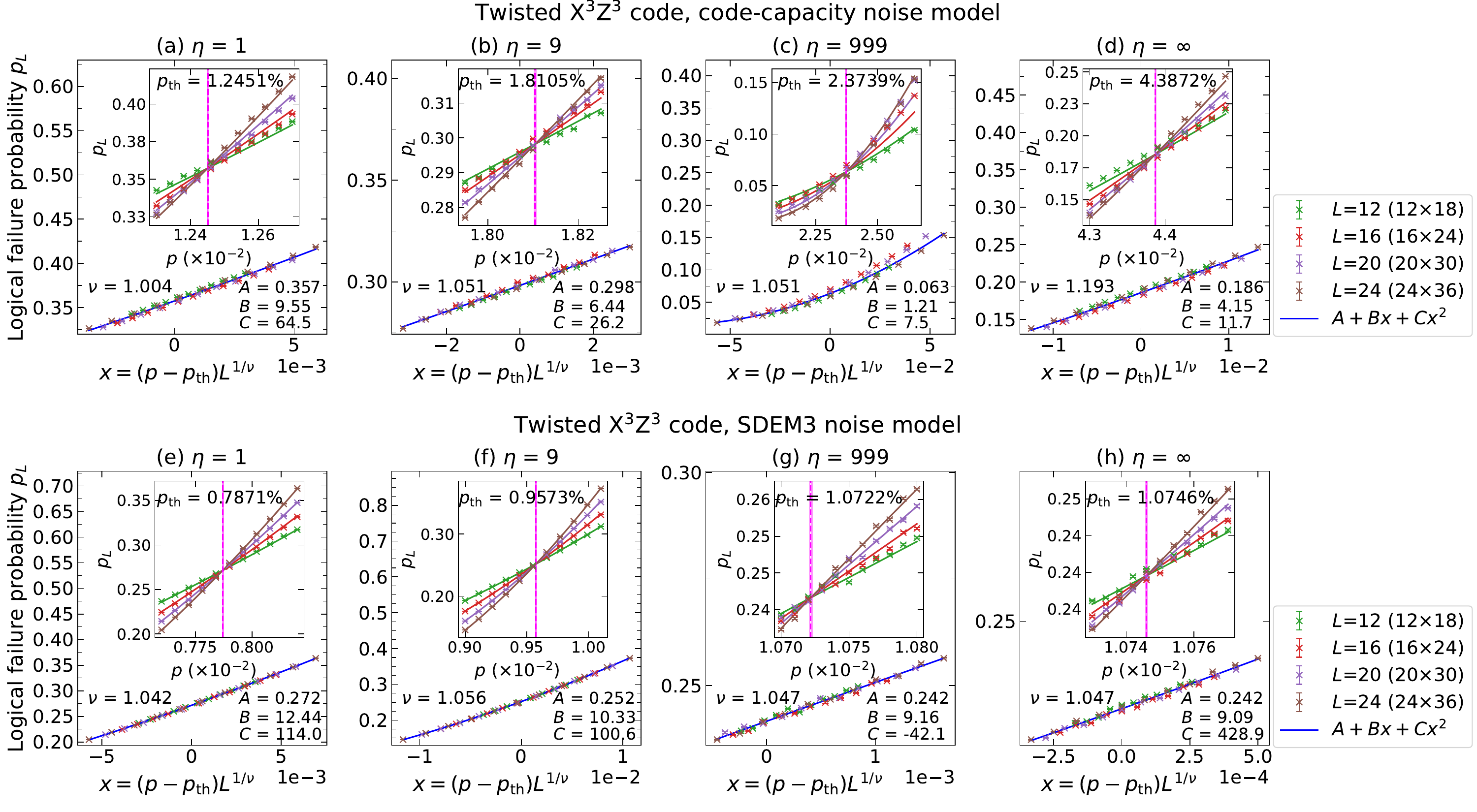}
\caption{\textbf{Logical failure probability $p_L$ of the X$^3$Z$^3$ Floquet code with twisted periodic boundary conditions as a function of $x = (p-\pth)L^{1/\nu}$.} Data shown are for different lattice lengths $L$ (lattice sizes $L\times 3L/2$). The plots are calculated for two different noise models: code capacity (upper panels) and SDEM3 (lower panels), with various bias strengths: (a,e) $\eta = 1$, (b,f) $\eta = 9$,
(c,g) $\eta = 999$, and (d,h) $\eta = \infty$. We fit the results to the function $A+Bx+Cx^2$ (blue line). Insets: Logical failure probability $p_L$ vs physical error rate $p$; thresholds $\pth$ are shown by magenta vertical lines. Results for each lattice length $L$ and physical error rate $p$ are averaged over $10^6-10^7$ number of shots.}
\label{fig:Harr_fit_twist_X3Z3}
\end{figure*}

In this section, we present simulation results obtained for the X$^3$Z$^3$ code with a twisted boundary condition. The twisted boundary refers to the case where the qubits at the top row of the honeycomb lattice are connected to qubits at the bottom row, shifted by two columns (see Supplementary Fig.~\ref{fig:twisted_code}). In the figure, we also show two (of the four) logical operators. The yellow logical operator, which consists of only $Z$ operators, has a length $L^2$/2 for a lattice with a size $L \times 3L/2$. This is longer than the length of the untwisted code logical operator, which is only $L$ (or $2L$) [cf. Fig.~1(c) of the main text]. As a result, the performance of the twisted code near infinite bias is expected to be better than the untwisted code since it will be harder for the pure dephasing errors to form such a long logical operator. 

\begin{figure*}
\includegraphics[width=0.75\linewidth]{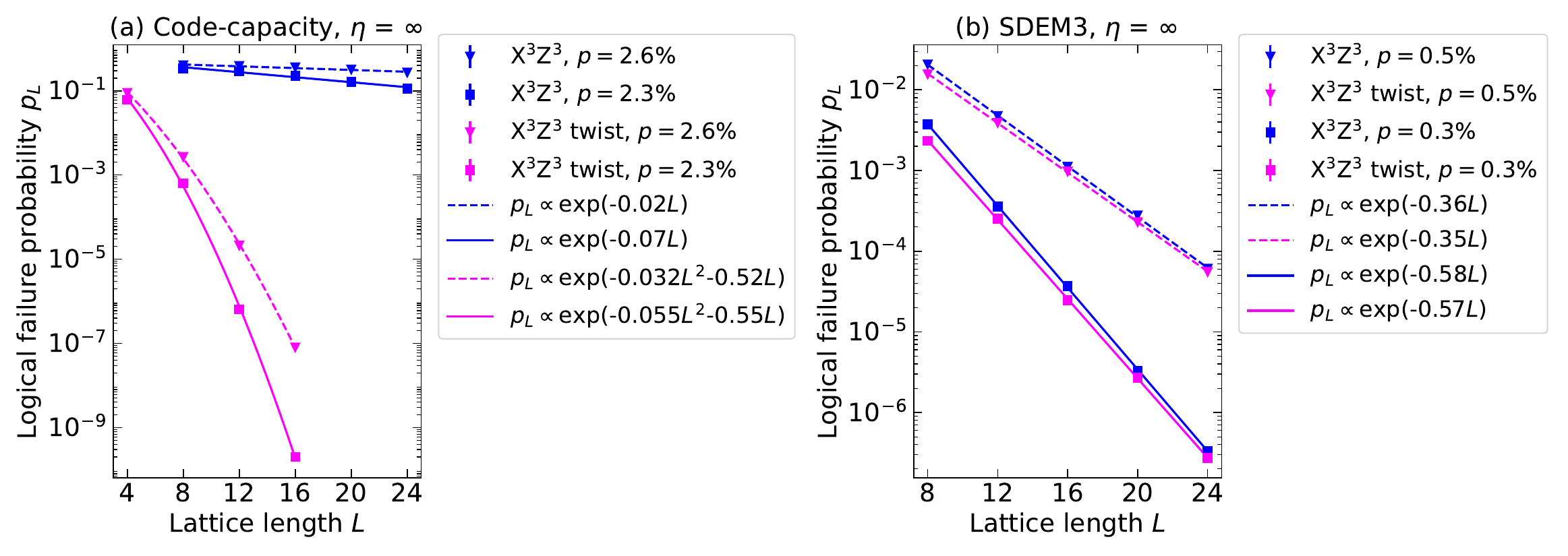}
	\caption{\textbf{Sub-threshold logical failure probabilities $p_L$ vs lattice length $L$ for X$^3$Z$^3$ Floquet codes with different periodic boundary conditions.} Data for untwisted and twisted code are shown in blue and magenta, respectively. Results are calculated for two different noise models: (a) code-capacity and (b) SDEM3 noise at infinite bias $\eta = \infty$. They are also computed with various physical error rates $p $. All curves can be fitted to decay functions $f \propto \mathrm{exp}(-\gamma_2 L^2 -\gamma_1 L)$ or $f \propto \mathrm{exp}(-\gamma_1 L)$ where $\gamma_{1,2}$ are increasing functions of $(p_{\mathrm{th}} - p)$. Each  data point is averaged over $10^5-10^{10}$ number of shots. }\label{fig:Error_distance_twist}
 \end{figure*}

To support this argument, we present the thresholds of such a code in Supplementary Fig.~\ref{fig:Threshold_twist_X3Z3}, where we compare them to the thresholds of the untwisted code. Data used to obtain the thresholds are presented in Supplementary Fig.~\ref{fig:Harr_fit_twist_X3Z3}. We find that there is a stronger finite-size effect for the twisted code compared to the untwisted one as can be seen from the finite-size collapse plots. As shown in Supplementary Fig.~\ref{fig:Threshold_twist_X3Z3}, the twisted code has better thresholds than the untwisted code only in the regime close to infinite bias. For the small noise bias regime ($\eta \lesssim 10$), the thresholds of the twisted code are roughly similar to those of the untwisted code. However, in the intermediate bias regime, the thresholds of the twisted code are lower than those of the untwisted code. 
 
Besides the thresholds, we also compute the logical failure probability of the twisted X$^3$Z$^3$ Floquet code as a function of the lattice length $L$. As can be seen from Supplementary  Fig.~\ref{fig:Error_distance_twist}, at infinitely phase-biased code-capacity noise the logical failure probability of the untwisted code decays as $p_L \propto \mathrm{exp} (-\gamma_1 L)$, 
whereas that of the twisted code decays as $p_L \propto \mathrm{exp} (-\gamma_2 L^2 -\gamma_1 L)$. 
However, when the SDEM3 noise model is considered, both the twisted and untwisted code have only an exponential decay. We note that $\gamma_1$ and $\gamma_2$ become larger for smaller subthreshold physical error rates. 

The super-exponential and exponential decay dependences of the twisted X$^3$Z$^3$ code's logical failure probability with respect to the lattice length $L$ under the code-capacity and SDEM3 noise models, respectively, are due to the fact that the code distances scale quadratically and linearly with the lattice length under these two different noise models. From the ``shortest graph-like error" generated via Stim~\cite{gidney2021stim}, we find that the measurement and Pauli errors in the SDEM3 model can conspire to produce logical error strings with lengths that scale linearly with the lattice length. Using Stim, we also calculate the size of the shortest graph-like error (i.e., the code distance) of the twisted X$^3$Z$^3$ code under the infinitely $Z$-biased noise and find the following relations:
\begin{subequations}
\begin{align}
d_{\mathrm{twist}} &= \frac{d_{\mathrm{untwist}}^2}{2} = \frac{L^2}{2}, \quad \\
d_{\mathrm{eff,twist}} &= 2d_{\mathrm{eff,untwist}}-1 = L-1, 
\end{align}
\end{subequations}
for the code-capacity and SDEM3 noise models, respectively.
Here $d_{\mathrm{twist}}$ and $d_{\mathrm{eff,twist}}$ are the code distances for the twisted X$^3$Z$^3$ code under the code-capacity and SDEM3 noise, respectively. Similarly, $d_{\mathrm{untwist}}$ and $d_{\mathrm{eff,untwist}}$ are the code distances defined for the untwisted code, while $L$ is the length of the honeycomb lattice with size $L \times 3L/2$. 
We note that for $\eta < \infty$ of both noise models, the size of the shortest graph-like error calculated from Stim is the same for both the untwisted and twisted codes.

In addition to the subthreshold scaling, as can be seen from Supplementary  Fig.~\ref{fig:Error_distance_twist}, for a fixed physical error rate under the infinitely phase-biased code-capacity noise, the logical failure probabilities of the twisted code are much lower than those of the untwisted case while they are roughly similar for the case of  SDEM3 noise. The reason is because the infinite-bias code-capacity threshold for the twisted code is higher than that of the twisted code while they are the same for the SDEM3 case, as shown in Supplementary  Fig.~\ref{fig:Harr_fit_twist_X3Z3}. 

\bibliography{References_NPJQI}